\documentclass[a4paper,11pt]{article}
\pdfoutput=1 

\usepackage{jheppub} 

\usepackage[T1]{fontenc} 

\newcommand{\refeq}[1]{(\ref{#1})}
\newcommand{\ssf}{\mathsf{s}}
\numberwithin{equation}{section}
\newcommand{\nn}{\nonumber \\}
\newcommand{\beq}{\begin{equation}}
\newcommand{\eeq}{\end{equation}}
\newcommand{\ba}{\begin{array}}
\newcommand{\ea}{\end{array}}
\newcommand{\bea}{\begin{eqnarray}}
\newcommand{\eea}{\end{eqnarray} }
\newcommand{\bal}{\begin{align}}
\newcommand{\eal}{\end{align}}

\newcommand{\ra}{\rangle}
\newcommand{\la}{\langle}
\newcommand{\Lcal}{\mathcal{L}}
\newcommand{\Bcal}{\mathcal{B}}
\newcommand{\Ocal}{\mathcal{O}}
\newcommand{\Dcal}{\mathcal{D}}
\newcommand{\Fcal}{\mathcal{F}}
\newcommand{\Rcal}{\mathcal{R}}
\newcommand{\Scal}{\mathcal{S}}
\newcommand{\Tcal}{\mathcal{T}}
\newcommand{\Sgra}{S^G}
\newcommand{\Pcal}{\mathcal{P}}
\newcommand{\Kcal}{\mathcal{K}}

\renewcommand{\d}{\partial}

\newcommand{\eps}{\epsilon}
\newcommand{\ssum}[2]{\scriptstyle\sum\limits_{#1}^{#2}\displaystyle}

\newcommand{\BC}[1]{\mathbb{BC}_{#1}}

\title{\boldmath Holographic renormalisation group flows and renormalisation from a Wilsonian perspective}

\author[a]{J. M.\ Lizana,}
\author[b]{T.\ R.\ Morris,}
\author[a]{and M.\ P\'erez-Victoria}

\affiliation[a]{CAFPE and Departamento de F\'{\i}sica Te\'orica y del Cosmos,  Universidad de Granada, Campus de Fuentenueva, E-18071, Granada, Spain}
\affiliation[b]{STAG Research Centre \& School of Physics and Astronomy, University of Southampton, Highfield, Southampton, SO17 1BJ, U.K.}

\emailAdd{jlizan@ugr.es}
\emailAdd{T.R.Morris@soton.ac.uk}
\emailAdd{mpv@ugr.es}


\arxivnumber{1511.04432}

\abstract{From the Wilsonian point of view, renormalisable theories are understood as submanifolds in theory space emanating from a particular fixed point under renormalisation group evolution. We show how this picture precisely applies to their gravity duals. We investigate the Hamilton-Jacobi equation satisfied by the Wilson action and find the corresponding fixed points and their  eigendeformations, which have a diagonal evolution close to the fixed points. The relevant eigendeformations are used to construct renormalised theories. We explore the relation of this formalism with holographic renormalisation. We also discuss different renormalisation schemes and show that the solutions to the gravity equations of motion can be used as renormalised couplings that parametrise the renormalised theories. This provides a transparent connection between holographic renormalisation group flows in the Wilsonian and non-Wilsonian approaches. The general results are illustrated by explicit calculations in an interacting scalar theory in AdS space.}

\begin{document} 
\maketitle
\flushbottom

\section{Introduction}
\label{sec:i} 
One of the most basic and interesting features of gauge/gravity dualities \cite{Maldacena:1997re,Witten:1998qj,Gubser:1998bc} is the holographic RG (renormalisation group), which relates the radial flow of classical gravity solutions in asympotically anti-de Sitter spaces and the RG evolution of their field-theory duals in the large-$N$ approximation \cite{Susskind:1998dq,Akhmedov:1998vf,Alvarez:1998wr,deBoer:1999xf}. The regions near the boundary of the space on which the gravity theory is defined correspond to the UV (ultraviolet) of the field theoy, while the deep interior of that space is related to its IR (infrared).\footnote{Whenever we use in this paper the terms UV and IR in the gravity theory they refer to the regimes in the field-theory dual and thereby to regions near and far from the boundary, respectively.} 
On the field theory side, the deepest understanding of renormalisation and the RG comes from a Wilsonian perspective~\cite{Wilson:1973jj,Polchinski:1983gv}, and therefore one might hope to understand holography itself at a deeper level through this framework (in the line, for instance, of~\cite{Heemskerk:2009pn,Albash:2011nq,Swingle:2012wq,Lee:2013dln,Balasubramanian:2013lsa,Behr:2015yna}). A number of attempts have been made to formulate holographic the RG in Wilsonian terms, but making this map precise has proved challenging.  

A first proposal of a holographic Wilsonian RG was made in Ref.~\cite{Balasubramanian:1999jd}, with the Wilson action given by the gravity action with an IR boundary cutoff, evaluated on solutions to the bulk equations. The solutions are determined by specific boundary conditions at the UV and IR ends of the space. As nicely explained in Ref.~\cite{Balasubramanian}, this is not yet a truly Wilsonian approach, as this Wilson action depends on physics below the IR cutoff. In Ref.~\cite{MPV}, one of us proposed to use as an effective action the cutoff gravity action evaluated on solutions with given UV conditions and Dirichlet conditions on the IR boundary. This object, which we call boundary action in this paper, is a functional of the restrictions of the bulk fields to the IR boundary. It only depends on UV data and can be used to calculate observables at large N by integration of the remaining degrees of freedom. The boundary action is the gravity counterpart of the Wilson action in field theory. The RG evolution of the sliding boundary action was studied in Ref.~\cite{Lewandowski:2004yr}.

Major progress has been made more recently by Heemskerk and Polchinski in Ref.~\cite{Polchinski}. These authors argue that the Wilson action itself, written as a functional of single-trace operators, is an integral transform of the boundary action. In the large-N limit, it reduces to a Legendre transform.\footnote{In Ref.~\cite{MPV} it was already argued that the analogous boundary actions obtained by integration of the IR degrees of freedom are related by a Legendre transform to the correlation functions of operators in the field theory with a UV cutoff. This is an IR version of the conjugate relation of the Wilson and boundary actions, which are UV objects.} Moreover, in Refs.~\cite{Polchinski} and~\cite{Faulkner:2010jy} it was shown that the holographic boundary and Wilson actions obey Hamilton-Jacobi equations that describe their dependence on the position of a sliding cutoff surface. (Beyond large N, they obey a Schr\"odinger equation.) This is a holographic formulation of the genuine Wilsonian RG. However,  as emphasised in Ref.~\cite{Polchinski}, the nature of the boundary cutoff on the field-theory side remains unknown.

In this paper, we explore in greater detail the precise relation between the Wilsonian RG in both sides of the holographic correspondence, in the strong 't Hooft coupling and large-N limits. We find fixed points of the RG/Hamilton-Jacobi evolution\footnote{Physically-relevant fixed-points were already found perturbatively in Ref.~\cite{Polchinski} in the potential approximation. Here we find them (also perturbatively in a field expansion) to all orders in derivatives. We also discuss the existence of non-analytic fixed points.} and study small deformations of them. The relevant deformations are used to construct holographic renormalised trajectories, following a standard field-theoretical treatment. We make an explicit connection between Wilsonian and renormalised (Gell-Mann-Low) RG flows and match the corresponding beta functions. In particular, this allows us to give a precise interpretation of the solutions of the gravity theory as running couplings in a specific renormalisation scheme.\footnote{A related interpretation, which we will discuss, was given in Ref.~\cite{Balasubramanian}.}  We also calculate perturbatively the renormalised boundary action in a scalar theory with a cubic interaction, and the Wilsonian and renormalised beta functions. The explicit calculations illustrate the general formalism. We pay special attention to certain subtle cancellations of the subdivergences in the three-point functions. The method of holographic renormalisation~\cite{deHaro:2000xn,Bianchi:2001kw,Skenderis:2002wp} plays a major role in many of these developments.

Wilsonian renormalisation group transformations relate a given Wilson action to another one with a lower cutoff. They involve two steps:  integration of the UV degrees of freedom and rescaling of all length scales in terms of the new cutoff. We examine the effect of this rescaling in the Hamilton-Jacobi evolution and show that it can be absorbed in a modified Hamiltonian. This allows us to extend the formalism to space-time dependent couplings. On the other hand, a strong limitation in this paper is that we work in a fixed AdS background. Therefore, we neglect the backreaction of the scalar fields on the geometry. This approximation necessarily breaks down in the IR of the renormalised theories, for any non-trivial theory. Nevertheless, we believe that our core insights in AdS already capture the essential features of the holographic Wilsonian RG in a complete treatment with dynamical gravity. In Section~\ref{sec:conclusions} we comment on some key ingredients that such a treatment will require.


The paper is organised as follows. We start in Section~\ref{sec:fieldtheory} with a review of the Wilsonian RG in the continuum context, the so-called exact renormalisation group \cite{Wilson:1973jj,Polchinski:1983gv,Morris:1998da,Latorre:2000qc, Bagnuls:2000ae} and show how various renormalisation schemes fit into this picture. Some of the observations we make here already seem to be new. 
Then, in Section~\ref{sec:holography}, we apply these ideas to a holographic Wilsonian description, starting with a review of Hamilton-Jacobi evolution, and paying particular attention to the inclusion of space-time dependent couplings and their derivatives. In Section~\ref{sec:schemes} we discuss various holographic renormalisation schemes beyond the UV scheme already introduced, with special emphasis on the possibility of using field solutions as renormalised couplings. In Section~\ref{sec:examples} we illustrate these ideas with explicit perturbative calculations in a scalar theory in AdS with a cubic interaction. Finally in Section~\ref{sec:conclusions} we draw our conclusions. Many general, somewhat technical results about fixed points and their deformations are derived and presented in the appendices.

\section{Wilsonian description of renormalisable theories: field theory}
\label{sec:fieldtheory} 

Let us consider a generic quantum field theory in $d$ Euclidean {dimensions. By definition we are therefore assuming that the description exists on all scales, {\it i.e.} has a continuum limit. However it is helpful to consider it} defined at some UV cutoff $\Lambda$ by a classical quasi-local Wilsonian action $S$, which can be written as
\beq
S = \int d^d x g^\alpha(x) \Ocal_\alpha(x),
\eeq
where $\Ocal_\alpha$ are $\Lambda$-independent local operators made out of the relevant quantum fields $\omega$ and their derivatives, with definite engineering mass dimension $\delta_{(\alpha)}$. They may obey symmetry constraints. We will concentrate on Lorentz scalar operators. We consider space-time dependent couplings, which in particular allows to extract correlations functions.
The partition function is obtained by functional integration of the fields $\omega$ with the UV cutoff $\Lambda$:
\beq
Z_\Lambda[g] = \int \left[\Dcal \omega \right]^\Lambda \, e^{-S_\Lambda(g)} . \label{partitionfunction}
\eeq
For the moment, we do not specify the nature of the cutoff (we will do it in the gravity side). 
    In the case of degenerate vacua for certain values of $g$, 
an extra condition can be imposed in \refeq{partitionfunction} to select a particular vacuum. A cutoff and a set of couplings defines in this way a particular quantum field theory, but this description is redundant.  To any pair of cutoff and couplings $(\Lambda_0,g_0)$, we can associate a one-dimensional family of pairs $(\Lambda,g)$ by integrating out the ``intermediate" degrees of freedom: if $\Lambda<\Lambda_0$,
\beq
e^{-S_\Lambda(g)} = \int \left[\Dcal \omega \right]^{\Lambda_0}_{\Lambda} \,e^{-S_{\Lambda_0}(g_0)}. \label{integrateout}
\eeq
If $\Lambda>\Lambda_0$, we just exchange in this equation $(\Lambda_0,g_0)$ and $(\Lambda,g)$ (assuming that $S_\Lambda(g)$ exists, which will generically be the case providing $\Lambda/\Lambda_0$ is not too large). 
The notation in the measure indicates that the path integral is performed with a UV cutoff $\Lambda_0$ and an IR cutoff $\Lambda$, chosen such that $\left[\Dcal \omega \right]^\Lambda \left[\Dcal \omega \right]^{\Lambda_0}_{\Lambda} = \left[\Dcal \omega \right]^{\Lambda_0}$ . This condition ensures that
\beq
Z_{\Lambda}[g]=Z_{\Lambda_0}[g_0] \label{invariantZ}
\eeq
in any vacuum, and thus all physical observables are the same. Besides the integration in~\refeq{integrateout}, the Wilsonian RG transformations involve another ingredient: scaling the cutoff back to its original size. Simpler and equivalent is to measure all variables in units of the cutoff. For this purpose, we   
write in the following the Wilson action as a functional of dimensionless couplings $\bar{g}$, which are functions of the dimensionless coordinates $\bar{x}^\mu = \Lambda x^\mu$:
\begin{align}
S(g) & = \bar{S} (\bar{g}) \nn 
& = \int d^d \bar{x} \bar{g}^\alpha(\bar{x}) \bar{\Ocal}_\alpha(\bar{x}) \nn
& \equiv \bar{g} \cdot \bar{\Ocal}. \label{action}
\end{align}
where we have introduced a basis $\{\bar{\Ocal}_\alpha\}$ of dimensionless operators,\footnote{As above we sum over pairs of repeated indices. The brackets, $(\beta)$, indicate the dependence on the index $\beta$ without the pairing. The pairing and thus summation is then indicated by the other two instances of $\beta$. (Occasionally, we will raise or lower indices with an implicit Euclidean metric.)}
\beq
\label{op-basis}
\bar{\Ocal}_\alpha(\bar{x})=
c_{\alpha}^{\beta} \, \Lambda^{-\delta_{(\beta)}} \Ocal_{\beta}(\bar{x}/\Lambda),
\eeq
When written in terms of fields made dimensionless with $\Lambda$, the operators $\bar{\Ocal}$ do not depend explicitly on $\Lambda$. Among these operators we include the identity operator, which contributes to the vacuum energy and will be useful in our formalism to absorb a local part of the breaking of scale invariance. We label this operator and its coupling with the index $\alpha=0$.
We will assume that some distance can be defined in the theory space given by all possible couplings, which we leave implicit.
We also redefine the partition function as
\beq
\bar{Z}[\bar{g}] = Z_{\Lambda}[g] . 
\eeq 
A change of variables in the functional integral shows that the left-hand side does not depend explicitly on $\Lambda$. Observe that, in terms of the new variables, any change of cutoff automatically involves a dilatation. In the remaining of this section we always work with dimensionless variables but drop the bars to simplify the notation. 

In terms of dimensionless variables~\refeq{invariantZ} reads
\beq
Z [g] = Z[g_0], \label{simpleZ}
\eeq
with $g$ and $g_0$ related by the dimensionless version of Eq.~\refeq{integrateout}.  This is the statement of RG invariance. The relation between $g$ and $g_0$ defines a Wilsonian RG flow in theory space, 
\beq
g^\alpha=f^\alpha_{\Lambda/\Lambda_0}(g_0). \label{RGflow}
\eeq
Here, $f^\alpha_{\Lambda/\Lambda_0}$ are quasilocal functionals, i.e. $f^\alpha_{\Lambda/\Lambda_0}(g_0)(x\Lambda_0/\Lambda)$ can be expanded as a infinite power series of $g_0^{\alpha}$ and their spatial derivatives in $x$. All RG flows are generated by the vector field 
 \beq
 \beta^\alpha _W(g) = \left. \frac{\d}{\d t} f^\alpha_t(g) \right|_{t=1}. \label{betaW}
 \eeq
 Eq.~\refeq{simpleZ} implies the differential RG equation
\beq
\beta_W(g) \cdot \frac{\delta Z[g] }{\delta g} = 0. \label{WRGeq}
\eeq

The fixed points $g_*$ of the flow have $\beta_W(g_*)=0$. Close to them, we can linearise the Wilsonian beta functions, which read
\beq
\beta^\alpha_W(g_*+\delta g) = -\lambda_{(\alpha)} \delta g^\alpha-D \delta g^\alpha+ O\left((\delta g^\alpha)^2\right) \label{betaWlinear},
\eeq
where $D$ is an infinitesimal dilatation, $[Df](x)=x^\mu \d_\mu f(x)$. Here and from now on we use a basis of couplings that diagonalizes the linearised flow around the fixed point of interest. The {operators \eqref{op-basis} are then the eigenperturbations such that}  $\lambda>0$, $\lambda=0$ and $\lambda<0$ correspond to relevant, marginal and irrelevant operators, respectively.

In this framework, the description of renormalisable theories is simple and intuitive. The simplest cases correspond to fixed points of the flow, which describe scale-invariant physics. More interesting renormalisable theories result from the linear combination of relevant and exactly marginal or marginally relevant  eigen-operators in \eqref{op-basis} about a particular fixed point which, modulo total derivative terms,\footnote{In other words, considering two operators equivalent if they differ by a total derivative, it is the quotient space that has dimension $r$.} span a vector space of finite dimension $r$. 
In these cases Wilsonian actions $S_\Lambda(g)$ exist no matter how large $\Lambda$ is taken, and thus describe the ``continuum limit''.  
The set of points that can be reached from these perturbed theories under RG evolution towards the IR, form the renormalised manifold $\Rcal$ of the given fixed point.\footnote{We mean the ``renormalised trajectories'' but we
will loosely regard the space $\Rcal$ as a manifold of dimension $r$, keeping in mind that singular behaviours such as boundaries are quite possible far from the fixed point or at the fixed point itself.}  Each integral curve of $\beta_W$ with image in $\Rcal$ defines a particular renormalisable theory, with definite physical predictions that do not depend on any cutoff. 

Consider a coordinate system of $\Rcal$, with dimensionless coordinates $g_R^a,~a=1,\ldots,r$. They will play the role of renormalised couplings. Different parametrisations $g(g_R)$ define different renormalisation schemes. Any RG integral curve in $\Rcal$ can be written as  $f_{\Lambda/\mu}(g(g_R))$ for some scale $\mu$ and renormalised couplings $g_R$. Therefore, $(\mu,g_R)$ defines a renormalised theory. Writing the points of these curves in terms of the coordinates, a flow $g_R^a \to \Fcal^a_{\Lambda/\mu}(g_R)$ is induced with local functionals evaluated at a space-time point $x$:
\beq
g(\Fcal_{\Lambda/\mu}(g_R)) = f_{\Lambda/\mu}(g(g_R)) . \label{inducedflow}
\eeq
The corresponding vector fields
\beq
\beta^a(g_R) = \left. \frac{\d}{\d t} \Fcal^a_{t}(g_R) \right|_{t=1}
\eeq
are local versions of the Gell-Mann-Low beta functions of the renormalised theory. They differ from the standard ones by the fact that they include the effect of the dilatation. These renormalised beta functions are related to the (local) Wilsonian ones, for points on $\Rcal$, by the chain rule:
\beq
\beta_W^\alpha(g(g_R)) =  \frac{\delta g^\alpha(g_R)}{\delta g_R} \cdot{} \beta(g_R). \label{betarelation}
\eeq
The renormalisation scale $\mu$ is required for dimensional reasons when a dimensionful cutoff $\Lambda$ is employed. 
Different choices of $\mu$ just amount to different parametrisations of the integral curves. A change in renormalisation scale $\mu\to \mu^\prime$ can be compensated by a change $g_R\to g_R^\prime = \Fcal_{\mu^\prime/\mu}(g_R)$ such that, thanks to the group property $\Fcal_{t} \circ \Fcal_{t^\prime} = \Fcal_{t t^\prime}$, the same integral curve is obtained. In this context, the functions $\Fcal$ play the role of running constants of the renormalised theory. 
The partition function of the renormalised theory, given by
\beq
Z^R [g_R] =  Z[g(g_R))], \label{renormalisedpartition}
\eeq
is invariant under renormalised RG flows:
\begin{align}
Z^R[\Fcal_{\Lambda/\mu}(g_R)] 
   & = Z[f_{\Lambda/\mu}(g(g_R))]  \nn
   & = Z[g(g_R)] \nn
   & = Z^R[g_R].
\end{align}
The differential version of this property is the Callan-Symanzyk equation for the renormalised theory,
\beq
\beta(g_R) \cdot \frac{\delta Z^R[g_R]}{\delta g_R} = 0.\label{CSeq}
\eeq
Note that the usual $\mu \d/\d \mu$ term in the Callan-Symanzyk equation is already taken into account in our formalism by the automatic dilatations.\footnote{The standard version of the Callan-Symanzyk equation can be found undoing the rescaling of coordinates: Defining $Z^R_{\mu}[g_R]=Z^R[\bar g_R]$, with $\bar g_R(x)=g_R(x/\mu)$, Eq.~\refeq{CSeq} takes the form
\begin{equation}
\left[ \mu \frac{\partial}{\partial \mu}+\tilde\beta(g_R) \cdot \frac{\delta}{\delta g_R}\right] Z_{\mu}^R[g_R] = 0,
\end{equation} 
where $\tilde\beta(g_R)(x) =[\beta (\bar{g}_R)+D \bar{g}_R](x \mu)$ are the standard Gell-Mann-Low beta functions.
}
Furthermore, a possible conformal anomaly in this equation, emerging from our usage of local couplings, can be included in the local term $\beta^0 \delta/\delta g_R^0$, which is related to the vacuum energy. This also holds for the corresponding explicit breaking by the cutoff in \refeq{WRGeq}. 

A natural renormalisation scheme is to parametrise directly the integral curves along the renormalised manifold by their linearised rates as they leave the fixed point:
\beq
\Gamma^\alpha_{\Lambda/\mu}(g_{\mathrm{UV}})(x)= g^\alpha_*+  (\Lambda/\mu)^{-\lambda_{(a)}} \delta^\alpha_a g_{\mathrm{UV}}^a(x \mu/\Lambda),  \quad{\rm as}\quad
\Lambda/\mu \to \infty, ~~ \mathrm{(linearised)} \label{UVscheme}
\eeq
with $a$ indicating the relevant or marginal eigendirections.\footnote{Actually, depending on dimensions, this expansion should be treated more carefully, as we discuss below. This includes in particular marginal directions.} This induces a parametrisation of the renormalised manifold $g(g_{\mathrm{UV}})=\Gamma_1(g_{\mathrm{UV}})$. With this parametrisation, $\Gamma_t(g_{\mathrm{UV}})=f_t(g(g_{\mathrm{UV}}))$. We will call this the UV scheme (hence the label for this instance of $g_R$). It is purely Wilsonian, as it can be defined in a neighbourhood of the fixed point without integrating out the IR degrees of freedom. The flow of renormalised couplings in this scheme is extremely simple:
\beq
\Fcal^a_t(g_{\mathrm{UV}})(x) = t^{-\lambda_{(a)}} g_{\mathrm{UV}}^a(x/t). \label{UVrenflow}
\eeq
It is diagonal for all values of $t$. This simplicity reflects the fact that this scheme is only sensitive to the UV dynamics of the theory. The corresponding beta functions are, exactly, 
\beq
\beta^a(g_{\mathrm{UV}})=-\lambda_{(a)} g_{\mathrm{UV}}^a-D g_{\mathrm{UV}}^a. \label{UVbeta}
\eeq
This is equivalent to the statement that, even though the form \eqref{UVscheme} above is modified away from the limit $\Lambda/\mu\to\infty$, it remains the case that $g_{\mathrm{UV}}^a$ and $\mu$ always appear in the combination $\mu^{\lambda_{(a)}} g_{\mathrm{UV}}^a$.
For small $g_{\mathrm{UV}}$, where both Eqs.\ (\ref{betaWlinear}) and~(\ref{UVscheme}) can be used, it is easy to check explicitly that \refeq{betarelation} holds.


If the dimensions satisfy $\lambda_{(a)}+\lambda_{(b)}\le\lambda_{(c)}$, for some $a,b,c$ then generically as $\Lambda/\mu\to\infty$, there are higher order terms that are as important or more important than the linearised terms shown in \eqref{UVscheme}. In particular this is always true if $a$ or $b$ corresponds to a non-vanishing marginally relevant coupling. In the non-exceptional case where $n^a\lambda_{a}$ is not a non-negative integer, for any non-zero vector of integers $n^a$, the 
generalisation is readily treated. We just have to recognise that $\Gamma^\alpha_{\Lambda/\mu}$ is then a Taylor expansion in the $r$ small quantities $\epsilon_{(a)} = (\Lambda/\mu)^{-\lambda_{(a)}}$, which can be treated as independent since each term in the large $\Lambda/\mu$ expansion can be uniquely expressed as some monomial $\Pi_a \epsilon_{(a)}^{n^a}$. In mathematical terms, the terms in the expansion form an integral domain. Equation \eqref{UVscheme} then makes rigorous sense as an expansion in the leading terms for $\epsilon_{(a)}$. 
 
If $g^{a=m}_R$ corresponds to a marginal direction or the dimensions are exceptional, then such a general treatment is not possible. We will remark only on the leading term of a marginal direction, 
specialising to the case of space-time independent coupling. 

The treatment depends on whether the direction is marginally relevant or exactly marginal. (Non-perturbatively, marginally irrelevant directions are excluded, since they correspond to theories that do not flow into the fixed point as $\Lambda/\mu\to\infty$.) The definition of the coupling given in \eqref{UVscheme} is correct only for an exactly marginal direction.  If the perturbative $\beta$ function is non-vanishing,  \eqref{UVscheme} is replaced along direction $a=m$  by the leading logarithmic running 
 \beq
\Gamma^m_{\Lambda/\mu}(g_R)= g^m_*+ 
\left( n\beta_{(m)}\ln\Lambda/\Lambda_c\right)^{-1/n}\quad{\rm as}\quad
\Lambda/\mu \to \infty, \label{UVscheme-marginal-relevant}
\eeq 
where $\Lambda_c$ is a dynamically generated physical scale assumed finite on the scale of $\mu$,  and \eqref{UVbeta} is replaced by the leading term
\beq
\beta^m(g_R)=-\beta_{(m)} \left(g^m_R\right)^{n+1}\,,
\eeq
in the $\beta$ function, all the higher order terms being neglected as vanishingly small in the limit $\Lambda/\mu \to \infty$. The normalisation of $g_R(\mu)$ is set by $g^m_R(\mu) \approx \left( n\beta_{(m)}\ln\mu/\Lambda_c\right)^{-1/n}$ for $\mu\gg\Lambda_c$. Except for some occasional comments, we will ignore these exceptions in the following, to simplify our discussion. That is, we will consider {\em generic} cases with non-exceptional eigenvalues in the relevant directions, in the way already described below \eqref{UVbeta}.

The usual mass-dependent schemes used in quantum field theory are defined in terms of correlation functions of the elementary fields. They require the integration of all the quantum fluctuations. In this paper we are interested in the gravity duals of gauge theories, which are manifestly gauge-invariant, so the correlation functions of elementary fields do not have a gravity counterpart. However, we can define a similar renormalisation scheme in terms of other observables, like Wilson loops or correlation functions of gauge-invariant operators. This requires the intermediate usage of another renormalisation scheme, such as the UV scheme above, in order to calculate them. For example we can \emph{choose to define} the Yang-Mills coupling $g_\mathrm{YM}$ through the expectation of a Wilson loop $\langle W(\mathcal{C})\rangle$ in general, by setting it equal to the exact formula for ${\cal N}=4$ Yang-Mills at large 't Hooft coupling $N g^2_\mathrm{YM}$ \cite{Maldacena:1997re} even when the theory no longer corresponds exactly to ${\cal N}=4$ Yang-Mills in this limit. At least for small perturbations away from such a theory, we can expect this definition of $g_\mathrm{YM}$ to remain sensible.
An interesting property of such {\em physical} schemes is that the beta functions are sensitive to IR details, such as mass thresholds or the choice of vacuum state, if degenerate.

A natural scheme for defining renormalised couplings in Wilsonian flows is by \emph{projection}, by which we mean that they are defined through the coefficient of the natural operator in the Wilsonian effective action. Thus we pick a natural subset of the $g^\alpha$ defined in eqn. \eqref{action} (possibly reparametrised)  to play the role of the renormalised couplings.
An example should make this clearer. In Yang-Mills theory a natural way to define $g_\mathrm{YM}$ directly from the Wilsonian action is to define the coefficient of the field-strength squared term in the Wilsonian action to be $F^2 / 4 g^2_\mathrm{YM}(\Lambda)$. This defines a coupling that runs with $\Lambda$ under eqn. \eqref{RGflow}. It can be considered to be renormalised if it is chosen to be finite when the integrating out is continued down to values of $\Lambda$ corresponding to finite energies.  Once we are on $\Rcal$, all the couplings $g^\alpha$ then become functions of these renormalised couplings. In this example we would have $g^\alpha\equiv g^\alpha(g_\mathrm{YM})$. Clearly, this scheme breaks down when the projection is not injective. The evolution of renormalised couplings in projection schemes is sensitive to the dynamics of the theory at the probed scales. However, unlike the physical schemes, they are of Wilsonian nature and the value of the renormalised couplings at a given finite renormalisation scale does not depend on lower scales.

\begin{figure}[t!]
\begin{center}
 \includegraphics[width=12cm]{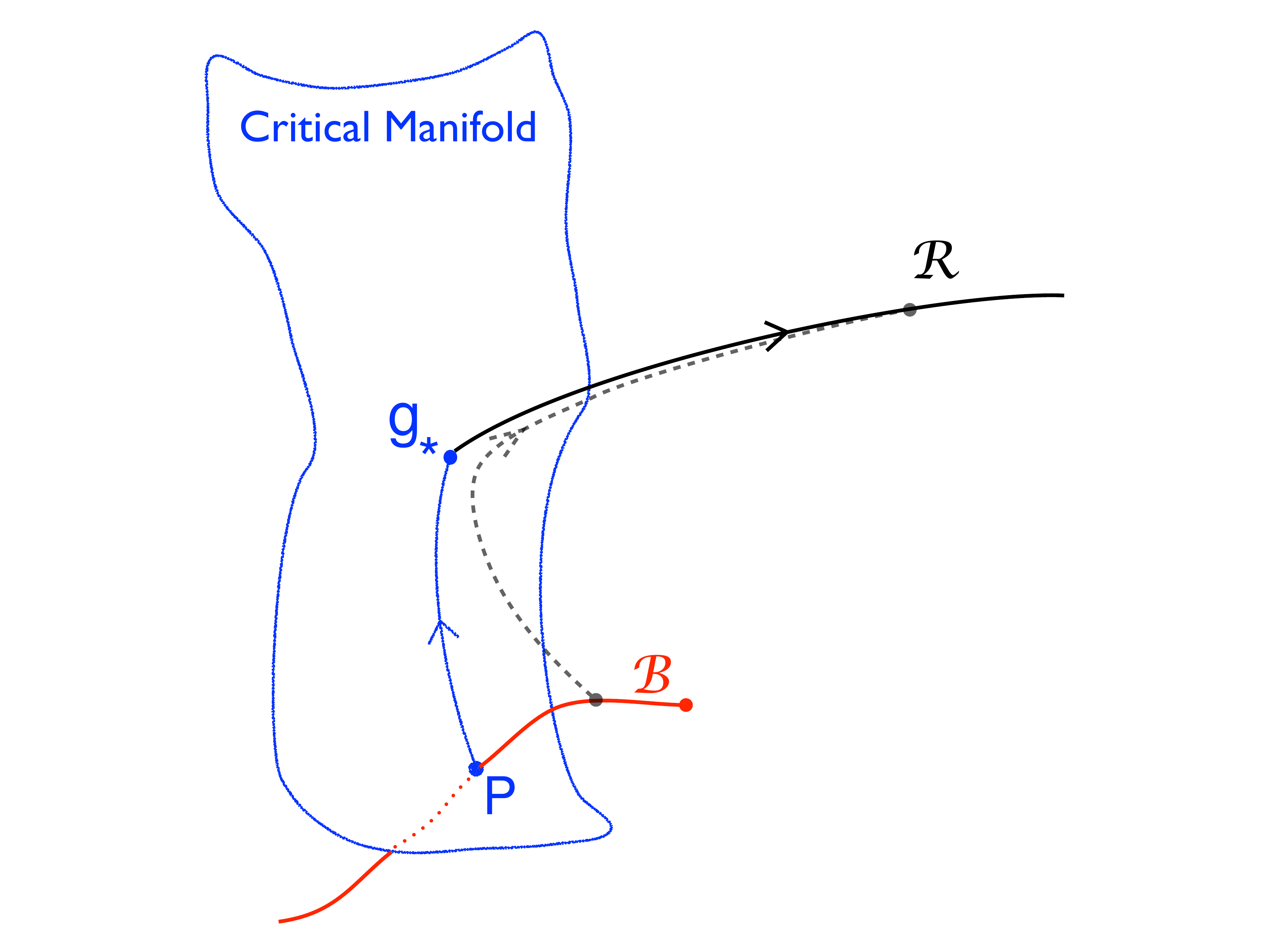}
\caption{Points lying in the critical manifold are shown in blue. The bare manifold $\Bcal$ cuts the critical manifold at a single point $P$. The dashed grey curve illustrates the action of combined RG evolution and renormalisation as in \eqref{hcondition}, finishing at a finite point in $\Rcal$.}
\label{f:RGflows}
\end{center}
\end{figure}
So far we have described a direct approach to renormalised theories, which only uses the renormalised manifold. However, in practice, it is often easier to follow a renormalisation procedure based on counterterms or, equivalently, bare couplings. For this, we choose some bare action at scale $\Lambda_0$ which depends on these $r$ tunable parameters. There is a great deal of freedom in the form of this action, equivalently in the dependence of $g$ on these parameters. (This is a statement of universality.) 
Now let us review the procedure in the Wilsonian language. The critical manifold is the set of points that under RG evolution towards the IR, reach the fixed point. We choose a manifold $\Bcal$ of dimension equal to or larger than $r$, in the same sense as above, that cuts the critical manifold at a point $P$. The RG curves of points close to $P$ will approach $g_*$ and, before they reach it, leave the critical manifold along the relevant directions, approximately, and stay (at least for a while) close to $\Rcal$. Let $h_{\Lambda_0/\mu}(g_R)$ be curves in  $\Bcal$ that, when $\Lambda_0/\mu \to \infty$, approach $P$ at a rate characterized by $g_R^b,~b=1,\ldots,r$, with the condition that 
\beq
\lim_{\Lambda_0 \to \infty} f_{\Lambda/\Lambda_0}(h_{\Lambda_0/\mu}(g_R)) \in \Rcal .     \label{hcondition}
\eeq
This defines a renormalised theory:
\begin{align}
Z^R[g_R] &=\lim_{\Lambda_0 \to \infty} Z[h_{\Lambda_0/\mu}(g_R)] \nn
&= Z[\lim_{\Lambda_0 \to \infty} f_{\Lambda/\Lambda_0}(h_{\Lambda_0/\mu}(g_R))] .
\end{align}
One possible parametrisation of the renormalised manifold is 
\beq
g(g_R)= \lim_{\Lambda_0 \to \infty} f_{\mu/\Lambda_0}(h_{\Lambda_0/\mu}(g_R)). \label{Rparam}
\eeq 
Then,
\beq
\label{hola}
\lim_{\Lambda_0\to\infty} f_{\Lambda/\Lambda_0}(h_{\Lambda_0/\mu}(g_R)) = \lim_{\Lambda_0\to\infty} f_{\Lambda/\Lambda_0} (h_{\Lambda_0/\Lambda}(\Fcal_{\Lambda/\mu}(g_R)) )
\eeq 
It follows that 
\beq
0=\lim_{t_0\to\infty} \left[\left. \frac{\delta f^\beta_{t_0^{-1}}}{\delta g} \right|_{g=h_{t_0}(g_R)} \cdot \left( t_0 \frac{\d h_{t_0}(g_R)}{\d t_0} 
- \beta(g_R) \cdot  \frac{\delta h_{t_0}(g_R)}{\delta g_R}\right)\right] \label{barebeta}.
\eeq
The renormalised beta function can be obtained from the asymptotic behaviour of the bare couplings $h_{t_0}$. This is possible, even if each of the two terms inside the parenthesis approaches zero in the limit $t_0\to \infty$, because the first factor $\d f / \d g$ diverges in this limit at the same rate. The reason is that both sides of \refeq{hola} are, by definition, finite points in the renormalised manifold.
The rate at which the curves $h_{t_0}$ approach the point $P$ will determine the rate at which an RG trajectory passing through $g(g_R)$ leaves the fixed point. 
Indeed for $\Lambda\gg\mu$ such that we are sufficiently close to keep just the first order perturbation,  we have that condition \eqref{hcondition} implies, at the linearised level,
\beq
\label{hcondition-high}
\lim_{\Lambda_0\to \infty} 
f_{\Lambda/\Lambda_0}^\alpha(h_{\Lambda_0/\mu}(g_R))(x) =  g_*^\alpha + \delta^\alpha_a (\Lambda/\mu)^{-\lambda_{(a)}}C^a(g_R)(x\mu/\Lambda),~~\mathrm{(linearised)}
\eeq
for some dimensionless functions of the renormalised couplings $C^a(g_R)$.
We have used \eqref{betaWlinear} and the existence of a limit to recognise that the $\Lambda$ dependence  must take this form.\footnote{\label{foot:sumRGdims}  At the non-linear level the non-exceptional case case follows the treatment given below \eqref{UVbeta}.}

By comparison with \eqref{UVscheme} we see that the renormalised couplings in the Wilsonian UV scheme are given by $g_{\mathrm{UV}}^a = C^a(g_R)$. If we choose in particular $C^a(g_R)=c_{(a)} g^a_R$, it follows that the beta functions of the two schemes have the same functional form. Although again we are displaying equations only for relevant or exactly marginal directions, it is clear from the discussion below \eqref{UVscheme} that a similar identification holds true also for the remaining case of marginally relevant directions.  

Furthermore, expanding the left hand side of \eqref{hcondition-high} about $h=P$, we obtain a Taylor series with terms of the form $\frac1{n!}\,\partial^n f_{\Lambda/\Lambda_0}(g)/\partial g^n\, (\Delta h)^n$, where only $\Delta h = h_{\Lambda_0/\mu}(g_R)-P$ carries dependence on $\mu$ and only the Taylor series coefficient carries dependence on $\Lambda$. Consider first the case in which all the eigenvalues of relevant directions fulfil the condition $\lambda_{(a)}+\lambda_{(b)} > \lambda_{(c)}$. Without some special tuning, already the first order term will contribute, fixing the $\mu$ dependence of $\Delta h$ and ensuring that actually only the first order term contributes in the $\Lambda\gg\mu$ regime:
\beq
\lim_{\Lambda_0\to \infty} f_{\Lambda/\Lambda_0}^\alpha(h_{\Lambda_0/\mu}(g_R)) = g_*^\alpha + \lim_{\Lambda_0\to \infty} \left[ 
  \left.  \frac{\delta f^\alpha_{\Lambda/\Lambda_0}(g)}{\delta g}\right|_{g=P}\!\!\cdot{} \left(h_{\Lambda_0/\mu}(g_R)-P\right) \right] , \label{firstorder}
\eeq
and that 
\beq
M^a(x)\cdot{} \left(h_{\Lambda_0/\mu}(g_R) -P\right)= C^a(g_R)(x\mu/\Lambda_0) \left(\frac{\Lambda_0}{\mu}  \right)^{-\lambda_{(a)}}\quad{\rm as}\quad \Lambda_0/\mu \to \infty,  \label{barecoupling}
\eeq
where 
\beq
M^a_{\ \beta}(x,y) = \lim_{\Lambda_0\to \infty} \left(\frac{\Lambda}{\Lambda_0}  \right)^{\lambda_{(a)}}  \left.\frac{\delta f^a_{\Lambda/\Lambda_0}[g(x)]}{\delta g^\beta (y)}\right|_{g=P}.
\eeq
When we withdraw the condition on the relevant eigenvalues, the second and possibly higher-order terms can give contributions that are more important than the right-hand side of \refeq{firstorder}, and also the $\mu$ behaviour of the left-hand side is modified. Then, to obtain the correct $\mu$ dependence of the left-hand side $h_{\Lambda_0/\mu}$ in \refeq{barecoupling} must be corrected. Again the result is the same for the non-exceptional case, interpreted as the leading terms in an $\epsilon_{(a)}$ expansion. At the non-linear level,  schematically,
\begin{align}
M^a(x) \cdot{} \left(h_{\Lambda_0/\mu}(g_R) -P\right)= C^a(g_R)(x\mu/\Lambda_0) \left(\frac{\Lambda_0}{\mu}  \right)^{-\lambda_{(a)}} + \sum_j \alpha^a_j(g_R)(x\mu/\Lambda_0) \left( \frac{\Lambda_0}{\mu}\right)^{-\lambda_{j}}\nn
\quad{\rm as}\quad \Lambda_0/\mu \to \infty,  \label{barecoupling2}
\end{align}
where the exponents $\lambda_j$ are sums of eigenvalues with $\lambda_j < \mathrm{Max} \{\lambda_{(a)}\}$  and the coefficients $\alpha^a_j$ depend on $C^b(g_R)$.

\section{Holographic Wilsonian description of renormalisable theories.}
\label{sec:holography} 

\subsection{Hamilton-Jacobi evolution}

Let us consider an asymptotically-AdS space in $d+1$ Euclidean dimensions. In some neighbourhood of the boundary, we can use Fefferman-Graham coordinates and write the metric as
\beq
ds^2=\frac{L^2}{z^2} dz^2+ h_{\mu\nu}(z,x) dx^\mu dx^\nu.
\eeq
The boundary is located at $z=0$. In this paper we consider a fixed metric, i.e. we neglect the backreaction of other fields on the geometry. Furthermore, below we will specialise to AdS space. These are strong restrictions and we comment on the complete treatment with a fluctuating geometry in the final discussion.

In agreement with holography and the UV/IR connection~\cite{Susskind:1998dq}, all the information about the ultraviolet of the gauge theory is encoded in the dual picture near the boundary of the asymptotically-AdS space. A natural way to do it is by enforcing boundary conditions on the degrees of freedom of the gravity theory. On the other hand, as in field theory, a regularisation is required to make quantities such as correlation functions well-defined. The standard regularisation used in the literature of gauge/gravity duality is to cut the space off close and parallel to the boundary~\cite{Freedman:1998tz}. Then, the boundary conditions must be imposed at the new boundary, i.e. the cutoff position. More generally, we can consider placing this cutoff boundary (which we keep parallel to the conformal boundary) at larger values of the radial coordinate $z$, which correspond to a lower UV cutoff in the field theory. A consistent way of imposing these boundary conditions is to add an action that depends on the fields restricted to the UV boundary.\footnote{We always employ field-theoretical language, which should be appropriately translated to string-theory analogues in precise formulations beyond the low-energy field-theory approximation, valid at large 't Hooft coupling.} It is then natural to identify the cut-off partition function in \refeq{partitionfunction}, for some boundary action $S^B_l$, with
\beq
Z_{1/l}[g] = \int \left[\Dcal \phi\right]_l e^{-S^B_l(g)[\phi(l)]-\Sgra[\phi]},   \label{cutoffduality}
\eeq
where $\left[\Dcal \phi\right]_l$ indicates functional integration in the fields $\phi$ of the gravity theory, with support restricted to $z\ge l$, and $\Sgra$ is the classical gravity action. The space-time integrals inside the functional integral are always understood to be restricted to the support of the fields.
 The boundary action $S^B_l$ is a differentiable functional of the fields restricted to the boundary and of the gauge-theory couplings. We will often not display explicitly the arguments of $S^B_l$. For definiteness, the fields $\phi$ are assumed to be $l$-independent dimensionless functions of the dimensionful space-time variables. This can be achieved with some dimensionful constant of the gravity theory, such as the AdS curvature. Then, the $l$ dependence of $S^B_l$ is dictated by dimensional analysis, similarly to \refeq{action}. It is useful to distinguish the boundary values of the fields and write \refeq{cutoffduality} in the following equivalent form:
\beq
Z_{1/l}[g] = \int \Dcal \varphi \, e^{-S^B_l(g)[\varphi]}\int \left[\Dcal \phi\right]_{l,\varphi} e^{-\Sgra[\phi]}. \label{splitcutoffduality}
\eeq
Here, 
$\left[\Dcal \phi \right]_{l,\varphi}$ indicates a path-integral measure for fields with support $z\ge l$ and such that $\phi(l,x)=\varphi(x)$. The usage of \refeq{cutoffduality} (or \refeq{splitcutoffduality}) entails a particular definition of the cutoff procedure in the gauge theory~\cite{MPV,Polchinski,Faulkner:2010jy}. It is not clear at all that this regularisation can be formulated, for arbitrary $l$, in an independent form in terms of the field-theory degrees of freedom, and we will not attempt here to find such a correspondence.
At any rate, \refeq{cutoffduality}  allows to formulate holographically all the field-theoretical Wilsonian formalism reviewed in the previous section in terms of the dual gravity theory. The relation between the gravity boundary action at $l$ and its corresponding Wilson action at cutoff $1/l$ will be examined below.

To any pair $(l_0,g_0)$, RG invariance associates a flow $g^\prime=f_{l_0/l}(g_0)$ given, for $l_0<l$, by
\beq
e^{-S^B_l(g^\prime)[\varphi]} = \int \left[\Dcal \phi \right]_{l_0}^{l,\varphi} e^{-S^B_{l_0}(g_0)[\phi(l_0)]-\Sgra[\phi]}. \label{SBflow}
\eeq
Here, $\left[\Dcal \phi \right]_{l_0}^{l,\varphi}$ indicates a measure for fields $\phi(z)$ with support $l_0\le z\le l$ such that $\phi(l)=\varphi$. In the following we work in the large N limit (with fixed large 't Hooft coupling), which is dual to gravity in the classical field-theory approximation and allows for a saddle-point calculation of the path integrals.\footnote{The $N$ global factors in the actions and in the normalisation of the operators that are necessary for a well-defined large-N limit are implicit in our equations.} In this limit, the gravity action can be written in terms of a local Lagrangian $\Sgra[\phi]=\int dz d^{d}x \Lcal(\phi(z,x),\d{\phi}(z,x),z)$, and the
path integrals in \refeq{cutoffduality} or~\refeq{splitcutoffduality} are obtained by extremizing the exponent, subject to the
UV boundary conditions
\beq
\BC{l} := \left\lbrace\Pi_c(l,x) = \left. \frac{\delta S^B(g)[\varphi]}{\delta \varphi^c(x)} \right|_{\varphi=\phi(l)} \right\rbrace, \label{bc}
\eeq
where we have defined the momenta
\beq
\Pi_c(z,x) = \frac{\d \Lcal}{\d \dot{\phi^c}(z,x)} \, ,
\eeq
with $\dot{\phi}=\d_z \phi$. Likewise, $S^B(g^\prime)$ is obtained by inserting in $S^G+S^B(g)$ the solutions $\phi_\mathrm{cl}$ to the equations of motion with boundary conditions $\BC{l_0}$ and $\phi(l,x)=\varphi(x)$:
\beq
S^B_l(g^\prime)[\varphi] = S^G[\phi_\mathrm{cl}]+S^B_{l_0}[\phi_\mathrm{cl}(l_0)] . \label{principalSB}
\eeq
The condition that $Z_{1/l}[f_{l_0/l}(g)]$ be independent of $l$ implies again the Wilson RG equation~\refeq{WRGeq}. This time, because we have a specific cutoff procedure, we can write $\d/\d l \, Z_{1/l}$ more explicitly. 
In fact, $S^B_l(g^\prime)$ in \refeq{principalSB} is defined exactly as a Hamilton's principal function in classical mechanics. As shown in~\cite{Polchinski,Faulkner:2010jy}, differentiation of \refeq{principalSB} with respect to $l$, with fixed $\varphi$, gives a Hamilton-Jacobi equation for the sliding boundary action $S^B_{\langle l,l_0\rangle}(g)[\varphi]=S^B_l(f_{l_0/l}(g))[\varphi]$:
\beq
\frac{\d}{\d l} S^B_{\langle l,l_0\rangle}(g)[\varphi] = -H_l[\varphi,\frac{\delta S^B_{l/l_0}(g)[\varphi]}{\delta \varphi}] , \label{HJB}
\eeq
with the Hamiltonian
\beq
H_z[\phi(z),\Pi(z)] =  \Pi(z) \cdot \dot{\phi}(z) - \int d^d x \Lcal(\phi(z,x),\d{\phi}(z,x),z) .   
\eeq

Let us now specialize to a fixed AdS$_{d+1}$ background. We work in the Poincar\'e patch and use Poincar\'e coordinates, with
\beq
ds^2 = \frac{L^2}{z^2} \left(\eta_{\mu\nu} dx^\mu dx^\nu+dz^2 \right)
\eeq
The AdS isometry allows us to write the Hamilton-Jacobi equation as an autonomous differential equation. Indeed, in the dimensionless coordinates 
$\bar{x}^\mu=x^\mu \, /l$,  $\bar{z}=z/l$, the induced metric on the sliding cutoff surface is just $L^2 \eta_{\mu\nu}$, the gravity Lagrangian has the same form and the equation reads
\beq
t \frac{\d}{\d t} \bar{S}^B_{\langle t \rangle}(g)[\bar{\varphi}]  = -\hat{H}[\bar{\varphi},\frac{\delta \bar{S}^B_{\langle t \rangle}(g)[\bar{\varphi}]}{\delta \bar{\varphi}}],  \label{rHJB} \\
\eeq
with the following definitions:
\begin{alignat}{2}
& \bar{\varphi}(\bar{x}) =  \varphi(x), \quad && \bar{\pi}(\bar{x})=l^d \pi(x) \nn 
& \bar{H}[\bar{\varphi},\bar{\pi}] = l H_l[\varphi,\pi],   \quad && \hat{H}[\bar{\varphi},\bar{\pi}] = \bar{H}[\bar{\varphi},\bar{\pi}] + \bar{\pi} \cdot D\bar{\varphi}, \label{Hhat} \\
& \bar{S}^B_{\langle l/l_0 \rangle}(g)[\bar{\varphi}]=S^B_{\langle l,l_0 \rangle}(g)[\varphi]. \quad && \mbox{}  \nonumber
\end{alignat}
Remember that $[Df](x)=x^\mu \d_\mu f(x)$. The point of these redefinitions is that the functional $\hat{H}$ in Eqs.~(\ref{rHJB}) does not depend on $l$. This modified Hamiltonian generates motions along the logarithmic radial coordinate, accompanied by a dilatation:
\begin{align}
& l \frac{\d}{\d l} \hat{\phi}_l = \frac{\delta \hat{H}[\hat{\phi}_l,\hat{\Pi}_l]}{\delta\hat{\Pi}_l} ,  \label{motion1} \\
& l \frac{\d}{\d l} \hat{\Pi}_l = - \frac{\delta \hat{H}[\hat{\phi}_l,\hat{\Pi}_l]}{\delta\hat{\phi}_l} , \label{motion2}
\end{align}
with $\hat{\phi}_l(\bar{x}) = \phi(l,\bar{x} l)$, $\hat{\Pi}_l(\bar{x}) = l^d \Pi(l,\bar{x}l)$ and $\phi$, $\Pi$ solutions of the original equations of motion derived from $H$.
The energy associated to $\hat{H}$ is conserved along this motion. On the other hand, $\bar{S}^B$ only depends on $l/l_0$, by dimensional analysis. In the following we only use barred quantities, but drop the bars to avoid cluttering the notation too much. (On the other hand, we keep hats explicit whenever they appear; for instance, $\phi$ and $\Pi$ refer to the the original definitions of the 5D fields and momenta as functions of the dimensionful coordinates.) The solution to~\refeq{rHJB}, analogous to~\refeq{SBflow}, reads in path-integral notation
\begin{align}
e^{-S^B_{\la t \ra}(g)[\varphi]} & = \int \left[\Dcal \phi \right]_{l_0}^{l_0 t,\tilde{\varphi}_{l_0 t}} e^{-S^B_{\la1 \ra}(g)[\hat{\phi}_{l_0}]-\Sgra[\phi]} \nn
& = \int \left[\Dcal \phi \right]_{l_0 t^{-1}}^{l_0,\tilde{\varphi}_{l_0}} e^{-S^B_{\la 1 \ra}(g)[\hat{\phi}_{l_0 t^{-1}}]-\Sgra[\phi]} , \label{dimensionlessSBflow}
\end{align}
where $\tilde{\varphi}_l(x) = \varphi(x/l)$ and $l_0$ is a dummy length introduced for dimensional reasons, due to our choice of 5D fields. The boundary condition imposed by the redefined $S^B(g)$ reads
\beq
\overline{\BC{l}} := \left\lbrace \hat{\Pi}_{l\,c}(x) = \left. \frac{\delta S^B(g)[\varphi]}{\delta \varphi^c(x)} \right|_{\varphi=\hat{\phi}_l} \right\rbrace, \label{barBC}
\eeq
Inserting this boundary condition for the sliding boundary action into \refeq{motion1}, we see, as in standard Hamilton-Jacobi theory, that a given solution $S^B_{\la t \ra}$ to the Hamilton-Jacobi equation generates solutions satisfying the first-order differential equation
\beq
l \frac{\d}{\d l} \hat{\phi}_l = \left. \frac{\delta \hat{H} [\hat{\phi}_l,\hat{\Pi}_l]}{\delta\hat{\Pi}_l} \right|_{\hat{\Pi}_l=\frac{\delta S^B_{l \mu}[\hat{\phi}_l]}{\delta \hat{\phi_l}} } ,  \label{HJmotion}
\eeq
where $\mu$ is an arbitrary reference scale. The Hamilton-Jacobi equation ensures that \refeq{motion2} is also satisfied.

It is easy to establish a connection with the evolution of coupling constants, once the functional $S^B[g]$ associated to the theory is known. Using the chain rule, 
\beq
\left. t\frac{\d}{\d t}  S^B_{\langle t \rangle}(g)[\varphi] \right|_{t=1} = - \beta_W (g) \cdot \frac{\delta S^B(g)[\varphi]}{\delta g}  ,
\eeq
so the Hamilton-Jacobi equation implies the relation
\beq
\beta_W (g) \cdot \frac{\delta S^B(g)[\varphi]}{\delta g} = \hat{H}[\varphi,\frac{\delta S^B(g)[\varphi]}{\delta \varphi}], \label{betaSB}
\eeq
which can be used to obtain the Wilsonian beta functions from a given $S^B(g)$.
Since the latter is calculated with an IR cutoff, we can expand it in derivatives:
\begin{align}
S^B(g)[\varphi] & =\int d^d x  \mathcal{S}^B(g(x))(\varphi(x),\d\varphi(x),\d^2 \varphi(x),\ldots) \nn
& = \int d^d x \sum_{n=0}^\infty \mathcal{S}^{B(n)}(g(x))(\varphi(x),\d \varphi(x), \ldots,\d^n \varphi(x)) \label{derivativeexpansionSB}
\end{align}
with $\mathcal{S}^{B(n)}$ containing $n$ derivatives and $\d^k = \d_{\mu_1}\ldots \d_{\mu_k}$.
On the other hand, writing $H[\varphi,\pi]=\int d^d x \mathcal{H}(\varphi(x),\pi(x))$, we have
\begin{align}
\hat{H}[\varphi,&\frac{\delta S^{B}(g)[\varphi]}{\delta \varphi}] = \int d^d x \left[ \mathcal{H} + x^\mu \left(\frac{\d \mathcal{S}^{B}(g)}{\d \varphi^c} - \d_\nu \frac{\d \mathcal{S}^{B}(g)}{\d \d_\nu\varphi^c} +\ldots \right) \d_\mu \varphi^c \right] \nn
& = \int d^d x \left[ \mathcal{H} + x^\mu \d_\mu \mathcal{S}^{B}(g) + \sum_{n=1}^\infty n \frac{\d \mathcal{S}^{B}(g)}{\d \d^n \varphi^c} \d^n \varphi^c -x^{\mu}\d_{\mu} g \frac{\d \mathcal{S}^{B}(g)}{\partial g}\right] \label{densityH} \\ 
& =  \int d^d x \left[ \mathcal{H} - d \mathcal{S}^{B}(g) + \sum_{n=1}^\infty n \frac{\d \mathcal{S}^{B}(g)}{\d \d^n \varphi^c} \d^n \varphi^c-x^{\mu}\d_{\mu} g \frac{\d \mathcal{S}^{B}(g)}{\partial g} \right] \nonumber ,
\end{align}
with
\beq
\mathcal{H}=\mathcal{H}\left(\varphi(x),\sum_{n=0}^\infty (-1)^n \d^n \frac{\d \mathcal{S}^B(g)}{\d [\d^n \varphi(x)]} \right) .
\eeq
The third term of the last line in \refeq{densityH} just counts the number of derivatives of each term of $\mathcal{S}^B$. Due to explicit derivatives and the fact that $\mathcal{H}$ also contains derivatives, each step of the RG evolution adds derivatives to $\mathcal{S}^B$. A derivative-independent $\mathcal{S}^B$ is not stable under RG. Actually, using \refeq{derivativeexpansionSB} and \refeq{densityH} we see that the derivative expansion of \refeq{rHJB} has a triangular form, with $\mathcal{S}^{B(n)}$ not entering in the equation for $\mathcal{S}^{B(m)}$ with $m<n$. At the leading order and for constant couplings, the Hamilton-Jacobi equation for  $\mathcal{S}^{B(0)}_{\la t \ra}$ is 
\beq
t \frac{\d}{\d t} \mathcal{S}^{B(0)}_{\la t \ra}(\varphi) = - \mathcal{H}^{(0)}\left(\varphi,\frac{\d \mathcal{S}^{B(0)}_{\la t \ra}(\varphi)}{\d \varphi}\right) + d \mathcal{S}^{B(0)}_{\la t \ra}(\varphi) , \label{constantpart}
\eeq
where $\mathcal{H}^{(0)}$ is the derivative-independent part of $\mathcal{H}$ and we have left the $g$ dependence implicit.
To be more explicit,  we shall often consider a gravity theory with a set of real active scalar fields $\{\Phi^i, i=1,\ldots M\}$ and Lagrangian density of the form
\beq
\Lcal = \sqrt{g} \left[\frac{1}{2} g^{MN} {\d_M \Phi^i}{\d_N \Phi_i}  + U(\Phi) \right] . \label{scalarL}
\eeq
To apply our equations, we define dimensionless fields $\phi = L^{(d-1)/2} \Phi$ and potential $V(\phi)=L^{d+1} U(\Phi)$.
The Hamiltonian density is 
\beq
\mathcal{H}(\varphi,\pi)= \frac{1}{2} \pi^i \pi_i - \frac{1}{2} \d_\mu \varphi^i \d_\mu \varphi_i - V(\varphi). \label{scalarH}
\eeq
In this case, \refeq{constantpart} reads
\beq
t \frac{\d}{\d t} \mathcal{S}^{B(0)}_{\la t \ra} = d \mathcal{S}^{B(0)}_{\la t \ra} - \frac{1}{2} \left(\d_i \mathcal{S}^{B(0)}_{\la t \ra} \right)^2 + V,
\eeq
with $\d_i=\d/\d\varphi^i$.

So far, we have expressed all the holographic Wilsonian formalism in terms of the boundary action $S^B_l$. In order to make precise contact with the standard formulation in terms of the field-theory degrees of freedom, we need a relation between $S^B_l$ and the field-theory Wilson action $S_{1/l}$. Such a relation has been proposed in Ref.~\cite{Polchinski} by Heemskerk and Polchinski as a generalization of the usual dynamical statement of the AdS/CFT correspondence for deformations of the conformal theory with single-trace operators $\Ocal_s$~\cite{Witten:1998qj,Gubser:1998bc}. Let us explain this proposal. Recall first that the elementary fields $\phi$ in the dual gravity theory are associated to single-trace primary operators $\Ocal_s$ (possibly with additional restrictions from supersymmetry) in the gauge theory.  A general Wilson action at scale $\Lambda$ can be written as a functional of the couplings and the primary single-trace operators:
\beq
S_\Lambda(g)=S(g)[-\Ocal_s^\Lambda].
\eeq
(The minus sign in this definition is just to make some formulas below look more natural.)
In particular, an action $S^s$ without multi-trace operators will be a bilinear functional
\beq
S^s(g_s)[\Ocal^{\Lambda}_s] = - g_s \cdot \Ocal^\Lambda_{s} \label{singletraceaction} ,
\eeq
since the derivatives in descendants can be absorbed in the space-time dependence of $g_s$.
Heemskerk and Polchinski postulate that the partition function associated to $S^s$ with some cutoff is equal to the gravity partition function with Dirichlet conditions at a UV boundary:
\beq
Z^s_{1/l}[g_s] = \int \left[\Dcal \phi\right]_{l,\tilde{g}_{s,l}} e^{-\Sgra[\phi]}, \label{singletracecutoffduality}
\eeq
with $\tilde{g}_{s,l}(x)=g_s(x/l)$.
This equation entails a choice of field variables in the gravity theory and of operators in the field theory.
In a neighbourhood of  $l=0$, this regularised version of the correspondence has been discussed and validated against particular field-theoretical calculations in Ref.~\cite{MPV}.  Furthermore, it is consistent with the success of the method of holographic renormalisation~\cite{deHaro:2000xn}, to be discussed below. For finite $l$, on the other hand, we simply take it as a definition of the cutoff procedure in the field-theory side. The assumption is then that such a cutoff can be formulated in terms of the field-theory degrees of freedom.  Combining  \refeq{splitcutoffduality} and \refeq{singletracecutoffduality}, and using the particular expression~\refeq{singletraceaction}, it follows that the partition function in~\refeq{partitionfunction} is reproduced if we choose a Wilson action $S(g)$ defined, as a functional of single-trace operators, by
\beq
e^{-S(g)[\pi]} \equiv \int \Dcal \varphi \, e^{-S^B(g)[\varphi]+\varphi\cdot \pi}.  \label{transform}
\eeq
Therefore, the Wilson action is given (at least for a static gravitational background) by a simple functional-integral transform of the boundary action. Note that the latter should be bounded from below for this definition to make sense.

In the large-$N$/classical-gravity limit  Eq.~\refeq{transform} reduces to a Legendre-Fenchel transform. One general property of the Wilson action defined in this manner is that it is concave as a functional of the single-trace operators. The Legendre-Fenchel transform is not invertible in general. To be more explicit, when using this transform we will assume that $S^B(g)$ is convex in $\varphi$.\footnote{It is of course perfectly possible that these properties hold only in some regions of theory space and/or only when the possible values of $\varphi$ and $\pi$ are restricted. A careful study of these basic issues would be interesting, but we will not pursue this course here. We simply note in this regard that our restriction to quasilocal Wilson actions and \refeq{transform} require $S^B(g)$ to be strictly convex at $\varphi_0$, the $\varphi$ value dual to $\pi=0$, which is an extremum of $S^B$.} In this case, the Wilson and boundary actions are related by the invertible Legendre transform
\beq
S(g)[\pi] = S^B(g)[\varphi] - \pi \cdot \varphi ,~~~~~\mbox{}~
\pi  =  \frac{\delta S^B(g)[\varphi]}{\delta \varphi} . \label{Legendre} 
\eeq
Observe that, when used near $l=0$, \refeq{Legendre} is nothing but Witten's prescription for deformations with multi-trace operators~\cite{Witten:2001ua}. We will also consider limit cases with $S(g)[\Ocal_s]$ linear in the variables $\Ocal_s$, for which Eq.~\refeq{Legendre} is singular. In fact, \refeq{transform} gives a linear Wilson action when $\exp\{-S^{sB}(g_s)[\varphi]\}=\delta(\varphi-g_s)$, which can be considered as a singular boundary action that imposes a Dirichlet boundary condition. Note that this is consistent with the initial assumption, Eq.~\refeq{singletracecutoffduality}. 

With this relation, all the equations above involving $S^B$ can be equivalently formulated in terms of the Wilson action.
Using Legendre conjugates, the boundary condition \refeq{barBC} reads
\beq
\overline{\BC{l}}^\prime := \left\lbrace \hat{\phi}_l^c(x) = - \left. \frac{\delta S(g)[\pi]}{\delta \pi_c(x)} \right|_{\pi(x)=\hat{\Pi}_l(x)} \right\rbrace \, .
\eeq
The flowing Wilson action $S_{\langle t \rangle}(g)=S(f_{t}(g))$ obeys a dual Hamilton-Jacobi equation,
\beq
t \frac{\d}{\d t} S_{\langle t \rangle}(g)[\pi] = - \hat{H}[-\frac{\delta S_{\langle t \rangle}(g)[\pi]}{\delta \pi},\pi]. \label{rHJW}
\eeq
The counterpart of \refeq{betaSB} is
\beq
\beta_W (g) \cdot \frac{\delta S(g)[\pi]}{\delta g} = - \hat{H}[-\frac{\delta S(g)[\pi]}{\delta \pi},\pi].  
\eeq
We will be interested in cases in which $S^B(g)[\varphi]$ is analytic in $\varphi$ in some region, where it can be written in the quasilocal form
\beq
S^B(g)[\varphi]=q\cdot Q \label{localB}.
\eeq
Here, $Q_\alpha(x)$ are linear combinations of products of fields $\varphi$ and their derivatives at $x$ while the (dimensionless) dual couplings $q_\alpha(x)$ are $x$-dependent functionals of $g$. The RG flow can then be equivalently described by a flow $\kappa_t(q)$ in the dual theory space, with
\beq
\kappa_{t\,\alpha}(q(g))=q_\alpha(f_t(g)). 
\eeq
The corresponding tangent vectors are
\beq
\beta^B_{\alpha}(q) = \left. \frac{\d}{\d t} \kappa_{t\,\alpha}(q) \right|_{t=1}.
\eeq
These boundary beta functions are related to the Wilsonian ones by
\beq
\beta^{B}_{\alpha}(q(g)) = \frac{\d q_\alpha(g)}{\d g^\beta} \beta_W^\beta(g). 
\eeq
Using this relation, Eq.~\refeq{betaSB} can be written in a quite explicit form:
\beq
\beta^B(q)\cdot Q[\varphi] = \hat{H}[\varphi,q\cdot \frac{\delta Q[\varphi]}{\delta \varphi}].
\eeq

\subsection{Renormalisable theories}

We now proceed to study how renormalisable field theories are described in this Wilsonian holographic framework. Let us introduce the following convention, which will save some writing: the indices $i,j,k$ label the fields; $a$ labels relevant (and marginal) directions and $\hat{a}$ irrelevant ones; the index $0$ labels the identity/vacuum-energy direction; and $b$ labels relevant (and marginal) directions different from 0. The first step is to look for fixed points of the flowing boundary action $S^B_{\la t \ra}$, which are also fixed points of the flowing Wilson action $S_{\la t \ra}$. From \refeq{rHJB}, the fixed-point condition is simply 
\beq
\hat{H}[\varphi,\frac{\delta S^B_*}{\delta \varphi}]=0, \label{Hequal0}
\eeq
where $S^B_*=S^B(g_*)$.  Let us consider the theory \refeq{scalarL}. 
The fixed-point equation in this case is analysed in detail in Appendix~\ref{sec:AppendixA}. Here, we just give the main results of this analysis (some of them appear also in~\cite{Polchinski}).  We consider solutions $S^B_*$ of \refeq{Hequal0} that are analytical at a point $\varphi_0$ where $\delta S^B_*[\phi_0]/\delta \phi^i_0 =0$.  This conditions guarantees a discrete set of linearly independent perturbations. It can be satisfied simultaneously only if $\varphi_0$ is also a critical point of the scalar potential, i.e. $\d_i V(\varphi_0)=0$. Then there are in general $2^M$ such solutions: 
%
\beq
\mathcal{S}^B_*(\varphi) = -\frac{v_0}{d} + \frac{1}{2} \Delta_{(i)} (\varphi^i-\varphi^i_0)^2 + O((\varphi-\varphi_0)^3) + \mathrm{derivatives}, \label{SBfixed}
\eeq
where $v_0=-d(d-1)/2$ is the AdS cosmological constant in units of $L$ and $\Delta_{(i)} = \Delta_{(i)}^{\pm} = d/2 \pm \sqrt{d^2/4+m_{(i)}^2}$, with $m_{(i)}$ the mass of $\phi_i$, also in units of $L$. The two possible values $\Delta_{(i)}^\pm$ for each $i$ correspond to the dimension of the operator dual to $\phi_i$ in the standard (upper sign) and alternate (lower sign) quantisations, as discussed in Ref.~\cite{Klebanov:1999tb}. We only consider in the following cases with non-integer values of $\sqrt{d^2/4+m_{(i)}^2}\equiv  \nu_{(i)}$.\footnote{Protected integer conformal dimensions are ubiquitous in supersymmetric theories and can be easily dealt with, but we make this restriction to avoid distinguishing multiple cases and thus keep the discussions as simple as possible.}
The set of chosen signs determines the higher order terms and characterizes each fixed point. The values $\Delta_{(i)}<d/2 -1$, only possible with the alternate quantisation, correspond to non-unitary quantum field theories in the continuum limit~\cite{Klebanov:1999tb,Breitenlohner:1982bm,Breitenlohner:1982jf}, so they should be excluded. In Appendix~\ref{sec:AppendixA}, we give the derivative terms at order $(\varphi-\varphi_0)^2$ in a closed form, and provide recursion formulas to obtain the higher order terms in the expansion about $\varphi_0$.
Observe in \refeq{SBfixed} that, consistently with our assumptions, $\mathcal{S}^B_*$ is convex at $\varphi_0$. In the following we take $\varphi_0^i=0$. This entails no loss of generality, as it just amounts to working in terms of fields without tadpoles in the gravity theory. 

At the quadratic level, \refeq{SBfixed} imposes the boundary condition
\beq
\Delta_{(i)} \phi^i(l,x) = l \frac{\d}{\d l} \phi^i(l,x) ,
\eeq
when $l p \ll 1$ (here $p$ is the $d$-dimensional dimensionful momentum of $\phi$). The solutions close to the boundary have the general form 
\beq
\phi^i(z,x) = \left(\frac{z}{l_0}\right)^{d-\Delta_{(i)}} \left(A_-^i(x)+O(z^2) \right) +  \left(\frac{z}{l_0}\right)^{\Delta_{(i)}} \left(A_+^i(x)+O(z^2) \right),~~~z \sim 0 ,\label{asymptoticsolution}
\eeq
when the dimensions are generic. The boundary condition requires $A_-^i=0$ and thus selects the solutions $\phi(z,x)$ that go like $z^\Delta$ when $z\sim 0$. Because the field solutions then approach zero in the limit $z\to 0$, the nonlinear corrections are suppressed and the same conclusion holds for the complete $S^B_*$. 

The Legendre transform of~\refeq{SBfixed} gives the fixed-point Wilson density action, which is analytic at $\pi=0$ and contains no single-trace operators, except the identity:
\beq
\mathcal{S}_*(\pi) = -\frac{v_0}{d}-\frac{1}{2\Delta_{(i)}}(\pi_i)^2 + O(\pi^3) + \mathrm{derivatives}. \label{Sfixed}
\eeq
More details are given in Appendix \ref{sec:AppendixA}.

Once we have understood the structure of the possible fixed points, we are ready to study small deformations of a given fixed point $S^B_*$. The Hamilton-Jacobi equation for the perturbation $S^{\prime B} = S^B-S^B_*$ reads
\beq
t\d_t S^{\prime B}_{\la t \ra}[\varphi] = H[\varphi,\frac{\delta S^B_*}{\delta \varphi}] - H[\varphi,\frac{\delta (S^B_*+S^{\prime B}_{\la t \ra})}{\delta \varphi}] - \frac{\delta S^{\prime B}_{\la t \ra}}{\delta \varphi}\cdot D \varphi,
\eeq
At the linearised level, 
\begin{align}
t\d_t S^{\prime B}_{\la t \ra}[\varphi] &= - \left( \left. \frac{\delta H[\varphi,\pi]}{\delta \pi}\right|_{\pi=\frac{\delta S^B_*}{\delta \varphi}} + D \varphi \right) \cdot \frac{\delta S^{\prime B}_{\la t \ra}}{\delta \varphi} . \nn
& \equiv \Psi S^{\prime B}_{\la t \ra}
\end{align}
This equation is studied in Appendix~\ref{sec:AppendixB}. As shown there, the eigenvectors of $\Psi$ can be constructed from {\em basic\/} functions of the form
\beq
\Tcal^i(\varphi) = \varphi^i + O(\varphi^2) + O(\d \varphi),
\eeq
which are themselves density eigenvectors (to be integrated in $d$ dimensions) with eigenvalue $\lambda_{(i)}=d-\Delta_{(i)}$. The detailed form of these basic functions is given in Appendix~\ref{sec:AppendixB}. If $Q$ and $Q^\prime$ are arbitrary density eigenvectors with eigenvalues $\lambda=d-\Delta$ and $\lambda^\prime=d-\Delta^\prime$, respectively, then $\d^n Q$ is a density eigenvector with eigenvalue $\lambda-n$, while $Q Q^\prime$ is a density eigenvector with eigenvalue $d-\Delta-\Delta^\prime$, in agreement with large-N factorisation. Therefore, general analytical density eigenvectors can be constructed as products of a finite number of basic functions $\Tcal^i$ and their derivatives $\d^n \Tcal^i$. Relevant, exactly marginal and irrelevant perturbations have eigenvalues $\lambda>0$, $\lambda=0$ and $\lambda<0$, respectively.
We see that the number of independent relevant directions is finite, as expected in field theory. Actually, at the fixed point with standard quantisation for all fields, the only relevant eigendeformations are given by the $\Tcal^i$ themselves, with $\Delta_{(i)}\leq d$. In fixed points with non-standard quantisation for some fields, there are also eigenvectors formed by products of two $\Tcal^i$ (and more, depending on $d$), possibly with derivatives. In all cases, there exists a trivial relevant eigendeformation with eigenvalue $\lambda=d$: a constant term in $\mathcal{S}^B$, which can be interpreted as a vacuum energy. Even though such a constant, which is dual to the identity operator in $S$, does not modify boundary conditions, it will be interesting to keep track of it. 

Much as we did for the Wilson actions, we choose in the following a basis in the space of boundary actions in which the operators $Q^\alpha$ in \refeq{localB} are eigenperturbations around the fixed-point of interest. The perturbed boundary action reads
\beq
S^{\prime B}(g)[\varphi] = (q-q_*) \cdot Q,
\eeq
with $q_*=q(g_*)$.
The Legendre transform between $S^B$ and $S$ preserves the eigendirections at the fixed point in the following sense: the Wilson action associated to $\Scal^B_*[\varphi]+ Q[\varphi]$, with $Q$  an eigenoperator, is, to linear order in $Q$, $\Scal_*[\pi]+\Ocal[\pi]$, with $\Ocal$ an eigenperturbation of $S_*$. Explicitly, $\Ocal[\pi]=Q[\varphi_*[\pi]]$,
where $\varphi_*[\pi]$ is a solution of the equation $\pi = \delta S_*[\varphi_*]/\delta \varphi_*$.  
The eigenvalues of $Q$ and $\Ocal$ are the same. In particular, the Legendre conjugates of the basic eigenperturbations $\Tcal^i$ have the form
\beq
\Ocal_i[\pi] = \Delta_{(i)}^{-1} \pi_i + O(\pi^2) + O(\d \pi).
\eeq
Therefore, we see that basic eigenperturbations are associated, to lowest order, to single-trace operators, but also that they involve a tower of multi-trace operators. Note that our choices of basis imply
\beq
q_\alpha(g)-q_\alpha^*=  c_{(\alpha)} \delta_{\alpha \beta} \left( g^\beta-g^\beta_* \right) + O\left((g-g_*)^2\right) \label{neighbourduality}
\eeq
in the neighbourhood of the fixed point, with the constant $c_{(\alpha)}$ depending on the normalization of the perturbations. For the basic eigenperturbations $Q^i=\Tcal^i$, we have $c_{(i)}=\Delta_{(i)}$. 

As explained in the previous section, renormalisable theories can be intrinsically described in terms of the renormalised space formed by the actions that can be reached, under RG evolution, from relevant or marginal deformations of a given fixed-point action. Each particular renormalised theory is given by an integral curve of the Wilsonian beta functions along the renormalised manifold, which in the gravity picture corresponds to a solution to the Hamilton-Jacobi equation that approaches the fixed point towards the UV. A parametrisation of these solutions defines a renormalisation scheme.  In the space of boundary actions, these solutions are integral curves $q=\bar{\Gamma}^\alpha_t(g_R)$ of the vectors $\beta_B(q)^\alpha$ that leave the fixed point along relevant or marginal directions. 
For instance, the UV scheme $ \refeq{UVscheme}$ introduced in the previous section is defined holographically by
\beq
\bar{\Gamma}_{t\,\alpha}(g_{\mathrm{UV}}) = q^*_\alpha + \delta_{\alpha a} c_{(a)} g_{\mathrm{UV}}^a t^{\lambda_{(a)}} + ~\mathrm{nonlinear}, ~~ t \to 0,  \label{integralB}
\eeq
with $a$ running over relevant directions. Nonlinear corrections are treated as discussed in the previous section.  The renormalised manifold can be parametrised by $q(g_R)=\bar{\Gamma}_1(g_R)$. The flow of renormalised couplings and the corresponding beta functions are identical to the ones in \refeq{UVrenflow} and \refeq{UVbeta}. Note that an implicit renormalisation scale $\mu$ is necessary to write these equations in terms of a dimensionful cutoff $\Lambda=t \mu$.
The ``perfect'' boundary actions $\bar{\Gamma}_t(g_R) \cdot Q$ can be used to calculate the partition function in terms of the renormalised parameters, for any position of the sliding cutoff. They impose modified boundary conditions on the fields associated to relevant directions. In the quadratic approximation, $\Tcal^i = \varphi^i$ and the basic perturbations $Q=\Tcal^i$ are dual to single-trace deformations. For these, the boundary condition imposed by the boundary action fixes the coefficient of the asymptotic term $z^{d-\Delta_{(i)}}$ to be proportional to the renormalised UV coupling: $A_-^i=g_{\mathrm{UV}}^i \, \Delta_{(i)}/(d-2\Delta_{(i)})$.  This works both for standard and alternate quantisation, with the corresponding values $\Delta_{(i)}=\Delta_{(i)}^\pm$. We note in passing that the factors $\Delta_{(i)}/(d-2\Delta_{(i)})$ in these relations are akin to the correction factors first found in Ref.~\cite{Freedman:1998tz}. For a deformation $Q = (\Tcal^i)^2$, for instance, which can be relevant if $\phi^i$ is quantised non-standardly at the fixed point, we find instead $A^i_-/A^i_+ = g_{\mathrm{UV}}^i c_{(i)}/(d-2\Delta_{(i)})$.

\subsection{Holographic renormalisation}

The renormalisation procedure via bare couplings can also be carried out in the gravity side. We examine in the following how to implement it in the Wilsonian picture. First, we need to define a space of bare theories that cuts the critical manifold at least at one point and, close to this critical point, has a dimension equal to the number of relevant directions of the fixed point of interest. One obvious choice is to choose bare boundary actions of the form
\beq
S^B_\mathcal{B} = S^B_* + q^\mathcal{B}_a Q^a,
\eeq
i.e. \ $q_\alpha = q^*_\alpha + \delta_\alpha^a q^\mathcal{B}_a$. Instead of the $Q_a$, one can use their first orders in the momentum expansion. (How many orders depends on the theory at hand.) This bare subspace cuts the critical manifold precisely at the fixed point, when $q_\mathcal{B}^a=0~\forall a$. Comparing with \refeq{integralB}, it is clear that the curves $q = \texttt{h}_t(g_R)$ renormalise the theory if we choose
\beq
\texttt{h}_{t\,\alpha}(g_R) = q_* + \delta_{\alpha a} C^a (g_R) t^{\lambda_{(a)}} + ~\mathrm{nonlinear} ,
\eeq 
which is the dual version of \refeq{barecoupling}. 
The relation with the Wilsonian UV scheme is $c_{(a)} g_{\mathrm{UV}}^{a} = C^a(g_R)$. One advantage of this renormalisation procedure is that it works in the same manner for standard and alternate quantisations, including multitrace relevant directions.

A simpler holographic renormalisation method~\cite{deHaro:2000xn,Bianchi:2001kw,Skenderis:2002wp,Papadimitriou:2004ap,Papadimitriou:2004rz} exists in the case of standard quantisation.\footnote{In \cite{Papadimitriou:2007sj}, this method is generalised to include the alternate quantisation and multi-trace deformations.} In this case the relevant directions are given by the basic perturbations $\Tcal^b$, and can be associated to the scalar fields $\phi^b$ with negative squared mass (satisfying the Breitenlohner-Freedman bound~\cite{ Breitenlohner:1982bm,Breitenlohner:1982jf}), which we will call relevant fields.  The remaining relevant direction is the constant term in $S^B$. The bare manifold $\mathcal{B}$ in holographic renormalisation is defined by singular boundary actions that impose Dirichlet boundary conditions for all fields: $\hat{\phi}^i_{l_0}=g^i$.
The boundary actions in $\mathcal{B}$ are conjugate to linear Wilson actions, which contain only the identity and single-trace operators. Since, as stressed in Ref.~\cite{Polchinski}, the Hamilton-Jacobi equation generates multi-trace operators, $\Bcal$ is not stable under RG evolution (unlike the critical and the renormalised manifolds).

The space $\Bcal$ so defined works as a good bare manifold for renormalisable theories emanating from the fixed point with {\em standard quantisation} for all fields. The reason, in the Wilsonian language, is that $\Bcal$ cuts the critical manifold of that particular fixed point. One point in the intersection is $P:~g^i=\varphi^i_0=0,~ g^0= -v_0/d$. 
To show that $P$ belongs to this critical manifold, let us prove that it flows under RG evolution towards the standard-quantisation fixed point $g_*$, that is to say, that $\lim_{t\to \infty} f_t(0)=g_*$. 
The boundary action after a finite RG evolution is given, in path integral notation, by
\beq
e^{-S^B_{\la t \ra}}[\varphi]  = \int \left[ \Dcal \phi \right]_{l_0 t^{-1},0}^{l_0,\tilde{\varphi}_{l_0}}  e^{\int d^d x \frac{v_0}{d} t^d -S^G[\phi]} . \label{criticaltofixed}
\eeq
$S^B_{\la t \ra}$ is thus obtained from solutions of the $S^G$ equations of motion that vanish at $l_0 t^{-1}$. If we now take the limit $t\to \infty$, this condition forces the solutions to approach zero as fast as possible when $z\to 0$. The quadratic approximation to $S^G$ is then valid in the near-boundary region and we can use \refeq{asymptoticsolution}.
The $t\to \infty$ boundary condition requires that the coefficient of the leading term $z^{d-\Delta^+_{(i)}}$ vanishes. This agrees with the boundary condition imposed by the all-standard-quantisation fixed boundary action.\footnote{This also means that $\Bcal$ cannot be used to renormalise theories with alternate quantisations.} Note also that $-v_0/d$ is precisely the constant term of the fixed-point boundary action. Therefore, in the limit $t\to \infty$, \refeq{criticaltofixed} is just the trivial RG evolution of the standard fixed point, which simply gives the very same fixed point, as claimed. 
An illuminating explicit check is performed in Section~\ref{sec:examples}. Let us now have a quick look at the RG trajectories that initiate in $\Bcal$. We see in~\refeq{asymptoticsolution} that the solutions for the relevant fields $\phi^b$ vanish on the AdS boundary. Consequently, a constant boundary condition $\hat{\phi}^b_{l_0 t^{-1}}=C^b\neq 0$ would give rise, in the limit $t_0\to \infty$, to divergent coefficients and divergent solutions, and thus to a divergent action. The need of renormalisation is thus clear. Note in contrast that the irrelevant fields $\phi^{\hat{a}}$ have particular solutions (at the quadratic level) that diverge at the boundary. If we choose a point with $\hat{\phi}^b_{l_0 t_0^{-1}}=0$ and $\hat{\phi}^{\hat{a}}_{l_0 t_0^{-1}}=C^{\hat{a}} \neq 0$, this component of the solution will vanish in the limit $t\to \infty$, so the RG evolution will end at the fixed point, just as in the case of $P$. This shows that these points also lie on the critical manifold.  
 
The main step in holographic renormalisation is to find a family of curves $h_{t_0}(g_R)$ in $\Bcal$ that approach the critical manifold at the right rate, such that RG evolution takes them past the fixed point and into arbitrary points of the renormalised manifold, as specifed in Eq.~\refeq{hcondition} and illustrated in Fig.~\ref{f:RGflows}. 
The theory at each point of the curves is defined by the Dirichlet conditions $\hat{\phi}^i_{1/(\mu t_0)}=h^i_{t_0}(g_R)$ plus the vacuum-enrgy density $g^0=h^0_{t_0}$. We will demand that the curves approach the particular point $P$: $\lim_{t_0\to \infty} h^i_{t_0}(g_R)=0$ and $\lim_{t_0\to \infty} h^0_{t_0}=-v_0/d$.  A stronger condition comes from the requirement that the limit
\beq
e^{-S^B_R(g_R)}=\lim_{t\to \infty} e^{-S^B_{\la t \ra}(g_R,t)} = \lim_{t\to \infty} \int \left[ \Dcal \phi \right]_{l t^{-1},{\left.\tilde{h}_{t/(l \mu)}(g_R)\right|}_{l t^{-1}}}^{l,\varphi}  e^{-\int d^d x t^d \left. \tilde{h}^0_{t/(l\mu)}(g_R) \right|_{l t^{-1}} -S^G[\phi]} \label{finiteSB}
\eeq 
be finite. Here, $\tilde{h}_{\ldots} |_{\epsilon}(x) = h_{\ldots}(x/\epsilon)$. An essential observation here is that $S^B_{\la t \ra}(g_R,t)$ does not follow an RG trajectory as $t$ changes, due to the explicit $t$-dependence of the boundary condition (see Fig.~\ref{f:RGflows}). Therefore, the limit in~\refeq{finiteSB} does not give a fixed point, in general. Once the limit is taken, a change in the renormalisation scale $\mu$ does correspond to an RG transformation of the boundary action. This follows from dimensional analysis, as in the general field-theoretical case. 
The finiteness of \refeq{finiteSB} constrains the asymptotic behaviour of the bare couplings $h^a_{t_0}$. Conversely, the renormalised action $S^B_R(g_R)$ depends only on the asymptotics of $h^i_{t_0}(g_R)$. The explicit form of $h^i_{t_0}$ has been studied for several tensor fields and interactions in Refs.~\cite{deHaro:2000xn,Bianchi:2001kw}. We find them explicitly in Section~\ref{sec:examples} for the case of several interacting scalars with arbitrary masses. In this section, we concentrate on the relation with the Wilsonian formalism.

A necessary condition to obtain a finite result in \refeq{finiteSB} is that the solutions stay finite in the limit. This can be achieved by taking as $h^b_{t_0}$ an arbitrary solution to the equations of motion (including dilatations and with $t_0^{-1}$ the dimensionless radial coordinate), for any non-singular IR condition ~\cite{deHaro:2000xn}.\footnote{An important advantage of this method is that it preserves relevant symmetries.~\cite{Bianchi:2001kw}} The reason is that the UV Dirichlet condition will then give rise to the same solution for any $t_0$. The limit will then be trivially finite. Different parametrisations of the solutions give rise to different renormalisation schemes. One possibility, further explored in the next section, is to use as renormalised couplings the values of the solutions at a given $t_0^{-1}$. Another one, which we use in Section~\ref{sec:examples}, is to parametrise $h_{t_0}$ by the leading term of the asymptotic solutions of the relevant fields in the free approximation:
\beq
h^b_{t_0}(g_R)(x)=C^b(g_R(x/t_0)) t_0^{\Delta_{(b)}-d},  ~~t_0 \to \infty ~~ \mbox{(no interactions)} . \label{bare_holoren}
\eeq
With a faster approach to $P$, the limit in~\refeq{finiteSB} would end into the fixed point, while a slower approach would give a divergent result. It is easy to check that the Dirichlet boundary condition with \refeq{bare_holoren} used at $l t_0^{-1} \sim 0$, selects the leading coefficient of the asymptotic solutions: $A_-^b = C^b(g_R)$ if we set $l_0=\mu^{-1}$. This agrees with the boundary condition imposed asymptotically by the perturbed boundary action $S^B_*+ (d-2 \Delta_{(b)}) C^b(g_R) \varphi^b$ in standard quantisation. When interactions are taken into account, the asymptotic behaviour of the solutions is corrected. In certain cases, which we study in detail in Section~\ref{sec:schemes}, the correction terms are more important than the ones shown in \refeq{bare_holoren}. Such terms must then be taken into account for a correct renormalisation: 
\beq
h^b_{t_0}(g_R)(x)=C^b(g_R(x/t_0)) t_0^{\Delta_{(b)}-d} +   \sum_{j} \alpha^b_j(g_R(x/t_0)) t_0^{-\lambda_j}~~t_0 \to \infty  , \label{int_holoren}
\eeq
where $\lambda_j< \mathrm{Max}\,\{d-\Delta_{(b)} \}$ and the coefficients $\alpha_j^b$ depend only on the set $C(g_R)$ and are independent of the IR conditions on the solutions. 
The renormalised theory is defined by the finite boundary action of the limit~\refeq{finiteSB}. Close to the UV and at the linearised level, this theory is given by the fixed point plus a linear combination of relevant basic eigenperturbations $\mathcal{T}^a$, which can be parametrised by the UV scheme~\refeq{integralB}. In the free-field approximation, we have $\mathcal{T}^b=\varphi^b+O(\partial \varphi^b)$, which gives a boundary condition satisfied by \refeq{bare_holoren}. When interactions are turned on, the renormalised action close to the UV gives a boundary condition satisfied by the asymptotic form of the full solution, \refeq{int_holoren}. (This is explained in more detail in the next section.) Therefore, the renormalised action is in fact given by ~\refeq{integralB} with $g_{\mathrm{UV}}^b = [(d-2\Delta_{(b)})/\Delta_{(b)}] C^b(g_R)$. In Section~\ref{sec:examples} we will check this conclusion by explicit calculations. 

After the field renormalisation we have just described, the limit $\refeq{finiteSB}$ is still divergent. The divergent terms are $\varphi$-independent functions of the renormalised couplings $g_R$. Furthermore, power counting shows that the degree of divergence for deformations with relevant operators is smaller than $d$. Therefore, a finite renormalised boundary action can be obtained by choosing an adequate $h^0_{t_0}(g_R)$ that approaches $-v_0/d$ as $t_0 \to \infty$ and cancels the divergences out when combined with the $t_0^d$ . It turns out that this counterterm is a local function of the $C^b(g_R(x))$~\cite{deHaro:2000xn,Skenderis:2002wp}, in agreement with standard renormalisation theory.  We will examine an example in Section~\ref{sec:examples}. As we will see there, in some cases the $\alpha$ terms in \refeq{int_holoren} are essential to be able to cancel all divergences with local counterterms.

\section{Holographic renormalisation schemes}
\label{sec:schemes} 
As in field theory, different renormalisation schemes can be used in the holographic description of renormalised theories. We have already discussed two of them, which are essentially equivalent: the UV scheme and the leading-term parametrisation of $h_{t_0}$ in holographic renormalisation. They are both insensitive to the IR dynamics. The renormalised beta functions are very simple for generic (non-marginal) dimensions. More physical schemes, sensitive to the deep IR, could in principle be defined based on correlation functions of local operators or expectation values of non-local operators, such as Wilson loops. We do not consider them in this paper. Instead, in the following two subsections we discuss, respectively, renormalisation schemes based on projections of the Wilsonian or boundary actions and, in greater detail, the popular scheme based on solutions to the field equations of motion.

\subsection{Projections}
Several schemes can be naturally defined in holography by projections of either the Wilsonian or the boundary action into convenient subspaces of the same dimension as $\Rcal$. In this subsection we briefly comment on them.

The most obvious possibility is to project into the subspace tangent to the fixed point that is spanned by the relevant operators. If we use the Wilson action, this subspace is given by $S=S_*+ g_R^a \Ocal_a^\Lambda$ and we identify the renormalised coupling $g_R^a$ with the point in $\Rcal$ that has coordinates $g^a = g_*^a +  g_R^a$ along the relevant directions. This is a good parametrisation in a neighbourhood of $g_*$, but it may break down further away if there are different points in $\Rcal$ with the same $g^a$. The relation between renormalised and Wilsonian beta functions is:
\beq
\beta^\alpha_W(g(g_R)) = \delta^\alpha_a \beta^a(g_R) + \delta^\alpha_{b} \frac{\delta g^{b}}{\delta g_R^a} \beta^a(g_R).  \label{projectbeta}
\eeq
Note that this renormalisation scheme is very different from the UV scheme for points far from the fixed point.
Analogously, we could use the relevant tangent subspace for boundary actions, $S^B=S_*^B + q^R_a Q^a$, with $s_0=1$ and identify the dual renormalised couplings $q^R_a$ with the point in $\Rcal$ with coordinates $q^R_a = q^*_a + q^R_a$ along the relevant directions. Points in this tangent subspace are given by a Legendre transformation of points in the tangent subspace defined above {\em only\/} when they are infinitesimally close to the fixed point. The boundary beta functions and dual renormalised beta functions are related as in \refeq{projectbeta}.

In the case of the completely standard-quantisation fixed point, instead of using the eigenoperators $\Ocal^\Lambda_a$ to define the projection subspace, it is possible to use their single-trace components $\Ocal^\Lambda_{s\, a}$ (including the identity). Namely, we associate $g_R^a$ with the point in $\Rcal$ for which the coefficient of the single-trace operator $\Ocal^\Lambda_{s\, a}$ is $g_*^a+g_R^a$. Shifting away the fixed point, with no effect on the parametrisation, the projection subspace is given by $S=g_R^a \Ocal^\Lambda_{s\, a}$. This is the exactly the same as the bare space of standard holographic renormalisation, with boundary actions that impose Dirichlet conditions on the fields. Even if this renormalisation scheme is quite different from the ones naturally defined in holographic renormalisation, we will see in the next subsection that a relation can be established by a non-trivial reparametrisation of the Wilson action. Even if the projection subspaces are different, this renormalisation scheme is exactly the same as the one above (with Wilson action), since the coefficients of each $\Ocal^\Lambda_{s\, a}$ in the Wilson action are the same as the coefficients of $\Ocal^\Lambda_a$. The reason for this is that  $\Ocal^\Lambda_{s\, a}$ appears as a component of  $\Ocal^\Lambda_a$, but not of other operators. 
Similar remarks apply to a projection of $S^B$ into $g_R^a \phi^a$, but in this case there is no relation with the bare manifold of standard holographic renormalisation.

\subsection{Field solutions as renormalised couplings}
\label{sec:Fields}
In the standard non-Wilsonian approach to holographic RG flows, the solutions to the field equations of motion are interpreted as running couplings of the dual theory \cite{Akhmedov:1998vf}. These solutions are often obtained by the (fake) superpotential method~\cite{Skenderis:1999mm,DeWolfe:1999cp,Freedman:2003ax}, which is a version of Hamilton-Jacobi theory~\cite{deBoer:1999xf,Papadimitriou:2010as,Trigiante:2012eb} and can also be used in the approximation with fixed background. This method splits the problem of solving the second order differential equations in two steps, corresponding to two integrations of first-order equations. In the first one, a superpotential (or Hamilton's principal function) is selected. A given superpotential generates a class of solutions that satisfy the same first-order differential equation. This makes the RG interpretation  possible.

In this section, we investigate the relation between the RG flows based on field solutions and the Wilsonian RG flows. The main idea is that using the solutions as renormalised couplings amounts to choosing one particular renormalisation scheme. Therefore, the general relation between the RG evolution of Wilsonian and renormalised couplings, discussed in Section~\ref{sec:fieldtheory} and summarized in \refeq{inducedflow}, also holds in this case.

Let $S_{\la t \ra}^{\mathrm{IR}}=\int d^dx \mathcal{S}_t^{\mathrm{IR}}$ be a Hamilton's principal function of the AdS theory, i.e.\ a solution to the Hamilton-Jacobi equation
\begin{equation}
t \frac{\d}{\d t} S_{\la t \ra}^{\mathrm{IR}}[\varphi]=\hat{H}[\varphi,-\frac{\delta S_{\la t \ra}^{\mathrm{IR}}}{\delta \phi}] . \label{IRHJ}
\end{equation} 
The reason for the opposite sign in the momentum is that here we are considering evolution from the IR to the UV, in contrast to the case of the boundary actions.
Comparing with \refeq{rHJB}, we see that this is the same Hamilton-Jacobi equation obeyed by $-S^B_{\la t \ra}$.  Particular solutions can be found explicitly from the
IR counterpart of \refeq{dimensionlessSBflow},
\begin{equation}
e^{-S_{\la l/l_0 \ra}^{\mathrm{IR}}[\varphi]}=\int \left[ \mathcal{D} \phi \right]_{l,\tilde{\varphi}_{l}} e^{-S^G[\phi]},
\end{equation} 
when some boundary condition is specified in the far IR. In this equation $l_0$ (with $l_0>l$) is some scale introduced by the IR boundary condition. If the IR condition respects the AdS isometry, no scale is introduced and then the l.h.s.\ will be independent of $l$ and thus a fixed-point solution. In fact, in order to get a standard running of the couplings, with scale-independent beta functions, we restrict our attention to these fixed-point solutions of \refeq{IRHJ}, $S_{\la t \ra}^{\mathrm{IR}}=W$, which obey
\beq
\hat{H}[\varphi,-\frac{\delta W}{\delta \varphi}]=0 . \label{fixedIRHJ}
\eeq
The possible solutions are the same as the ones found in Appendix~\ref{sec:AppendixA} for $S^B_*$, up to a global minus sign. Note that for exceptional dimensions an analytic $W$ may not exist. In this case we can split it and take only the local part, as in~\cite{deBoer:1999xf,Martelli:2002sp,Papadimitriou:2004ap}, but the relation of the scheme with actual solutions will be lost. We continue with the study of generic dimensions.
Given a particular $W$, a class of solutions to the equations of motion can be found solving the first order equation,
\begin{equation}
z \d_{z}{\hat{\phi}_z^i}= \left. \frac{\delta \hat{H} [ \hat{\phi}_z,\hat{\Pi}_z ]}{\delta \hat{\Pi}^i_z}  \right|_{\hat{\Pi}_z= - \delta W[\hat{\phi}_z] / \delta \hat{\phi}_z} .
\label{gensol}
\end{equation} 
Remember that $\hat{\phi}_z(x)=\phi(z,x z)$, where we are taking $z$ dimensionful and $x$ dimensionless.

We want to interpret these solutions as renormalised couplings running with the scale $\mu=1/z$. In other words, we want the solutions to provide a 
parametrisation of the renormalised manifold in which the renormalised RG flows obey \refeq{gensol}. We consider only the renormalisable theory $\mathcal{R}$ associated to the fixed point with standard quantisation in all directions. 
As discussed in the previous section, the relevant eigendeformations of this fixed point are in one-to-one correspondence, apart from the constant term, with relevant fields $\phi^b$ (those with a negative squared mass). To any $S^B_{\la t \ra}$ in $\mathcal{R}$ we associate the set of fields $\varphi_t^i$ that extremizes the sum $S^B_{\la t \ra}[\varphi]+W$:
\beq
S^B_{\la t \ra} \to \varphi_t :~~ \frac{\delta}{\delta \varphi_t^i} \left( S^B_{\la t \ra}[\varphi_t]+W[\varphi_t] \right) = 0 .  \label{compatibility}
\eeq
Observe that $S^B_{\la t \ra}$ and $W$ are, respectively, the result of the UV and IR integrations at the classical level, and this extremization corresponds to the remaining integration over $\varphi$~\cite{MPV,Polchinski,Balasubramanian}. Note as well that \refeq{compatibility} can be understood as the requirement of compatibility of the boundary conditions imposed on the classical fields by $S^B_{\la t \ra}$ and $W$. Given a $\varphi_t$ that solves \refeq{compatibility}, we choose a reference scale $\mu$ and define $\hat{\phi}^i_{z}= \varphi^i_{\mu z}$. These $\hat{\phi}^i_{z}$ satisfy \refeq{gensol}, so they are solutions to the equations of motion. This can easily be checked using the $\varphi_t$ evolution derived from $S^B_{\la t \ra}$ together with \refeq{compatibility}. It is the result of connecting the UV evolution of on-shell fields with their IR evolution through the on-shell $\varphi^i$. 

For our parametrisation purposes, we need~\refeq{compatibility} to have a solution for every $S^B_{\la t \ra}$ in $\mathcal{R}$, and the solution to be unique. Let us see that this can be achieved by an adequate choice of $W$. First, we observe that the solution associated to the fixed point $S^B_*$ is constant in $z$, $\hat{\phi}_z=\hat{\phi}_*$. Then, for a Hamiltonian quadratic in momenta we must require that $W$ possesses a critical point at $\hat{\phi}_*$, such that \refeq{gensol} is satisfied for this solultion.\footnote{For any Hamiltonian without linear terms in the momenta, the existence of a critical point is a sufficient condition for \refeq{gensol} to be valid at $\hat{\phi}_*$.} In particular, we must identify $\hat{\phi}_*=\varphi_0=0$. We also require the solutions to be analytic in the field expansion and also in momenta (at $p=0$).\footnote{Usually, the regularity of fields in the deep interior is imposed as an IR boundary condition to calculate physical quantities in Euclidean AdS space. Using the associated $W$ to define a physical renormalisation scheme would lead to a non-local parametrisation.} As we have seen, for scalars in fixed AdS background there is only a finite number of analytic fixed-point solutions to the Hamilton-Jacobi equations with critical points, and all of them have as critical point an extremum of the potential. These solutions are labelled by the values $\Delta_{(i)}^\pm$. 
It is clear that we cannot choose $\Delta_{(i)}=\Delta_{(i)}^+$ for all $i$, since this would lead to infinite solutions at the fixed point, and no solution for perturbations around it. Moreover, imagine that we choose $\Delta_{(i)}=\Delta_{(i)}^+$ for some $i$. Then, \refeq{gensol} generates a solution $\hat{\phi}^i_z$ with a vanishing asymptotic mode $z^{\Delta_{(i)}^-}$. This solution cannot obey the boundary condition imposed by a perturbed density action $\mathcal{S}^B_*+\Tcal^i$. On the other hand, if this direction is not perturbed, \refeq{compatibility} does not fix the asymptotic behaviour of~$\hat{\phi}^i_z$ and the solution will remain undetermined in general. Therefore, we choose $\Delta_{(i)}=\Delta_{(i)}^-$ for all $i$,
\beq
W[\varphi] = \int d^d x \left[\frac{v_0}{d} - \frac{1}{2} \Delta_{(i)}^- \varphi^i \varphi_i + O(\varphi^3) + \mathrm{derivatives} \right].  \label{SIR}
\eeq
Because $W$ here is just a tool to define a scheme, we do not impose restrictions from unitarity on the values $\Delta_{(i)}^-$. Nevertheless, we should point out that below the unitarity bound singularities may arise at certain values of the momenta~\cite{Balasubramanian}. 
With this $W$, \refeq{compatibility} has a unique solution, at least in some neighbourhood of the fixed point. Then, we can define the renormalised constants associated to the renormalised action $S^B_{\la 1 \ra} \in \Rcal$ as 
\beq
g_R^b = \varphi^b_1 =\hat{\phi}^b_{1/\mu}. 
\eeq
The corresponding running couplings are
\beq
\mathcal{F}_t^b(g_R) = \hat{\phi}^b_{1/t\mu}.
\eeq
The renormalised constant parametrising the constant direction is defined as 
the (shifted) constant term in the boundary action:
\beq
g_R^0=\mathcal{S}^B_{\la 1 \ra}(0,0,\ldots) + \frac{v_0}{d} + \mbox{total derivatives}.  \label{gR0}
\eeq
The total derivatives are irrelevant for all purposes.
The map from $\Rcal$ to the space of renormalised couplings $g_R$ must be invertible. The inverse relation can be defined by means of holographic renormalisation. Consider a set of renormalised couplings $g_R^a$, choose a scale $\mu$ and let $\hat{\phi}_z$ be the solution of \refeq{gensol}, with $W$ given by \refeq{SIR}, that satisfies the following conditions: 
\begin{align}
& \hat{\phi}^b_{1/\mu}=g_R^b.  \nn
& \lim_{z\to 0} \hat{\phi}^{\hat{a}} = 0.  \label{solconditions}
\end{align}
Then, we perform the holographic renormalisation, at the same scale $\mu$, with the bare couplings
\beq
h_{t_0}^b(g_R) = \hat{\phi}^b_{1/(t_0 \mu)}.
\eeq
Note that the second condition \refeq{solconditions} ensures that the bare couplings $h^b_{t_0}$ approach the critical point $P$. 
We also need to add the counterterms to get a finite $S^B_R$. As shown in Ref.~\cite{Papadimitriou:2010as}, for our choice of $W$ they are given by
\beq
h_{t_0}^0(g_R) = -\mathcal{W}(\hat{\phi}_{1/(t_0 \mu)},\d \hat{\phi}_{1/(t_0 \mu)},\ldots) + g_R^0 t_0^{-d} + \mbox{total derivatives} , \label{counterterms}
\eeq
where $\mathcal{W}$ is the density of the fixed-point action: $W=\int d^dx \mathcal{W}$.

To show that this procedure does provide the inverse map, consider a particular solution $\hat{\phi}_z$ of \refeq{gensol} and the renormalised theory $S_{R}^B$ obtained from \refeq{finiteSB} with $h_{t_0}=\hat{\phi}_{(t_0 \mu)^{-1}}$. Motion in the radial direction $(\mu t_0)^{-1}$ can be described, with $t_0$ fixed, as a rescaling $\mu \to t \mu$. This rescaling induces the RG transformation $S_{R}^B \to S_{R\, \la t \ra}^B$. By construction, $\hat{\phi}_z$ obeys the boundary condition imposed by $S_{R\, \la t \ra}^B$ at any $z=(\mu t)^{-1}$. Then, $\varphi_t=\hat{\phi}_{t/\mu}$ must be a solution of \refeq{compatibility}. By the uniqueness requirement, solving \refeq{compatibility} with $S_{R\, \la t \ra}^B$ will give the solution $\hat{\phi}_z$ we have started with. 

Conversely, let us start with a given $S_{\la t \ra}^B$ and obtain the associated solution $\hat{\phi}_z$. Perform the holographic renormalisation  with $h_{t_0}=\hat{\phi}_{(t_0 \mu)^{-1}}$ and rescale $\mu \to t \mu$ to obtain $S_{R\, \la t \ra}^B$. The solution $\hat{\phi}_z$ obeys the boundary condition imposed by both $S_{\la t \ra}^B$ and $S_{R\, \la t \ra}^B$, and also the one imposed by $W$. Moreover, both $S_{\la t \ra}^B$ and $S_{R\, \la t \ra}^B$ are boundary actions of the renormalised manifold. Close to the fixed point, they are described by a linear combination of relevant eigenperturbations. From our previous study of the asymptotic behaviours for $t \gg 1$ we know that the coefficients of the relevant perturbations (excluding the constant term) are determined by the leading behaviour of the solution. Therefore, up to a constant term, $S_{\la t \ra}^B$ and $S_{R\, \la t \ra}^B$ have the same functional form for $t \gg 1$. Since they follow the same RG trajectory, they are actually equal for any value of $t$, up to the constant (which does not interfere in the RG evolution). Finally, it is easy to check that \refeq{counterterms} and~\refeq{gR0} are inverses of each other, up to unimportant total derivatives.

In this manner, we have defined a renormalisation scheme in which the running renormalised constants are solutions to the field equations of the gravity theory, and we have shown how to obtain the associated Wilson action.\footnote{This scheme may break down far from the fixed point if the solutions become singular or if they are not determined by \refeq{compatibility}.} This relation between Wilsonian and non-Wilsonian holographic RG flows (with our choice of $W$) precisely matches the general field-theoretical relation between Wilsonian and renormalised (Gell-Mann-Low) RG flows.  Our interpretation of this relation looks quite transparent and explicit to us, but we should point out that it is closely related to previous proposals in \cite{Polchinski} and \cite{Balasubramanian}, respectively. The first proposal~\cite{Polchinski} is a perturbative version of our renormalisation scheme, as can be readily checked by a field expansion of our equations. The second one requires more explanation. The authors of~\cite{Balasubramanian} define a modified Wilson action in which the RG evolution of the coefficients of single-trace operators is given by particular solutions to the equations of motion. Let us briefly review this procedure and show that it is equivalent to the renormalisation scheme presented in this subsection. Let us define a modified boundary action
\beq
S^{B\prime}(g)[\varphi] = S^B(g)[\varphi] + S^{ct}[\varphi],
\eeq
where the functional $S^{ct}$ is analytic in fields and momenta. Let $S^\prime(g)[\pi]$ be the Legendre transform of $S^{B\prime}(g)[\varphi]$ (we assume that the necessary properties of differentiability and convexity are preserved),
\begin{equation}
\begin{gathered}
S^\prime[\pi]=S^B[\phi(\pi)]+W[\phi(\pi)]-\pi \cdot \phi(\pi),\\
\pi_i[\varphi]=\frac{\delta}{\delta \varphi^i} ( S^B[\varphi]+W[\varphi] ), \\
\varphi^i[\pi]= - \frac{\delta S^\prime[\pi]}{\delta \pi_i}.
\end{gathered}
\end{equation} 
This modified Wilson action can be understood as a reparametrisation of the couplings, $S^\prime(g)=S(g^\prime)$. Obviously, the original partition function can be obtained using $S^{B\prime}(g)$ instead of $S^B(g)$ and adding at the same time $S^{ct}[\varphi]$ to the exponent of the integrand in~\refeq{cutoffduality}. The couplings of the single-trace terms of the modified Wilson action are given by
\begin{align}
g_s^i & = - \left. \frac{\delta S^\prime[\pi]}{\delta \pi_i} \right|_{\pi=0} \nn
 & = \varphi^i[0].
\end{align}
The value $\pi=0$ is conjugate to the stationary point $\varphi[0]$ of $S^{B\prime}(g)$. Now, the reason for the equivalence is that
the authors of~\cite{Balasubramanian} make the "maximal substraction" choice $S^{ct} = W$, where $W$ is the same as the one in~\refeq{SIR}.\footnote{A choice of $S^{ct}$ is called a ``renormalisation scheme'' in Ref.~\cite{Balasubramanian}. This should not be confused with the field-theoretical meaning we give to that term in this paper.} In this case, our condition \refeq{compatibility} is the same as the requirement that the modified boundary action $S^{B\prime}(g)$ be stationary. Therefore,
\beq
g_s^i = \varphi^i_{\mathrm{cl}}
\eeq
where $\varphi^i_{\mathrm{cl}}$ is the solution of \refeq{compatibility}. Using in these equations the modified sliding boundary actions
\beq
S^{B\prime}_{\la t \ra} =  S^B_{\la t \ra}+ W,
\eeq
we conclude that the RG evolution of the single-trace couplings of the modified Wilson action, $f_{s,t}^i(g)$, reproduces the solutions $\varphi_{\mathrm{cl}\,t}$ that are generated by $W$. This is true for any RG trajectory. If we start with $S^B$ in the renormalised manifold these solutions are forced to have a specific behaviour close to the AdS boundary, and can be described in terms of $r$ parameters (the number of relevant directions), which can work as renormalised parameters. So, the renormalisation scheme that uses the field solutions can alternatively be understood as a projection into the space of single-trace operators after a suitable reparametrisation of the Wilson action. Of course, this parametrisation carries non-trivial information about the dynamics of the gravity theory in the interior of AdS.


\section{Perturbative calculation of boundary action and beta functions}
\label{sec:examples} 

Let us consider once more a theory of $M$ real scalar fields in fixed $AdS_{d+1}$ space, given by~\refeq{scalarL}, with potential
\beq
V(\phi) = v_0 + \frac{1}{2} m_{(i)}^2\phi^i \phi_i + v_{ijk} \phi^i \phi^j \phi^k . \label{potential}
\eeq
We assume that all these fields are relevant, i.e.\ all of them have negative $m_{(i)}^2$. The other possible fields in the theory (including irrelevant ones) are assumed to decouple from the ``active'' ones in \refeq{potential}.
In this section we calculate explicitly, to cubic order in $\varphi$ and linear order in $v_{ijk}$, the general boundary actions $S^B_R[\varphi]$ that describe the renormalisable theories associated to the fixed point with standard quantisation for all the $M$ fields. We work perturbatively in the bare manifold $\mathcal{B}$ and holographically renormalise to reach points on the renormalised manifold. The results in this section provide a partial check of the more general ones in the appendices, which are obtained instead from the differential Hamilton-Jacobi equation. Moreover, they also probe the renormalised theory far from the fixed-point. Finally, we calculate the Wilsonian beta functions and the renormalised ones in different schemes. To shorten the discussion we consider, once more, generic dimensions and non-integer $\nu_{(i)}$.\footnote{The treatment of exceptional and integer dimensions is very similar, and can be recovered by analytical continuation in the dimensions~\cite{D'Hoker:2000dm}. We refer to Ref.~\cite{MPV}, in which two-point and three-point correlators were calculated in a theory with a UV cutoff, in both AdS and the CFT (using differential regularisation in position space) sides. It was shown there that, for certain exceptional dimensions, logarithms and double logarithms appear in the large UV-cutoff expansion. After renormalisation, these logs remain in the renormalised expressions of the correlation functions and give rise to conformal anomalies. These results for exceptional dimensions have been recovered with a different (momentum-space) CFT method and studied in detail in the new paper~\cite{Bzowski:2015pba}. } We will mostly work in momentum space. We use the following notation: The letter $q$ refers to dimensionless momenta, while the letter $p$ is employed for dimensionful momenta. The fields and solutions $\check{\phi}^i(\epsilon,p)$ refer to the Fourier transform with $p$ conjugate to the dimensionful coordinate, while $\check{\varphi}(q)$ 
and $\check{g}_R(q)$ are Fourier transforms of dimensionless variables. With these conventions, the Fourier transform of ${\hat{\phi}}_z(x)$ is $z^{-d} \check{\phi}(z, q/z)$.

To compute the boundary action we need to integrate out, at the classical level, the degrees of freedom between a UV boundary at $z=\epsilon$ and an IR boundary at $z=l$.  We will sometimes write $\epsilon = l/t$. At $l$ we impose the boundary condition $\phi(l)=\tilde{\varphi}_l$, while at $\epsilon$, following holographic renormalisation, we impose an $\epsilon$-dependent Dirichlet condition, $\phi(\epsilon) = \tilde{h}_{1/(\epsilon\mu)}(g_R)|_\epsilon$. Eventually we will take the limit $\epsilon \to 0$ with fixed $l$, i.e. $t \to \infty$. We perform perturbative calculations in a mixed position/momentum representation, writing the action as in~\refeq{SBmomentum}. Let us define the IR-boundary-to-bulk propagator $\mathcal{K}^{(i)}_{\epsilon,l}(z,p)$ and the UV-boundary-to-bulk propagator $\mathcal{P}_{\epsilon,l}^{(i)}(z,p)$ as solutions of the free theory with boundary conditions
\begin{alignat}{3}
& \mathcal{K}^{(i)}_{\epsilon,l}(\epsilon,p) = 0; ~~ & \mathcal{K}^{(i)}_{\epsilon,l}(l,p)=1; \nn
& \mathcal{P}_{\epsilon,l}^{(i)}(\epsilon,p) = 1; & \mathcal{P}_{\epsilon,l}^{(i)}(l,p)=0 .  
\end{alignat}
Explicitly,
\begin{align}
\mathcal{K}_{\epsilon,l}^{(i)}(z,p) &= \left(\frac{z}{l}\right)^{\frac{d}{2}} \frac{I_{-\nu_{(i)}}(\epsilon p)I_{\nu_{(i)}}(z p)-I_{\nu_{(i)}}(\epsilon p)I_{-\nu_{(i)}}(z p)}{I_{-\nu_{(i)}}(\epsilon p)I_{\nu_{(i)}}(l p)-I_{\nu_{(i)}}(\epsilon p)I_{-\nu_{(i)}}(l p)}\nn
&=\left(\frac{z}{l}\right)^{\frac{d}{2}}  \frac{I_{\nu_{(i)}}(z p)}{I_{\nu_{(i)}}(l p)}+O(\epsilon^2)+O(\epsilon^{2\nu_{(i)}}) ,
\end{align}
and
\begin{align}
\mathcal{P}^{(i)}_{\epsilon,l}(z,p) & =\left(\frac{z}{\epsilon} \right)^{\frac{d}{2}}\frac{I_{-\nu_{(i)}}(zp)I_{\nu_{(i)}}(lp)-I_{-\nu_{(i)}}(lp)I_{\nu_{(i)}}(zp)}{I_{-\nu_{(i)}}(\epsilon p)I_{\nu_{(i)}}(lp)-I_{-\nu_{(i)}}(lp)I_{\nu_{(i)}}(\epsilon p)}\nn
&=\epsilon^{-\Delta^-_{(i)}}\bigg[2^{-\nu_{(i)}}\pi \csc(\pi \nu_{(i)}) l^{\Delta^-_{(i)}}\left(\frac{z}{l}\right)^{d/2}(lp)^{\nu_{(i)}}\frac{I_{\nu_{(i)}} (l p) I_{-\nu_{(i)}} (z p)- I_{-\nu_{(i)}} (l p) I_{\nu_{(i)}} (z p)}{\Gamma(\nu_{(i)})I_{\nu_{(i)}}(lp)} \nn
& \mbox{} +O(\epsilon^2)+O(\epsilon^{2\nu_{(i)}})\bigg].
\end{align}
Both $z$ and the momentum $p$ in these expressions are dimensionful. 
\begin{figure}[t!]
\begin{center}
\includegraphics[width=3.5cm]{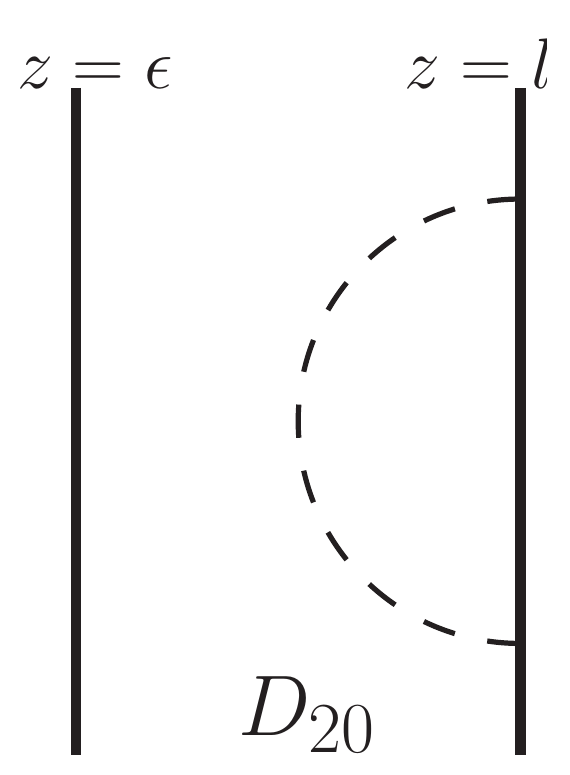}
\hspace{2cm}
\includegraphics[width=3.5cm]{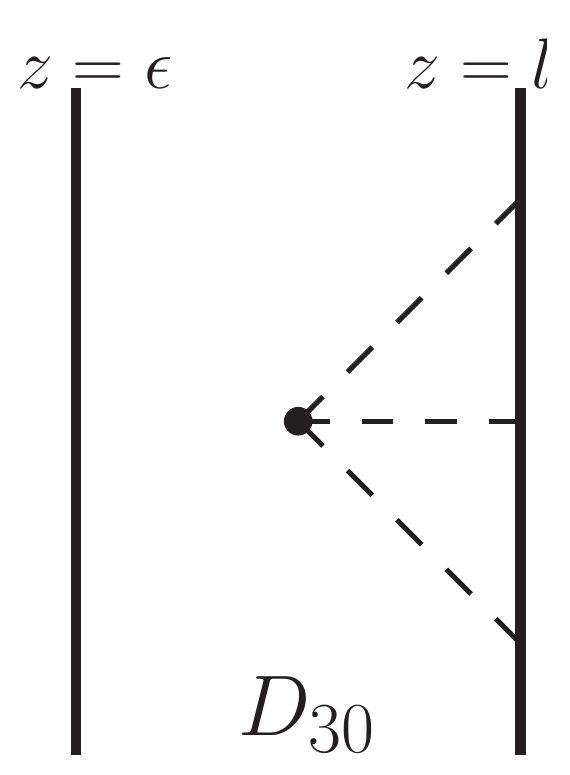}
\caption{Witten diagrams contributing to the boundary action of the fixed point.}
\label{fig:Wittenfixed}
\end{center}
\end{figure}
Let us first calculate the fixed-point action $S^B_*[\varphi]$ to third order in $\varphi$. To do this, we start at the critical point $P$. That is to say, we use $h^i_t(0) = 0$ and $h^0_t(0)=-v_0/d$. The cubic interaction induces terms in $S^B_*$ with arbitrary powers of $\varphi$, as shown in Fig.~\ref{fig:Wittenfixed}.  The constant, vacuum-energy part is given by the limit of the sum of $-\frac{v_0}{d} t^d$ and the integral of the cosmological term in $S^G$:
\begin{align}
\ssf_0 & = \lim_{t\to \infty}  \left[ -\frac{v_0}{d} t^d + l^{-d}\int_{l t^{-1}}^{l} dz \, z^{-d-1} v_0 \right] \nn
& = - \, \frac{v_0}{d}.
\end{align}
The quadratic term is given by a boundary term at $l$ after integration by parts in $S^G$. The two-point function is
\begin{align}
\ssf^{+}_{(i)}(l p) & =  l^d \lim_{\epsilon \to 0}\frac{1}{2} \left.  z^{-d+1} \d_z  \mathcal{K}_{\epsilon,l}^{(i)}(z,p) \right|_{z=l}  \nn
& = \frac{d}{4} + \frac{I'_{\nu_{(i)}}(p l)}{I_{\nu_{(i)}}(p l)},
\label{quadfixS}
\end{align}
with zero-momentum limit
\beq
\ssf^{+}_{(i)}(0) = \frac{1}{2} \Delta_{(i)}^+ .
\eeq
For the cubic term we need to perform a bulk integral. The three-point function is
\begin{align}
s_{ijk}(lp_1,lp_2,lp_3) & = v_{ijk} (2\pi)^d\delta(p_1+p_2+p_3)  \lim_{\epsilon\to0}\int_\epsilon^{l} \frac{dz}{z} z^{-d}\, \mathcal{K}_{\epsilon,l}^{(i)}(z,p_1) \mathcal{K}_{\epsilon,l}^{(j)}(z,p_2) \mathcal{K}_{\epsilon,l}^{(k)}(z,p_3) \nn
& = v_{ijk}  (2\pi)^d\delta(p_1+p_2+p_3) \int_0^{l} \frac{dz}{z} z^{-d}\, \mathcal{K}_{0,l}^{(i)}(z,p_1) \mathcal{K}_{0,l}^{(j)}(z,p_2) \mathcal{K}_{0,l}^{(k)}(z,p_3) . \label{fixedthree}
\end{align}
Note that all these calculations for the fixed-point action are directly finite and do not require any renormalisation.
We can directly see that, to order $\varphi^2$, the fixed-point action $S^B_*$ is identical to the one found in Appendix~\ref{sec:AppendixA} for standard quantisation. We have also checked that the first orders in the momentum expansion of the cubic terms \refeq{fixedthree} precisely agree. More generally, it should be the case that in the limit $\epsilon\to 0$ the Witten diagrams give an integral representation of the solutions to the recursive equations in the appendices.
As explained in Section~\ref{sec:holography}, we have found the fixed point with standard quantisation due to our choice of $P$, which lies on the critical manifold of this particular fixed point. 

\begin{figure}[t!]
\begin{center}
\includegraphics[width=3.5cm]{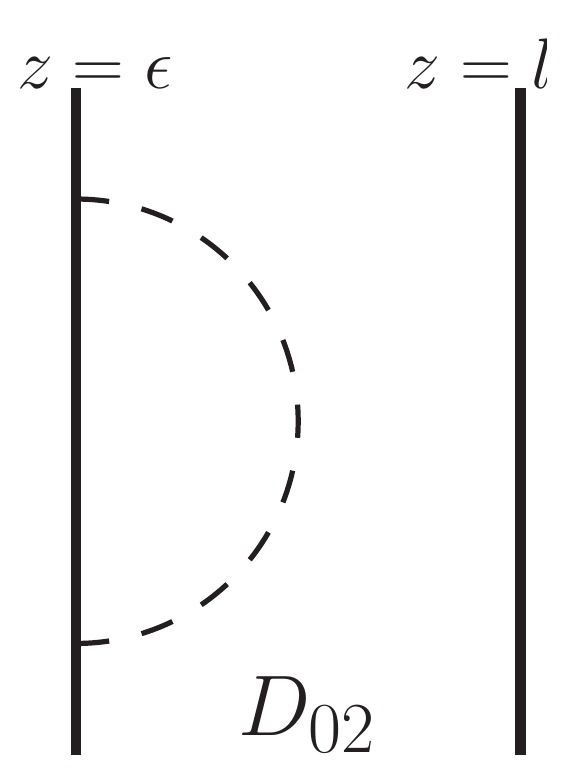}
\hspace{2cm}
\includegraphics[width=3.5cm]{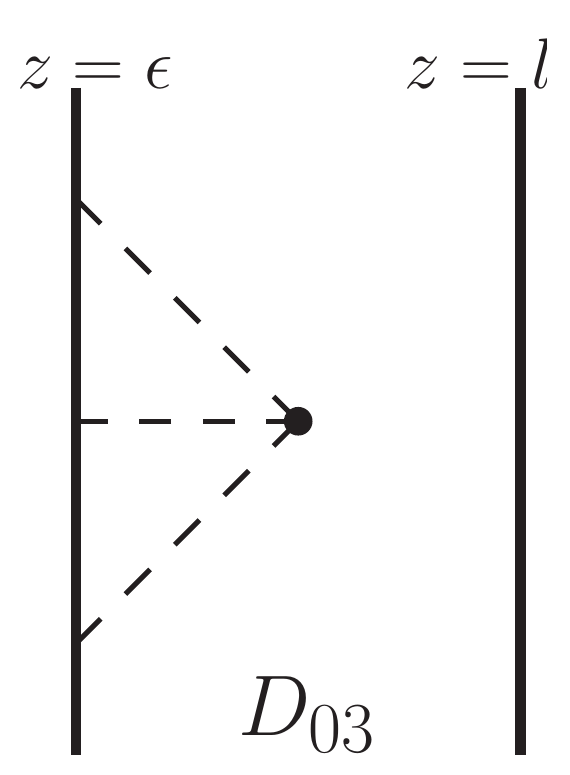}
\caption{Witten diagrams contributing to the vacuum energy of for the deformed theory up to order $O(v_{ijk})$.}
\label{fig:Witten0}
\end{center}
\end{figure}

\begin{figure}[t!]
\begin{center}
\includegraphics[width=3.5cm]{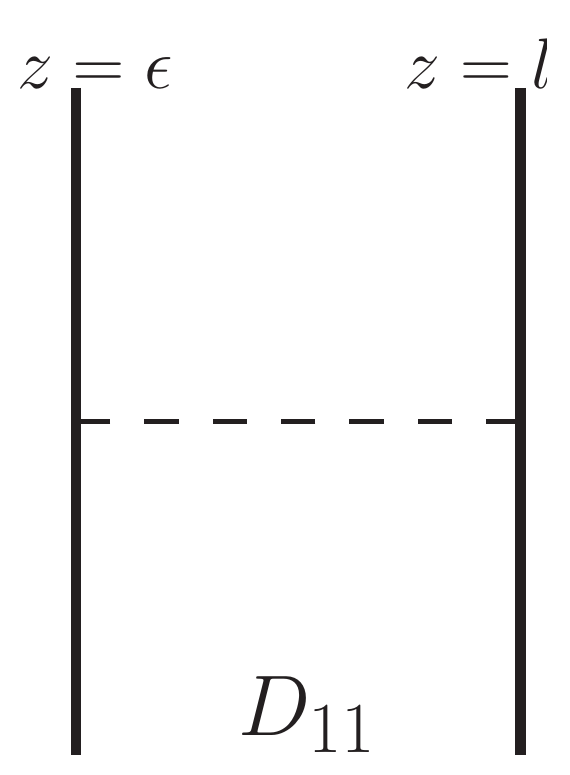}
\hspace{2cm}
\includegraphics[width=3.5cm]{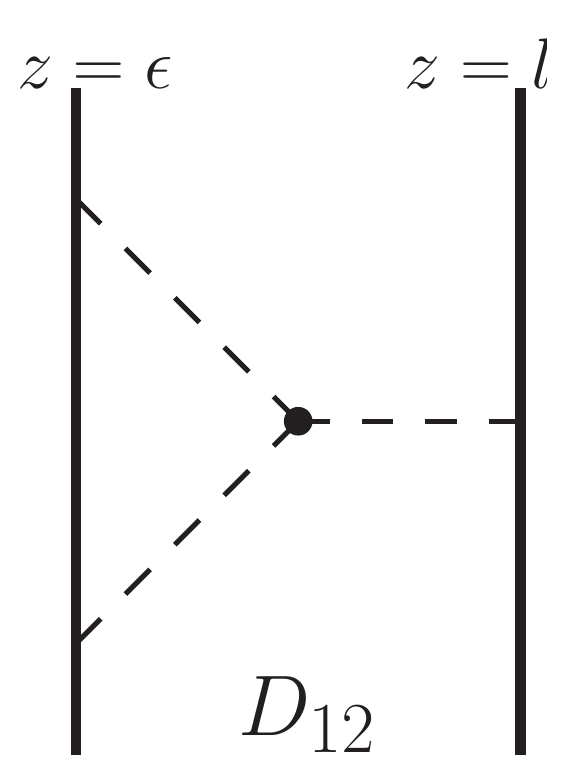}
\caption{Witten diagrams contributing to the linear term in $\check{\varphi}$ for the deformed theory up to order $O(v_{ijk})$.}
\label{fig:Witten1}
\end{center}
\end{figure}

\begin{figure}[t!]
\begin{center}
\includegraphics[width=3.5cm]{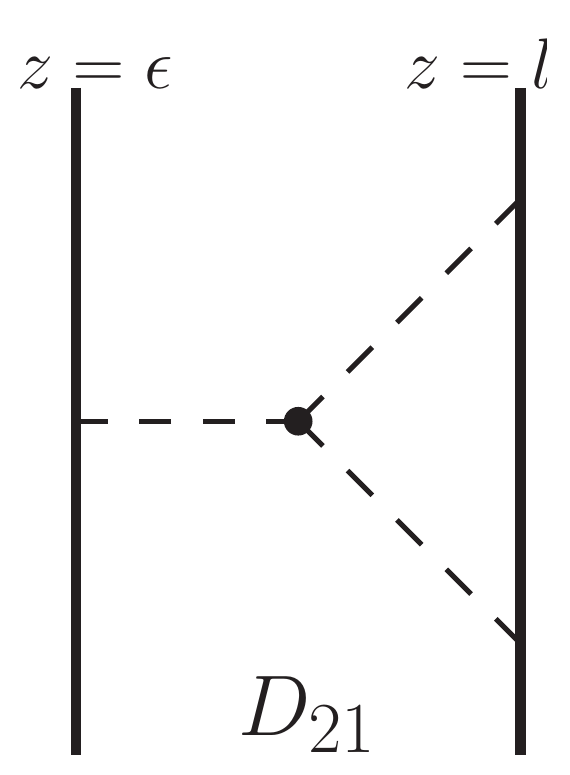}
\caption{Witten diagram contributing to the quadratic term in $\check{\varphi}$ for the deformed theory up to order $O(v_{ijk})$.}
\label{fig:Witten2}
\end{center}
\end{figure}

In order to reach the renormalizable manifold $\mathcal{R}$ of this fixed point, we need non-trivial bare couplings $h_t$. They need to be chosen in such a way that the divergences in the $t\to \infty$ limit be cancelled. As discussed in Sections~\ref{sec:holography} and~\ref{sec:schemes}, this cancellation will be guaranteed if we use as bare couplings adequate solutions of the equations of motion (with the dilatation included). For generic dimensions we can use the solutions generated by $W$ in \refeq{SIR}.\footnote{For exceptional dimensions a fixed-point action $W$ analytic in momenta may not exist. Then (and also in the generic case) we can separate the local part of $W$ and use it in \refeq{gensol} to define the (local) bare couplings, which will no longer be actual solutions of the field equations. This is the method used in~Ref.~\cite{deBoer:1999xf}.} If $\hat{\phi}_z$ is a solution of~\refeq{gensol}, then we take $h^i_t = \hat{\phi}^i_{1/(t \mu)}$. In practice, instead of working with $\hat{H}$ and~\refeq{gensol}, it is easier to undo the rescaling~\refeq{Hhat} and use the original equation
\begin{equation}
\d_{z}{\phi^i}(z,x)= \left. \frac{\delta H_z [\phi,\Pi]}{\delta \Pi_i(z,x)} \right|_{\Pi= - \delta W_z[\phi] / \delta \phi} .
\label{defhflow2}
\end{equation} 
The UV boundary condition is just to impose that the value of the scalar field $i$ at $\epsilon$ be equal to $\phi^i(\epsilon,x)$, with $\phi^i$ a solution of~\refeq{defhflow2}. This does not mean that the on-shell field will be the same as this solution for all values of $z$, as it obeys a different IR boundary condition. 
$W$ is just minus the special fixed point action with alternate quantisation for all fields. From Appendix \ref{sec:AppendixA} we have, to order $v_{ijk}$,
\begin{align}
W[\check{\varphi}] =& -\int \frac{d^dq}{(2\pi)^d} \sum_a \ssf^{-}_{(i)}(q)\check{\varphi}^i(q)\check{\varphi}_i(-q)\nn
& \hspace{-1cm} \mbox{} - \int \frac{d^dq_1}{(2\pi)^d}  \frac{d^dq_2}{(2\pi)^d}  \frac{d^dq_3}{(2\pi)^d} (2\pi)^d\delta(q_1+q_2+q_3)\ssf^-_{ijk} \check{\varphi}^i(q_1) \check{\varphi}^j(q_2) \check{\varphi}^k(q_3)+O(v_{ijk}^2) \nn
 =& - \int \frac{d^dq}{(2\pi)^d} \left[\frac{d}{4}+ q \frac{I'_{-\nu_{(i)}}(q) }{ 2I_{-\nu_{(i)}}(q)} \right] \check{\varphi}^i(q) \check{\varphi}_i(-q)\nn
& \hspace{-1cm}\mbox{} - v_{ijk}\int \frac{d^dq_1 d^dq_2 d^dq_3}{(2\pi)^{2d}}\delta(q_1+q_2+q_3) \check{\varphi}^i(q_1)\check{\varphi}^j(q_2)\check{\varphi}^k(q_3)\bigg[ \frac{1}{\Delta_{(i)}^-+\Delta_{(j)}^-+\Delta_{(k)}^--d} \nn
&\hspace{-1cm}\mbox{} + \frac{1}{2}\frac{\frac{q_1^2}{\nu_{(i)}-1}+\frac{q_2^2}{\nu_{(j)}-1}+\frac{q_3^2}{\nu_{(k)}-1}}{(\Delta_{(i)}^-+\Delta_{(j)}^-+\Delta_{(k)}^-+2-d)(\Delta_{(i)}^-+\Delta_{(j)}^-+\Delta_{(k)}^--d)}+O(q^4)\bigg]+O(v_{ijk}^2).\label{Wused}
\end{align}
$W_z[\phi]$ can be easily obtained from $W[\check{\varphi}]$ 
changing the Euclidean metric by the induced metric at $z$. 
To order $v_{ijk}$, (\ref{defhflow2}) is 
\begin{align}
&\check\phi^i(\epsilon,p)=(\epsilon \mu)^{\Delta^-_{(i)}}\frac{\Gamma(1-\nu_{(i)})}{2^{\nu_{(i)}}}(\epsilon p)^{\nu_{(i)}}I_{-\nu_{(i)}}(\epsilon p_i) \mu^{-d} \check g^i_R(p/\mu)
+ 3 v^i_{jk}(\epsilon \mu)^{\Delta^-_{(j)}+\Delta^-_{(k)}}\nn
&\times \int \frac{d^dp_{1} d^dp_{2}}{(2\pi)^d} \delta(p+p_{1}+p_{2})\mu^{-2d} \check g_R^{j} (p_{1}/\mu)\check g_R^{k}(p_{2}/\mu)\bigg[ \frac{1}{(\Delta^-_{(j)}+\Delta^-_{(k)}-\Delta^-_{(i)})(\Delta^-_{(j)}+\Delta^-_{(k)}-\Delta^+_{(i)})}  \nn
&\mbox{}+\epsilon^2\frac{4p^2-\left(\frac{p_{1}^2}{\nu_{(j)}-1}+\frac{p_{2}^2}{\nu_{(k)}-1}\right)(\Delta^-_{(j)}+\Delta^-_{(k)}-\Delta^-_{(i)})(\Delta^-_{(j)}+\Delta^-_{(k)}-\Delta^+_{(i)})}{4(\Delta^-_{(j)}+\Delta^-_{(k)}-\Delta^-_{(i)})(\Delta^-_{(j)}+\Delta^-_{(k)}-\Delta^+_{(i)})(\Delta^-_{(j)}+\Delta^-_{(k)}-\Delta^-_{(i)}+2)(\Delta^-_{(j)}+\Delta^-_{(k)}-\Delta^+_{(i)}+2)} \nn
&\mbox{} +O(\epsilon^4 p^4)\bigg].
\label{solution2}
\end{align}
The $O((v_{ijk})^0)$ part of the solution $\phi^i$, which we call $\phi^i_0$, has the following momentum expansion
\begin{equation}
\check \phi^i_0(\epsilon,p) 
=\mu^{-d}(\epsilon \mu)^{\Delta^-_{(i)}} \left[1+\frac{\epsilon^2p^2}{4-4\nu_{(i)}}+O(\epsilon^4p^4)  \right] \check g^i_R(p/\mu).
\label{solution}
\end{equation}
Depending on the set of dimension, the $O(v_{ijk})$ terms may give important contributions that cancel subdivergences. On the other hand, high-enough orders in $\epsilon$ will not contribute in the limit $\epsilon \to 0$. It will be useful to write the solutions as
\begin{equation}
\check \phi^i(\epsilon,p)=\check \phi^i_0(\epsilon,p)+\int\frac{d^dp_1 d^dp_2}{(2\pi)^d} \delta(p+p_1+p_2)\check \phi^j_0(\epsilon,p_1)\check \phi^k_0(\epsilon,p_2)\Omega^i_{~jk}(\epsilon p,\epsilon p_1, \epsilon p_2) +\mathcal{O}(v_{ijk}^2),
\label{iterativesol}
\end{equation}
where the function $\Omega_{ijk}$ is defined, to all orders in momenta, as the analytic solution at $q_m=0$ of the equation
\begin{align}
& \left(-2\ssf^{-}_{(i)}(q_1)+2\ssf^{-}_{(j)}(q_2)+2\ssf^{-}_{(k)}(q_3)\right)\Omega_{ijk}(q_1,q_2,q_3)+\sum_{m=1}^3 q_m\partial_{q_m}\Omega_{ijk}(q_1,q_2,q_3) \nn
& ~~~~ = 3 \ssf^-_{ijk}(q_1,q_2,q_3), \label{omegaeq}
\end{align}
where the $\ssf^-$ are the coefficients of the $\check{\varphi}$ series of the action $W$.
The first terms of its low-momentum expansion are 
\begin{align}
&\Omega_{ijk}(q,q_1,q_2) = 3 v_{ijk} \left[ \frac{1}{(\Delta^-_{(j)}+\Delta^-_{(k)}-\Delta^-_{(i)})(\Delta^-_{(j)}+\Delta^-_{(k)}-\Delta^+_{(i)})} \right. \nn
& +\frac{q^2+\left[\frac{q_{1}^2}{2(\nu_{(j)}-1)}+\frac{q_{2}^2}{2(\nu_{(k)}-1)}\right](d-\nu_{(j)}-\nu_{(k)}+2)}{(\Delta^-_{(k)}+\Delta^-_{(j)}-\Delta^-_{(i)})(\Delta^-_{(k)}+\Delta^-_{(j)}-\Delta^+_{(i)})(\Delta^-_{(k)}+\Delta^-_{(j)}-\Delta^-_{(i)}+2)(\Delta^-_{(k)}+\Delta^-_{(j)}-\Delta^+_{(i)}+2)}  \nn
& \left. \mbox{} +\mathcal{O}(q^4) \right].
\label{Omega}
\end{align}
To calculate $S^B_R(g_R)$ to linear order in $v_{ijk}$, we need to add to $S^B_*$ the contribution of the diagrams in Figs.~\ref{fig:Witten0},~\ref{fig:Witten1} and~\ref{fig:Witten2}, which contribute to the vacuum energy, to the linear term in $\check{\varphi}$  and to the quadratic term in $\check{\varphi}$, respectively.

To order $v_{ijk}$, the vacuum energy density $s_0$ receives corrections $\delta s_0$ with two and three legs on the UV boundary, as shown in Fig.~\ref{fig:Witten0}. They are respectively equal to
\begin{align}
D_{02} &= -  \frac{1}{2}  \int \frac{d^dp}{(2\pi)^d} \, \check \phi^i(\epsilon,p) \check\phi_i(\epsilon,-p) \left. z^{-d+1} \d_z \Pcal^{(i)}_{\epsilon,l}(z,p) \right|_{z=\epsilon} \nn
& =  \int \frac{d^dp}{(2\pi)^d}\,\check \phi^i(\epsilon,p) \check\phi_i(\epsilon,-p) \epsilon^{-d} \left[- \frac{d}{4}-\epsilon p\frac{I'_{-\nu_{(i)}}(\epsilon p)}{ 2I_{-\nu_{(i)}}(\epsilon p)}\right] + O(\epsilon^0) \nn 
& =  -   \int \frac{d^dp}{(2\pi)^d} \, \check\phi^i(\epsilon,p) \check\phi_i(\epsilon,-p) \epsilon^{-d} \ssf^{-}_{(i)}(\epsilon p) + O(\epsilon^0)
\label{D02}
\end{align}
and
\begin{align} 
D_{03} & = v_{ijk} \int \frac{d^dp_1}{(2\pi)^d} \frac{d^dp_2}{(2\pi)^d} \frac{d^dp_3}{(2\pi)^d}(2\pi)^d \delta(p_1+p_2+p_3) \check\phi^i(\epsilon,p_1) \check\phi^j(\epsilon,p_2) \check\phi^k(\epsilon,p_3) \nn
&~~~~~~~~~~~~~\mbox{} \times \int_\epsilon^l dz z^{-1-d}\, \Pcal^{(i)}_{\epsilon,l}(z,p_1) \Pcal^{(j)}_{\epsilon,l}(z,p_2)\Pcal^{(k)}_{\epsilon,l}(z,p_3) . \label{D03}
\end{align}
In these expressions the $\phi^i$ are not generic fields, but the solutions~\refeq{solution2}. Because they contain not only linear but also quadratic terms in the renormalised couplings $g_R$, proportional to $v_{ijk}$, $D_{02}$ will contribute to order $v_{ijk}$ to $\delta s_0^{[2]}$ (terms with two couplings $g^i_R$ and  $g^j_R$), and also to $\delta s_0^{[3]}$ (terms with three couplings $g^i_R$, $g^j_R$ and $g^k_R$). The diagram $D_{03}$ contributes only to $\delta s_0^{[3]}$ at this order. 

The limit $\epsilon \to 0$ of both $D_{02}$ and $D_{03}$ is divergent. However, all the non-local divergences are nicely cancelled out by the divergent terms in the $\phi^i$. Here and in the following ``local'', ``non-local'' and ``semi-local'' refer to terms that have these properties in the limit in which the IR cutoff is removed (with the IR cutoff, all divergences are actually local). These cancellations are not trivial. The completely non-local divergences in $\delta s_0^{[2]}$ and in $\delta s_0^{[3]}$ are cancelled, as is well-known, by the $\phi_0$ terms in $\phi$, which are linear terms in $g_R$. This widely-employed linear renormalisation of the sources is insufficient in some cases. Indeed, as found in Ref.~\cite{MPV}, $D_{03}$ contains semi-local divergent terms when, for some $i$, $j$, $k$ (not necessarily different), $\Delta_{(i)}^- + \Delta_{(j)}^- < \Delta_{(k)}^-$. These terms cannot possibly be cancelled by a linear renormalisation, and seem to require a non-local divergent counterterm. However, as mentioned above, the non-linear terms in $\phi$ that appear in $D_{02}$ give another contribution to $\delta s_0^{[3]}$. It turns out that this contribution precisely cancels the semi-local divergences. The remaining local divergence can then be cancelled by a local counterterm in the vacuum energy. Let us show all this explicitly. 

First, using $\phi \to \phi_0$ in \refeq{D02} we get
\begin{align}
&(\delta s_0^{[2]})_{D_{02}}(\epsilon p) = - \int \frac{d^dp}{(2\pi)^d} \, \check g_R^i(p/\mu) \check g_{R\,i}(-p/\mu) (\mu \epsilon)^{-2 \nu_{(i)}} \ssf^{-}_{(i)}(\epsilon p)\nn
&+\int \frac{d^dp}{(2\pi)^d} \,\check g_R^i(p/\mu) \check g_{R\,i}(-p/\mu)\left(\frac{p}{\mu}\right)^{2\nu_{(i)}}\frac{\Gamma(1-\nu_{(i)})^2\sin[\pi \nu_{(i)}]I_{-\nu_{(i)}}(lp)}{4^{\nu_{(i)}}\pi I_{\nu_{(i)}}(lp)}+O(\epsilon^{2\nu_{(i)}}).
\end{align}
The local divergence is to be cancelled by a counterterm in the vacuum energy, as we discuss below. The $O(v_{ijk})$ part of $D_{02}$, arising from nonlinear terms in $\phi$, is 
\begin{align}
(\delta s_0^{[3]})_{D_{02}}(\epsilon p_1,\epsilon p_2,\epsilon p_3) = &  -  \int \frac{d^dp_1}{(2\pi)^d} \frac{d^dp_2}{(2\pi)^d} \frac{d^dp_3}{(2\pi)^d} \, (2\pi)^d \delta(p_1+p_2+p_3)  \nn
& \mbox{} \times \check \phi_0^i(\epsilon, p_1) \check\phi_0^j(\epsilon, p_2) \check\phi_0^k(\epsilon, p_3) \epsilon^{-d} \Omega_{ijk}(\epsilon p_1,\epsilon p_2,\epsilon p_3) z\d_z \Pcal^{(i)}_{\epsilon,l}(\epsilon p_1).
\end{align}
The $z$ integral in~\refeq{D03} can be written as,
\begin{align}
v_{ijk} \int_{\epsilon}^{l_0} & dz   z^{-1-d}\mathcal{P}_{\epsilon,l}^{(i)}(z,p_1) \mathcal{P}_{\epsilon,l}^{(j)}(z,p_2) \mathcal{P}_{\epsilon,l}^k(z,p_3)\nn
\mbox{}  = &  \epsilon^{-d} \left[ \mathcal{Z}_{ijk}(\epsilon p_1, \epsilon p_2, \epsilon p_3)+ \mathcal{Y}_{ijk}(\epsilon p_1, \epsilon p_2, \epsilon p_3)+
 \mathcal{Y}_{jik}(\epsilon p_2, \epsilon p_1, \epsilon p_3) \right. \nn
& \phantom{\epsilon^{-d} } \left. \mbox{} + \mathcal{Y}_{kij}(\epsilon p_3, \epsilon p_1, \epsilon p_2) +
\mathcal{O}(\epsilon^{2\nu_{(j)}+2\nu_{(k)}}) + \mathcal{O}(\epsilon^{2\nu_{(i)}+2\nu_{(j)}}) + \mathcal{O}(\epsilon^{2\nu_{(i)}+2\nu_{(k)}}) \right]\nn
& \mbox{} + O(\epsilon^{-\Delta_{(i)}^- - \Delta_{(j)}^- - \Delta_{(k)}^-}),  \label{Zexpression}
\end{align}
with
\begin{align}
&\mathcal{Z}_{ijk}(\epsilon p_1,\epsilon p_2,\epsilon p_3)= -v_{ijk} \left.\int \frac{dz}{z} \left(\frac{z}{\epsilon}\right)^{d/2} \frac{I_{-\nu_{(i)}}(zp_1) I_{-\nu_{(j)}}(zp_2) I_{-\nu_{(k)}}(zp_3)}{I_{-\nu_{(i)}}(\epsilon p_1) I_{-\nu_{(j)}}(\epsilon p_2) I_{-\nu_{(k)}}(\epsilon p_3)}\right|_{z=\epsilon}\nn
&=- v_{ijk} \bigg[ \frac{1}{\Delta_{(i)}^-+\Delta_{(j)}^-+\Delta_{(k)}^--d} \nn
&\left. \mbox{}+\frac{\epsilon^2}{2}\frac{\frac{p_1^2}{\nu_{(i)}-1}+\frac{p_2^2}{\nu_{(j)}-1}+\frac{p_3^2}{\nu_{(k)}-1}}{(\Delta_{(i)}^-+\Delta_{(j)}^-+\Delta_{(k)}^-+2-d)(\Delta_{(i)}^-+\Delta_{(j)}^-+\Delta_{(k)}^--d)}+O(\epsilon^4 p_1^4)\right] \nn
& =- \ssf_{ijk}^- \,,
\end{align}
and
\begin{align}
&\mathcal{Y}_{ijk}(\epsilon p_1,\epsilon p_2,\epsilon p_3)=
\frac{\tilde{\Omega}_{ijk}(\epsilon p_1,\epsilon p_2,\epsilon p_3)}{3}\left[  z\d_z \Pcal^{(i)}_{\epsilon,l}(z,p_1)|_{z=\epsilon}- 2 \ssf^{-}_{(i)}(\epsilon p_1)  \right],\nn
&\tilde{\Omega}_{ijk}(\epsilon p_1,\epsilon p_2,\epsilon p_3)=- 3v_{ijk} \frac{\pi}{2 \sin(\pi \nu_{(i)})} \frac{1}{I_{-\nu_{(j)}}(\epsilon p_2)I_{-\nu_{(k)}}(\epsilon p_3)} \nn
&\times \left.\int\frac{dz}{z} \left(\frac{z}{\epsilon}\right)^{d/2} \left\{I_{-\nu_{(j)}}(zp_2 ) I_{-\nu_{(k)}}(zp_3 )\left[ I_{-\nu_{(i)}}(\epsilon p_1)I_{\nu_{(i)}}(zp_1)-I_{-\nu_{(i)}}(zp_1)I_{\nu_{(i)}}(\epsilon p_1)  \right]\right\}\right|_{z=\epsilon}
\end{align}
The indefinite integrals above and hereafter are defined as the primitive with vanishing constant term in the Laurent expansion at $z=0$. It can be shown that $\tilde{\Omega}$ satisfies the defining equation~\refeq{omegaeq}, so in fact $\tilde{\Omega}=\Omega$. When used in~\refeq{D03}, $\mathcal{Z}$ gives local divergences if $\nu_{(i)}+\nu_{(j)}+\nu_{(k)}>d/2$ . The same is true for the second term in $\mathcal{Y}$. On the other hand, the first term, $\Omega z\d_z \Pcal$, gives semi-local divergences when $\Delta^-_{(i)} + \Delta^-_{(j)} < \Delta^-_{(k)}$ for some $i,j,k$. But we see that this contribution cancels exactly against $\delta s_0^{[3]}$. From the arguments in Section~\ref{sec:holography} and previous work on holographic renormalisation we know this should be the case when we use solutions as bare couplings. All this agrees as well with the field-theoretical discussion in Section~\ref{sec:fieldtheory}. In particular, note that the same condition on the dimensions gives rise to the non-linear contributions.
Finally, the local divergence that remains in $\delta s_0$ can and should be cancelled by a counterterm, which can be chosen as
\beq
h_{1/(\epsilon\mu)}^0(x) =  - \mathcal{W}[\hat{\phi}_\epsilon(x),\d \hat{\phi}_\epsilon(x),\ldots] +  (\epsilon \mu)^d g_R^0 (\epsilon \mu\, x).\label{h0ren}
\eeq
In this equation, $g_R^0$ parametrises the renormalised vacuum energy. The integral of the first term,
\begin{align}
W[\hat{\phi}_\epsilon] = & - \int \frac{d^dp}{(2\pi)^d} \epsilon^{-d} \ssf^{-}_{(i)}(\epsilon p)\check\phi_0^i(\epsilon,p)\check\phi_{0\,i}(\epsilon,-p)\nn
&-\int \frac{d^dp_1}{(2\pi)^d} \frac{d^dp_2}{(2\pi)^d} \frac{d^dp_3}{(2\pi)^d} (2\pi)^d\delta(p_1+p_2+p_3) \epsilon^{-d} \check\phi_0^i(\epsilon,p_1) \check\phi_0^j(\epsilon,p_2) \check\phi_0^k(\epsilon,p_3) \nn
& ~~~~\mbox{} \times \left[ \ssf^-_{ijk}(\epsilon p_1,\epsilon p_2,\epsilon p_3) +2 \ssf^-_{(i)} (\epsilon p_1) \Omega_{ijk}(\epsilon p_1,\epsilon p_2,\epsilon p_3) \right] + O(v_{ijk}^2),
\end{align} 
manifestly cancels all the remaining divergences.

Even if we are using non-local bare couplings $h_t$, all the terms that survive in the limit $\epsilon \to 0$ are actually local. In fact, we are oversubstracting (much as the ``maximal substraction'' in~\cite{Balasubramanian}), but we could equivalently use the complete solutions or the first terms in the $\epsilon$ expansion, up to the necessary order, which are polynomial in momenta and thus local in position space.

Let us consider next the linear term $\int d^dq \delta s_i(q) \check{\varphi}(q)$ of $S^B_R$. Remember that $s_i=0$ in the fixed-point action. In the presence of sources, it is given to order $v_{ijk}$ by the diagrams in Fig.~\ref{fig:Witten1}, with either one or two legs on the UV boundary. The contribution of the first diagram to $\delta s_i(p l)$ is
\begin{align}
D_{11} & = \frac{1}{2}  \phi^i(\epsilon,-p)  \left(\left. z^{-d+1} \d_z \Pcal^{(i)}_{\epsilon,l}(z,p) \right|_{z=l} - \left. z^{-d+1} \d_z \Kcal^{(i)}_{\epsilon,l}(z,p) \right|_{z=\epsilon} \right) \nn
& = \phi^i(\epsilon,-p) B_{\epsilon,l}(p ) ,
\end{align}
where
\begin{align}
B_{\epsilon,l}(p)& =  \frac{1}{\pi}\frac{2 \epsilon ^{-d/2} l^{-d/2} \sin (\pi \nu)}{I_{-\nu_{(i)}}(lp) I_{\nu_{(i)}}(\epsilon p)-I_{-\nu_{(i)}}(\epsilon p) I_{\nu_{(i)}}(lp)} \nn
& = \epsilon^{-\Delta^-_{(i)}}l^{-\Delta^+_{(i)}}\left[\frac{-2\nu_{(i)}(lp)^{\nu_{(i)}}}{2^{\nu_{(i)}} \Gamma(\nu_{(i)}+1) I_{\nu}(lp)} +O(\epsilon^2)+O(\epsilon^{2\nu}) \right]. \label{Neumannbranetobrane}
\end{align}
This contributes to $\delta s_i^{[1]}$, with one renormalised coupling $g_R^j$ and ---through the non-linear terms in $\phi^i$--- to $\delta s_i^{[2]}$, with two renormalised couplings $g_R^j$ and $g_R^k$. In $\delta s_i^{[1]}$, the divergence is directly cancelled:
\beq
\delta s_i^{[1]}(q) = -2\nu_{(i)}\check g_{R\,i}(q) (l\mu)^{-\Delta^+_{(i)}} \frac{q^{\nu_{(i)}}}{2^{\nu_{(i)}} \Gamma(\nu_{(i)}+1) I_{\nu}(q)}
\eeq
The contribution to $\delta s_i^{[2]}$ is
\beq
(\delta s_i^{[2]})_{D_{11}}(lp) = \int \frac{d^dp_1 d^dp_2}{(2\pi)^d} \, \delta(p+p_1+p_2)\check \phi_0^j(\eps,p_1)\check \phi_0^k(\eps,p_2) \Omega_{ijk}(\eps p,\eps p_1,\eps p_2) B_{\epsilon,l}(p)
\eeq
The contribution of the second diagram to $\delta s_i(p l)$ is
\begin{align}
D_{12} = &3 v_{ijk}  \int \frac{d^dp_1}{(2\pi)^d} \frac{d^dp_2}{(2\pi)^d} \, (2\pi)^d \delta(p_1+p_2+p) \check \phi^j(\epsilon,p_1)\check \phi^k(\epsilon,p_2) \nn
&\times \int_{\epsilon}^{l} dz  z^{-1-d}\mathcal{P}_{\epsilon,l}^{(j)}(z,p_1) \mathcal{P}_{\epsilon,l}^{(k)}(z,p_2) \mathcal{K}_{\epsilon,l}^{(i)}(z,p).
\label{UV2IR1}
\end{align}
The completely non-local divergences can be easily seen to cancel out in this expression. But again, a semi-local divergence remains when, for some fixed $i,j,k$, with $\check{\varphi}^i$ the field on the IR boundary, the corresponding dimensions satisfy $\Delta_{(j)}^- + \Delta_{(k)}^- < \Delta_{(i)}^-$:
\begin{equation}
(\delta s_i^{[2]})_{D_{12}}(p l) = - \int \frac{d^dp_1 d^dp_2}{(2\pi)^d} \, \delta(p+p_1+p_2) \check\phi_0^j(\eps,p_1) \check\phi_0^k(\eps,p_2) \tilde{\Omega}_{ijk}(\eps p,\eps p_1,\eps p_2) B_{\epsilon,l}(p)
+ O(\epsilon^0).
 \label{divform}
\end{equation}
The cancellation of the divergences of $(\delta s_i^{[2]})_{D_{12}}$ and $(\delta s_i^{[2]})_{D_{11}}$ is manifest. After this, no divergences remain in $\delta s_i^{[1]}$ and $\delta s_i^{[2]}$, so we are directly left with the renormalised $\delta s_i$.

The last correction to the fixed-point action to order $v_{ijk}$ is quadratic in $\check{\varphi}$. The only corrrection to the two-point function, $\delta s_{ij}$, is given by the diagram in Fig.~\refeq{fig:Witten2} and reads
\begin{equation}
\delta s_{jk}(l p_1,l p_2)=3v_{ijk} \int d^dp \, \delta(p_1+p_2+p) \check\phi^i(\epsilon,p) 
 \int_{\epsilon}^{l} dz  z^{-1-d}\mathcal{P}_{\epsilon,l}^{(i)}(z,p) \mathcal{K}_{\epsilon,l}^{(j)}(z,p_1) \mathcal{K}_{\epsilon,l}^{(k)}(z,p_2).
\label{UV1IR2}
\end{equation}
The non-local divergence of the integral is cancelled by the $\phi$ (only the $\phi_0$ part contributes to $O(v_{ijk})$, and a finite expression remains, so that the limit $\eps\to 0$ of \refeq{UV1IR2} gives directly the renormalised $\delta s_{ij}(q_1,q_2)$.

We have found $S_R^B(g_R)[\varphi]$ in a particular scheme given by holographic renormalisation with solutions parametrised by their asymptotic behaviour. We have shown explicitly that it is finite. The fixed-point action is $S_*^B(0)$, which we have calculated at the beginning of this section. The renormalised manifold is tangent to the relevant directions at the fixed point. Therefore,
\beq
\left. \frac{\d S_R^B(g_R)[\varphi]}{\d g_R^a} \right|_{g_R=0} = \mathcal{N}^b_a Q_b .
\eeq
In the case we are studying, all the relevant egenperturbations are basic functions $\mathcal{T}^i=\varphi^i + \ldots$. Because the UV to IR propagator is diagonal, the matrix $\mathcal{N}$ is also diagonal in this basis. Comparing with $\delta s_i^{[1]}$ we also find the normalisation:
\beq
\left. \frac{\d S_R^B(g_R)[\varphi]}{\d g_R^i} \right|_{g_R=0} = (d-2\Delta^+_{(i)}) (\mu l)^{-\Delta^+_{(i)}} \mathcal{T}_i .
\eeq
We have checked that this equation is also satisfied at order $v_{ijk}$ for the first terms in the momentum expansion. 

The renormalised Wilson action $S_R(g_R)[\pi]$ to order $v_{ijk}$ can be readily found by a perturbative Legendre transform of $S_R^B(g_R)[\varphi]$. We do not do it here explicitly. Instead, we proceed to calculate the Wilsonian and renormalised beta functions for the scalar theory at hand.


As we have discussed in Section~\ref{sec:holography}, the Wilsonian beta functions of both the couplings and of the conjugate couplings can be directly computed from the Hamiltonian, written in terms of the Wilson and boundary action, respectively. To facilitate the comparison with renormalised beta functions below we choose to work with boundary variables. We continue working in (dimensionless) momentum space. Instead of working in the basis of eigenperturbations, it is simpler to use the basis given by products of $\varphi^i$. The conjugate couplings $s$ are then the coefficients of the Taylor expansion of $S^B[\varphi]$ in $\varphi$, 
\begin{align}
S^B(s)[\check{\varphi}]& = s_0 +\int \frac{d^dq_1}{(2\pi)^d} s_i(q_1)  \check{\varphi}^i(q_1)+ \int \frac{d^dq_1}{(2\pi)^d} \frac{d^dq_2}{(2\pi)^d} s_{i_1i_2}(q_1,q_2) \check{\varphi}^{i_1}(q_1) \check{\varphi}^{i_2}(q_2) \nonumber \\
& \mbox{} + \sum_{n\geq3} \int \frac{d^dq_{i_1}}{(2\pi)^d}...\frac{d^dq_{i_n}}{(2\pi)^d} s_{i_1...i_n}(q_1,...,q_n) \check{\varphi}^{i_1}(q_1)...\check{\varphi}^{i_n}(q_n),
\end{align}
The Wilsonian beta functions of the boundary couplings $s$ can be easily performed using Eq.~\refeq{betaSB}. We find
\begin{align}
\beta^B_{j_1...j_n}(s)(q_{j_1}...q_{j_n})&=\frac{1}{2}\sum_{k=0}^{n} (k+1)(n-k+1) \underset{\{(j_k,q_k)\}_{k=1}^n}{\mathrm{Sym}}\int \frac{d^dq}{(2\pi)^d}\left[ s^i_{~j_1...j_k}(-q,q_1,...,q_k) \nonumber \right. \\
&  \mbox{} \times \left. s_{j_{k+1}...j_ni}(q_{k+1},...,q_n,q)\right] +  \sum_{l=1}^{n} q_l^{\mu} \frac{\partial}{\partial q_l^{\mu}} s_{j_1...j_n}(q_1,...,q_n)\nonumber \\
& -\frac{1}{2}\delta_{2n}(2\pi)^d\delta(q_1+q_2)\left(q_1^2+\delta_{j_1 j_2}m^2_{(j_1)}\right)-\delta_{3n}(2\pi)^d\delta(q_1+q_2+q_3)v_{j_1j_2j_3}.
\label{wilsonianbeta}
\end{align}
where $\mathrm{Sym}$ symmetrises over the pairs $\{(j_k,q_k)\}_{k=1}^n$:
\begin{equation}
\underset{\{(j_k,q_k)\}_{k=1}^n}{\mathrm{Sym}}  A_{j_1...j_n}[q_1,...,q_n]= \frac{1}{n!}\sum_{\sigma \in S_n} A_{j_{\sigma(1)}...j_{\sigma(n)}}[q_{\sigma(1)},...,q_{\sigma(n)}].
\end{equation}
Here, $\beta^B_{j_1 \ldots j_n}$ is the Wilsonian boundary beta function in the direction of $s_{j_1 \ldots j_n}$.
Notice how, in general, the beta function for the coupling $s_{i_1...i_n}$ with $n$ indices is affected by the couplings with $n+1$ indices or less.
In Appendix~\ref{sec:AppendixA}, the fixed points are computed by requiring that these beta functions vanish. 
To order $v_{ijk}$ and in terms of the deviation of the couplings from their fixed-point value, $\delta s_{i_1...i_n}=s_{i_1...i_n}-\ssf^+_{i_1...i_n}(2\pi)^d\delta\left( \sum_{r=1}^n q_r \right) $, \refeq{wilsonianbeta} reduces to
\begin{align}
&\beta^B_0=\frac{1}{2}\int \frac{d^dq}{(2\pi)^d} \delta s^i(q)\delta s_i(-q)+O(v_{ijk}^2), \nn
&\beta^B_{j}(q)=\left[2\ssf^+_{(j)}(q)+q_{\mu}\frac{\partial}{\partial q_{\mu}}\right] \delta s_{j}(q)\nn
&\hspace{1.5cm}+2\int \frac{d^dq_1}{(2\pi)^d}  \delta s_{ji}(q,q_1)\delta s^i(q_1)+O(v_{ijk}^2),\nn
&\beta^B_{j_1j_2}(q_1,q_2)=\left[\sum_{k=1}^2 \left( 2 \ssf^{+}_{(j_k)}(q_k) + q^{\mu}_k \frac{\partial}{\partial q^{\mu}_k}\right)\right] (\delta s)_{j_1j_2}(q_1,q_2)\nn
&\hspace{2cm}+3 \delta s^i(-q_1-q_2)\ssf^+_{ij_1j_2}(-q_1-q_2,q_1,q_2)+O(v_{ijk}^2),\nn
&\beta^B_{j_1...j_n}(q_1,...,q_n)=O(v_{ijk}^2)\hspace{1cm}n\geq3.
\end{align} 
The boundary couplings representing points in the renormalised manifold are functions of the renormalised couplings and the renormalisation scale. We choose $\mu=1/l$ in the following to simplify the formulas. In the scheme given by holographic renormalisation with solutions parametrised by their asymptotic behaviour, the non-vanishing couplings of the renormalised boundary action to order $O(v_{ijk})$ have the form
\begin{align}
&s_0=- (2\pi)^{d} \frac{v_0}{d}\delta(0)+ \check g_R^0(0)+\int \frac{d^dq}{(2\pi)^d} R^{02}_{(i)}(q) \check g^i_R(q)\check g_{R\,i}(-q) \nn
&\hspace{0.9cm}+\int \frac{d^dq_1 d^dq_2 d^dq_3}{(2\pi)^{2d}} \delta(q_1+q_2+q_3) R^{03}_{;ijk}(q_1,q_2,q_3)\check g^i_R(q_1)\check g^j_R(q_2)\check g_R^k(q_3),\nn
&s_i(q)=R^{11}_{(i)}(q) \check g_{R\,i}(q)+ \int \frac{d^dq_1 d^dq_2}{(2\pi)^d}\delta(q+q_1+q_2) R^{12}_{i;jk}(q;q_1,q_2)\check g^j_R(q_1)\check g^k_R(q_2),\nn
&s_{ij}(q_1,q_2)=R^{20}_{(i)}(q_1)\delta_{ij}(2\pi)^d\delta(q_1+q_2)+\int d^dq\delta(q+q_1+q_2)R^{21}_{ij;k}(q_1,q_2;q) \check g_R^k(q), \nn
&s_{ijk}(q_1,q_2,q_3)=R^{30}_{ijk;}(q_1,q_2,q_3)(2\pi)^d \delta(q_1+q_2+q_3).\nn
\label{wilsoncoup}
\end{align}
The renormalised functions $R^{nm}$ are the finite pieces of the diagrams after subtracting the infinite part. The first superindex, $n$, refers to the number of indices of the corresponding coupling $s_{i_1,\ldots i_n}$  . The second superindex, $m$, indicates the number of subindices to be contracted with the renormalised coupling $g_R$. Once more, an index in brackets is used for diagonal elements of diagonal matrices. Some of these functions have actually been defined above or in the appendices: $R^{11}_{(i)}(q)$ is the eigenperturbation of the boundary action to order $O((v_{ijk})^0)$ in~\refeq{leading_t}, $R^{21}_{ij;k}(q)$ is its correction to $O(v_{ijk})$, whereas $R^{20}_{(i)}=\ssf^{+}_{(i)}$ and $R^{30}_{ijk;}=\ssf^{+}_{(ijk)}$ (the coefficients in the expansion of the boundary action of the standard fixed point).

Let us now compute the renormalised, Gell-Mann-Low beta functions. They are scheme dependent and can be calculated in two ways: from the bare couplings and requiring that the renormalised action be independent of the renormalisation scale. We follow the first method and continue using the same renormalisation scheme. Using \refeq{barebeta} with $h_t^a$ given by the solutions~\refeq{iterativesol} and ~\refeq{h0ren}, we find extremely simple beta functions:
\begin{align}
&\beta^{i}(q)=\Delta_{(i)}^+\check g^i_R(q)+q^{\mu} \frac{\partial}{\partial q^{\mu}} \check g_R^i(q), \\
&\beta^0(q)=q_\mu\frac{\d}{\d q_\mu} \check g_R^0(q).
\end{align}
These are the same as in the UV scheme. Remember that we are always considering generic dimensions. Note that $g^0_R(q)$ is always evaluated at $q=0$ in final expressions. The trivial $\beta^0$ reflects the fact that in this case there are no conformal anomalies. The beta functions are more involved, within the same scheme, in the case of exceptional dimensions, including marginal directions, and $\beta^0$ will be non-trivial due to the conformal anomalies~\cite{Bzowski:2015pba}.

These renormalised beta functions are related by the chain-rule relation~\refeq{betarelation} to the Wilsonian (or boundary) ones, restricted to the renormalised manifold. Since we also have the boundary couplings written in terms of the renormalised ones in \refeq{wilsoncoup}, it is possible to check this relation to $O(v_{ijk})$. The relation for $\beta^B_{j_1j_2}$ at $O(v_{ijk})$ reads
\begin{align}
& \left[\sum_{r=1}^2 \left( 2 \ssf^{+}_{(j_r)}(q_r) + q^{\mu}_{r} \frac{\partial}{\partial q^{\mu}_{ r}}\right)-\Delta_{(i)}^+\right] R^{21}_{j_1j_2;i}(q_1,q_2;-q_1-q_2)\nn
& \mbox{} +3R^{11}_{(i)}(q_1+q_2)R^{30}_{ij_1j_2;}(-q_1-q_2,q_1,q_2) =0 .
\end{align}
Taking into account the relation between the $R$ functions and some objects already defined, this is exactly the equation for the $O(v_{ijk})$-correction to the eigenperturbation $t^i_{j_1j_2}(q_1,q_2)$ at order $O(v_{ijk})$ in~\refeq{eigenpSB}.
The relation for $\beta^B_{j}$ at $O((v_{ijk})^0)$ is
\begin{align}
&\left[2\ssf_{(i)}^{+}(q)-\Delta^+_{(i)}+q_{\mu}\frac{\partial}{\partial q_{\mu}}\right]R^{11}_{(i)}(q)= 0.
\end{align}
This equation is nothing but~\refeq{eqeigenpSB}, the equation for the leading order of the perturbation in $O((v_{ijk})^0)$.
The relation for $\beta^b_0$ to $O((v_{ijk})^0)$ reads
\begin{equation}
\frac{1}{2}R^{11}_{(i)}(q) R^{11}_{(i)}(-q) =\left(2 \nu_{(i)}^--q^{\mu}\frac{\partial}{\partial q^{\mu}} \right) R^{02}_{(i)}(q).
\end{equation}
We have checked that this relation holds using the analytic solutions.
The rest of relations (the $O(v_{ijk})$ order for $\beta^B_{j}$ and for $\beta^B_{0}$) give similar relations between the $R$-functions. 
%

To finish, let us see how the renormalisation scheme in this section is related to the scheme of the field solutions studied in Section~\ref{sec:Fields}. Let us call $\bar{g}_R^i(q)$ the renormalised couplings of the latter scheme.  These are nothing but the solutions at $l$. Writing them in terms of the solution parametrised as in~\refeq{solution} and~\refeq{iterativesol}, we find the following relation between couplings:
\begin{equation}
\bar{g}_R^i(q)=\psi^i_0(q)+\int \frac{d^dq_1 d^dq_2}{(2\pi)^d} \delta(q+q_1+q_2) \psi^j_0(q_1) \psi^k_0(q_2)\Omega^i_{~jk}(q,q_1, q_2) +{O}(v_{ijk}^2),
\end{equation}
where,
\begin{equation}
\psi^i_0(q)= \frac{\Gamma(1-\nu_{(i)})}{2^{\nu_{(i)}}} q^{\nu_{(i)}}I_{-\nu}(q)\check g_R^i(q)=\left[1+\frac{q^2}{4-4\nu_{(i)}} +O(q^4) \right]\check g_R^i(q).
\end{equation} 
$\psi^i_0(q)$ is just a dimensionless version of $\phi_0^i(\epsilon,p)$, see~\refeq{solution}.
The chain rule gives the beta functions in the solution scheme:
\begin{align}
\bar{\beta}^{i}(\bar{g}_R)(q)&=  \left.\frac{\delta W}{\delta \phi_i}\right|_{\phi=\bar{g}_R(q)}+\left[d+q_{\mu} \frac{\partial}{\partial q_{\mu}}\right] \bar{g}_R^i(q) \nn
& =-2 \ssf^{-}_{(i)}(q)\bar{g}_R^i(q) - 3\int \frac{d^dq_1}{(2\pi)^d} \frac{d^dq_2}{(2\pi)^d} \ssf^{-i}_{jk}(q_1,q_2,q)  \bar{g}_R^j(q_1) \bar{g}_R^k(q_2)+{O}(v_{ijk}^2) \nn
& ~~ \mbox{}+\left[d  +q_{\mu} \frac{\partial}{\partial q_{\mu}} \right] \bar{g}_R^i(q).
\end{align} 
We have used the defining equation~\refeq{omegaeq}.


\section{Conclusions}
\label{sec:conclusions} 

In this paper we have developed some details of the Wilsonian holographic RG formalism proposed in Ref.~\cite{Polchinski} and have used this formalism to investigate the large-$N$ Wilsonian structure of renormalised theories dual to field theories in asympotically-AdS spaces. Our main purpose has been to show how the different features of holographic RG flows and renormalisation precisely fit within a standard field-theoretical Wilsonian picture. We have also put to work the general ideas and have obtained a few basic ingredients of the holographic Wilsonian description of renormalised theories. In particular, we have found fixed-points of the flow and the eigenperturbations of these fixed points that diagonalise the RG evolution at the linearised level. We have used two independent methods to achieve this: i) the study of the first-order differential Hamilton-Jacobi equation that dictates the RG evolution, performed in the appendices; and ii) the direct calculation of the renormalised Wilson actions (or rather, of their Legendre conjugates, the renormalised boundary actions), performed to leading order in the cubic interaction in Section~\ref{sec:examples}. The second method, already employed in~\cite{MPV}, directly provides the integrated solutions of the Hamilton-Jacobi equation. It can be used to find the renormalised actions at arbitrarily-low values of the cutoff. 

We have discussed different holographic renormalisation schemes, paying special attention to the scheme in which the renormalised and running couplings are given by particular solutions to the field equations of motion. We have written in detail the bijection between these renormalised couplings and the corresponding renormalised boundary actions. Even if the interpretation of field solutions as running couplings is quite standard, we believe that the connection we find with the Wilsonian couplings is valuable, as it gives a precise meaning to this interpretation.

Our formalism incorporates space-time dependent couplings and derivative terms in the holographic RG evolution. This requires a careful treatment of the dilatation associated to RG transformations. The dilatation is equivalent to measuring lengths with the induced metric at the sliding cutoff position. In AdS space, its isometry ensures that the Hamilton-Jacobi equation written in the position-dependent units is an autonomous differential equation. Then, just as in field theory, the Wilsonian beta functions do not depend on the scale, but only on the couplings. It is for this reason that we have only considered AdS space here. Departures from AdS background would introduce new scales into the problem and preclude our simple usage of dimensional analysis. However, this problem is automatically avoided when dynamical gravity is taken into account, as we comment below. Local couplings are known to lead to conformal anomalies. We have not investigated this issue here, partly because we have sticked to non-exceptional dimensions, but it would be interesting to study how they arise in the Wilsonian context.

A complete treatment of RG flows should in fact include the backreaction on the geometry, i.e.\ should treat the metric as a dynamical field. This is necessary to study realistically the IR of non-trivial renormalised theories,  since the size of relevant deformations increases towards the IR and their impact on the geometry cannot be neglected at arbitrarily low energies. Most of the work on non-Wilsonian holographic RG flows is actually based on complete solutions of the gravity-scalar coupled equations~\cite{Girardello:1998pd,Porrati:1999ew,Freedman:1999gp,Pilch:2000fu,Bianchi:2001de}. A holographic Wilsonian formalism that incorporates dynamical gravity has been sketched by Heemskerk and Polchinski in~Ref.~\cite{Polchinski}. A key point of the proposal is that the boundary action should not satisfy the Hamiltonian constraint. The constraint applies, on the other hand, when the boundary (or Wilson) action is used to calculate the partition function by integration of the IR degrees of freedom.\footnote{The (fake) superpotential in Refs.~\cite{Skenderis:1999mm,DeWolfe:1999cp,Freedman:2003ax} plays a role similar to $W$ in this paper (the symbol we use is no coincidence), generating classes of solutions.}
In this Wilsonian formulation, the treatment of the gauge-fixed metric (or any gauge field) is similar to the one of matter fields. Therefore, we expect that the analysis in this paper will qualitatively apply as well to an exact Wilsonian description with dynamical gravity. Of course, many details, such as the form of the Hamilton-Jacobi equation and the actual fixed-points and eigenperturbations will have to be modified (but the perturbative calculations of scalar correlators should stay the same at leading order in the couplings of the gravity-scalar theory).  A crucial and welcomed new ingredient is the invariance under diffeomorphisms, which implies the absence of absolute scales in the theory. Diffeomorphism invariance will play the role of the AdS isometry in this paper and guarantee that the RG equations are autonomous. Note also that, in contrast to the AdS/CFT correspondence in the continuous limit, the presence of a UV cutoff makes $d$-dimensional gravity dynamical on the field-theory side. This naturally leads to the local RG~\cite{Jack:1990eb,Osborn:1991gm}, which studies the response of the theory to Weyl transformations, rather than just rigid scale transformations.\footnote{See~\cite{Erdmenger:2001ja,Rajagopal:2015lpa} for a non-Wilsonian interpretation of the holographic RG in terms of the local field-theoretical RG.} Note also that to preserve manifest general covariance the metric appears non-linearly in the field theory, as usual in general relativity and in contrast to the couplings of scalar operators. The Legendre transform that relates the boundary and Wilson actions will then have to substituted by a more complicated transform. 

Once the holographic Wilsonian RG with dynamical gravity is developed in detail, we hope that the insights in this paper will be helpful to make more precise the field-theoretical interpretation of the holographic RG flows.

\paragraph{Note added.}

As we were finishing this paper, Ref.~\cite{Bzowski:2015pba} has appeared, with some overlap with our Section~\ref{sec:examples}. This work studies in detail the renormalisation of three-point functions in conformal field theories (without IR cutoff) and includes a sample AdS calculation. The authors focus on the cases of integer, marginal and exceptional conformal dimensions, in which conformal anomalies appear. These cases are orthogonal to the ones with generic dimensions discussed here. Nevertheless, many features, including the presence of non-local subdivergences and their cancellation in (holographic) renormalisation, are qualitatively the same. This is not surprising, as the results with integer, marginal and exceptional dimensions can be obtained from our results by analytic continuation in the dimensions. Some of these issues had also been addressed before, with different methods, in Ref.~\cite{MPV}.

\acknowledgments

It is a pleasure to thank Vijay Balasubramanian for useful discussions and Nick Evans for discussions at a very early stage of this work.
This work has been supported by the Spanish MICINN Projects FPA 2010-17915 and FPA 2013-47836-C3-2-P, 
and by the European Commission through the contract PITN-GA-2012-316704 (HIGGSTOOLS).

\appendix

\section{Fixed points of the Hamilton-Jacobi equation}
\label{sec:AppendixA}

\subsection{Boundary Action}
In this appendix we look for fixed points of the Hamilton-Jacobi evolution for a set of scalar fields $\phi^i$ living in AdS$_{d+1}$ space and subject to a potential $V(\phi)$. The Lagrangian and Hamiltonian densities are given, respectively, by Eqs.~\refeq{scalarL} and~\refeq{scalarH}. We first work in terms of boundary actions. 
As in the rest of the paper, we consider quasilocal actions $S^B$, which can be written as a integral over a function of the field and its derivatives at each point:
\begin{equation}
\begin{gathered}
S^B(g)[\varphi]=\int d^d x \mathcal{S}^B(g(x))\left( \varphi(x),\partial \varphi(x), \partial^2 \varphi(x),... \right).
\end{gathered}
\end{equation} 
This is consistent with the Hamilton-Jacobi evolution. We expand the density $\mathcal{S}^B$ in derivatives:
\begin{align}
\mathcal{S}^B &=\sum_{n=0}^{\infty} \mathcal{S}^{B(n)} \nonumber \\
& =\mathcal{W}^{(0)}(\varphi)+\mathcal{W}_{ij}^{(2)}(\varphi)\partial_{\mu} \varphi^i \partial^{\mu}\varphi^j+... ,
\end{align} 
where $\mathcal{S}^{B(n)}$ is a function in which a total of $n$ derivatives is distributed among the fields $\varphi^i$ and $\mathcal{W}^{(n)}$ depends only on the value of the fields. 

The fixed points $S^B_*$ are solutions of the equation
\beq
\hat{H}[\varphi,\frac{\delta S^B_*[\varphi]}{\delta \varphi} ] = 0,  \label{simplefixedpointeq}
\eeq
with $\hat{H}$ defined in \refeq{Hhat}. For our scalar theory, 
\begin{equation}
\hat{H}[\varphi,\frac{\delta S^B[\varphi]}{\delta \varphi}]= \int d^d x \left\{ \frac{1}{2} \frac{\delta S^B[\varphi]}{\delta \varphi^i} \frac{\delta S^B[\varphi]}{\delta \varphi_i} - \frac{1}{2} \d_\mu \varphi^i \d_\mu \varphi_i -V(\varphi) +\frac{\delta S^B[\varphi]}{\delta \varphi^i} x^{\mu} \partial_{\mu}\varphi^i \right\} .
\label{HJequation}
\end{equation} 
We look for fixed points with constant couplings.
Up to total derivatives, we can write the fixed-point equation at the integrand level:
\begin{equation}
0 = -\frac{1}{2} \frac{\delta S^B_*}{\delta \varphi^i} \frac{\delta S^B_*}{\delta \varphi_i} 
  +V(\varphi)  + \frac{1}{2} \d^\mu \varphi^i \d_\mu \varphi_i 
 +d \mathcal{S}^B_*- \mathcal{N}_\varphi \mathcal{S}^B_* ,
\label{fixedHJexpanded}
\end{equation}
where
\begin{equation}
\frac{\delta S^B_*}{\delta \varphi^i}  = \sum_{n=0}^{\infty} (-1)^n\partial^n \frac{\partial \mathcal{S}^B_*}{\partial [\partial^n \varphi^i]}
\end{equation} 
and we have defined the linear differential operator
\begin{equation}
\mathcal{N}_\varphi=\sum_{n=1}^{\infty}  n  \partial^n \varphi^i  \frac{\partial }{\partial  \left[ \partial^n \varphi_i \right]} .
\end{equation} 
Note that this operator simply counts the number of derivatives of each term in the derivative expansion:
\begin{equation}
\mathcal{N} \mathcal{S}^{B(m)}=  m \mathcal{S}^{B(m)}.
\label{logder}
\end{equation} 
%
We are ready to look for solutions to \refeq{fixedHJexpanded}. As a warm up, we start with the potential approximation (ignoring derivatives), which was also studied in Ref.~\cite{Polchinski}:
\begin{equation}
0=-\frac{1}{2} \left[\d \mathcal{W}_*^{(0)}(\varphi)  \right]^2+d \mathcal{W}_*^{(0)}(\varphi) + V(\varphi),
\label{potaproximation}
\end{equation} 
where $\partial_i=\partial/\partial \varphi^i$. This equation can be written as
%
\begin{equation}
 \left|\d_i \mathcal{W}_*^{(0)}(\varphi) \right|=\sqrt{2\left[V(\varphi)+d \mathcal{W}_*^{(0)}(\varphi)\right]}.  \label{offcritical}
\end{equation} 
Real solutions require 
\begin{equation}
\mathcal{W}_*^{(0)}(\varphi)\geq-\frac{1}{d}V(\varphi).  \label{inequality}
\end{equation}
At the points where this inequality is strict, the solutions will be analytic. Notice that the derivative does not vanish in these points for these solutions. On the other hand, even if the solutions are generically non-analytic at points where the inequality is saturated, we will see that analytic solutions exist about certain points. These are actually the solutions that lead to physically meaningful renormalisable theories. 
\begin{figure}[t!]
\begin{center}
 \includegraphics[width=11cm]{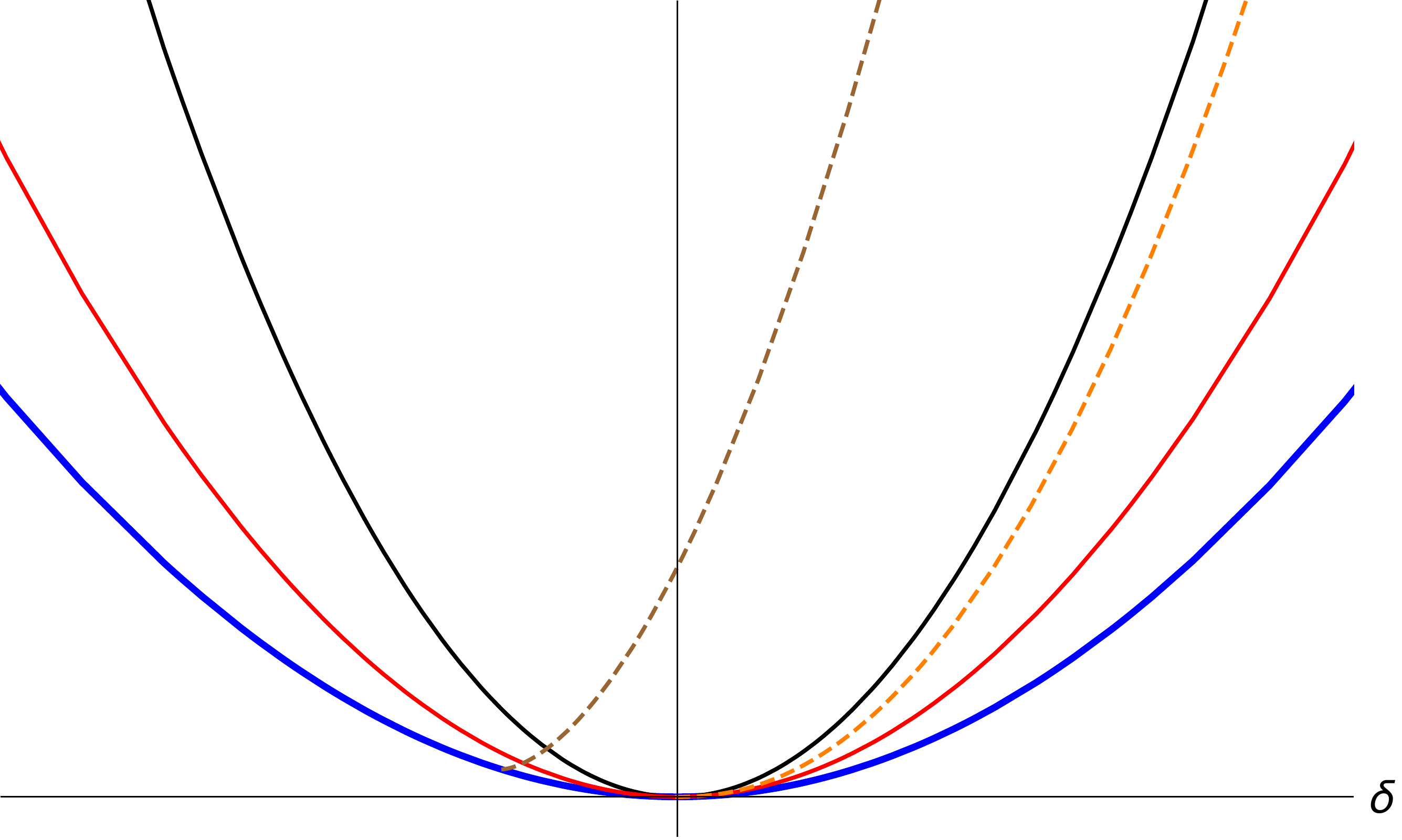}
\caption{Different numerical solutions of the one-dimensional~\refeq{offcritical}. The lowest blue curve corresponds to $-V(\delta)/d$, which gives a lower bound to the solutions. The other two solid curves are the only analytic solutions around the point where their derivative vanishes. From top to bottom, they are associated to the standard (black) and alternate (red) quantisation. The dashed curves are generic non-analytic solutions where their derivative vanishes. From left to right, they correspond to a solution with an asymptotically behaviour $\delta \mathcal{W}^{(0)}_*\sim (\delta-\delta_0)^{3/2}$ (brown curve), and $\delta \mathcal{W}^{(0)}_*\sim \frac{\Delta^-}{2} \delta^2+ w \delta^{d/\Delta^-}$ (orange curve), both around the point where their derivative vanishes.}
\label{exactsolutions}
\end{center}
\end{figure}

Let us look for analytic solutions of \refeq{potaproximation} about some point $\varphi_0$ and work in perturbation theory. We expand $V$ and $\mathcal{W}_*^{(0)}$ in powers of $\delta^i=\varphi^i-\varphi_0^i$,
\begin{align}
V(\varphi) &= v_{i_1...i_n} \delta^{i_1}...\delta^{i_n},\\
\mathcal{W}_*^{(0)}(\varphi) &= w_{i_1...i_n} \delta^{i_1}...\delta^{i_n},
\end{align}
and insert these expansions in (\ref{potaproximation}). Then we get the algebraic equations
\begin{equation}
 w_i w^i=2(v_0+d w_0)  \label{firstordereq}
\end{equation}
and
\begin{align}
& (n+1) w_{j_1...j_n i} w^i \nonumber \\
& =  v_{j_1...j_n}+dw_{j_1...j_n} - \frac{1}{2}\sum_{k=1}^{n-1} (k+1)(n-k+1)w^i_{~(j_1...j_{k}}w_{j_{k+1}...j_{n})i}, ~~~ n\geq 1, \label{higherordereq}
\end{align} 
where the parentheses around indices indicate their symmetrisation. 
If the inequality \refeq{inequality} is strictly satisfied at $\varphi_0$, \refeq{firstorder} has a discrete set of solutions $w_i\neq 0$, and for each of them the tower of equations \refeq{higherordereq} can be iteratively solved. At each order, the mutiplicity of the solutions increases. This is related to the fact that we are solving a non-linear partial differential equation, so the solution is not determined in general by a finite set of integration constants. In fact, as discussed in Section~\ref{sec:holography},  we are interested in power expansions at critical points of the boundary action, with $w_i=0$, which are conjugate to $\pi_i=0$. Both~\refeq{offcritical} and \refeq{firstordereq} show that $w_i=0$ if and only if the inequality~\refeq{inequality} is saturated at $\varphi_0$. The situation is pretty different in this case. Eq.~ \refeq{higherordereq} implies
\begin{align}
&v_i=0\hspace{1cm} (n=1),  \label{criticalpotential} \\  
& 2 w^i_{~(j_1}w_{j_2)i}-d w_{j_1j_2}=v_{j_1j_2} \hspace{1cm} (n=2), \label{w2} 
\end{align} 
Eq.~\refeq{criticalpotential} shows that there is no analytic solution about a point $\varphi_0$ with $\mathcal{W}_*^{(0)}(\varphi_0)=-\frac{1}{d}V(\varphi_0)$ unless $\varphi_0$ is a critical point of the potential. Eq.~\refeq{w2} can be easily solved if we use a base of fields diagonalizing the Hessian matrix $v_{ij}=\frac{1}{2} \delta_{ij} m^2_{(i)}$. If there are $M$ fields, we have $2^M$ solutions (a sign $\pm$ is to be chosen for every field):
\begin{align}
 &w_{ij}=\frac{\Delta^\pm_{(i)}}{2} \delta_{ij},\\
 &\Delta_{(i)}^\pm=\frac{d}{2}\pm \nu_{(i)}=\frac{d}{2}\pm \sqrt{\frac{d^2}{4}+ m_{(i)}^2}.
\end{align}
The remaining equations can then be written as,
\begin{equation}
\left(\sum_{i=1}^n \Delta_{(i)} -d \right) w_{j_1...j_n}=-\frac{1}{2}\sum_{k=2}^{n-2} (k+1)(n-k+1)w^i_{~(j_1...j_{k}}w_{j_{k+1}...j_{n})i}+v_{j_1...j_n} \hspace{0.5cm} n \geq 3.
\label{iterativeeq}
\end{equation} 
They can be solved iteratively. This shows there are exactly $2^M$ analytic solutions about a critical point of both the potential and $\mathcal{W}_*^{(0)}$.\footnote{For exceptional dimensions there is a subtlety: the main coefficient of the equation (\ref{iterativeeq}) may vanish for some $n$ when the alternate solution is taken for some field. The resulting equation has then no solution for generic potentials. However, solutions exist for specific potentials, as in the case of five-dimensional $\mathcal{N}=8$ gauged supergravity. Besides, non-analytic solutions exist that involve expansions with non-integer exponents or logarithms.}  This guarantees a well-defined boundary condition. In Fig.~\ref{exactsolutions} we plot the different kinds of solutions to~\refeq{offcritical}, obtained numerically, in the case of only one active scalar field.
The standard and alternate solutions are the only ones with the property of being analytic in the point where their derivative vanishes.  

Now, let us proceed and study \refeq{fixedHJexpanded} taking into account the (unavoidable) derivative terms. It is convenient to work in momentum space (with dimensionless momenta). The field expansion of a general $S^B$ in momentum space reads
\begin{align}
S^B(s)[\varphi]& = s_0 +\int \frac{d^dq_1}{(2\pi)^d} s_i(q_1)  \check{\varphi}^i(q_1)+ \int \frac{d^dq_1}{(2\pi)^d} \frac{d^dq_2}{(2\pi)^d} s_{i_1i_2}(q_1,q_2) \check{\varphi}^{i_1}(q_1) \check{\varphi}^{i_2}(q_2) \nonumber \\
& \mbox{} + \sum_{n\geq3} \int \frac{d^dq_{i_1}}{(2\pi)^d}...\frac{d^dq_{i_n}}{(2\pi)^d} s_{i_1...i_n}(q_1,...,q_n) \check{\varphi}^{i_1}(q_1)...\check{\varphi}^{i_n}(q_n).
\end{align}
where $\check{\varphi}$ is the Fourier transform of $\varphi$. For the fixed point solution, since the couplings are not space-time dependent, the couplings $s$ simplify, 
\begin{equation}
s_{i_1...i_n}(q_{1}...q_{n})=\ssf_{i_1...i_n}(q_{1}...q_{n}) (2\pi)^d\delta\left(\sum_{r=1}^n q_r\right) 
\end{equation} 
and
\begin{align}
S^B[\varphi]= &(2\pi)^{d} \ssf_0 \delta (0) + \ssf_i \check{\varphi}^i(0)+ \int \frac{d^dq_1 d^dq_2}{(2\pi)^d} \delta(q_1+q_2) \ssf_{i_1i_2}(q_1,q_2) \check{\varphi}^{i_1}(q_1) \check{\varphi}^{i_2}(q_2) \nonumber \\
& \mbox{} + \sum_{n\geq3} \int \frac{d^dq_{i_1}}{(2\pi)^d}...\frac{d^dq_{i_n}}{(2\pi)^d} (2\pi)^d\delta\left( \sum_k q_{k} \right) \ssf_{i_1...i_n}(q_1,...,q_n) \check{\varphi}^{i_1}(q_1)...\check{\varphi}^{i_n}(q_n), \label{SBmomentum}
\end{align}
The functional derivative at a definite momentum is
\begin{align}
&\frac{ \delta S^B [\varphi]}{\delta \check{\varphi}^i(q)}= (2\pi)^{\frac{d}{2}}\delta(0) \ssf_i+2\int \frac{d^dq}{(2\pi)^d} \ssf_{i_1i}(-q,q) \check{\varphi}^{i_1}(-q)\nn
& ~~ \mbox{}+ \sum_{n\geq2} n \int \frac{d^dq_1}{(2\pi)^d}... \frac{d^dq_n}{(2\pi)^d} (2\pi)^d\delta \left( q+\sum_j q_j \right) \ssf_{i_1...i_n i}(q_1,...,q_n,q) \check{\varphi}^{i_1}(q_1)...\check{\varphi}^{i_n}(q_n).
\end{align}
Inserting these expansions in the momentum-space version of (\ref{simplefixedpointeq}) we can write (\ref{HJequation}) perturbatively as
\begin{align}
&\sum_i \ssf_i^2=2 \left( v_0+d\ssf_0\right),\hspace{1cm}(n=0),\\
&0=-\frac{1}{2}\sum_{k=0}^{n} (k+1)(n-k+1) \underset{\{(j_k,q_k)\}_{k=1}^n}{\mathrm{Sym}}\left[ \ssf^i_{~j_1...j_k}(-\scriptstyle \sum\limits_{a=1}^k\displaystyle q_a,q_1,...,q_k) \nonumber \right. \\
& \hspace{5cm} \mbox{} \times \left. \ssf_{j_{k+1}...j_ni}(q_{k+1},...,q_n,-\scriptstyle\sum\limits_{a=k+1}^n\displaystyle q_a)\right] \nonumber \\
& \mbox{} +v_{j_1...j_n}+\left(d- \sum_{r=1}^n q_r^{\mu} \frac{\partial}{\partial q_r^{\mu}}\right)\ssf_{j_1...j_n}(q_1,...,q_n)+\frac{1}{2}\delta_{2n}q_1^2,\hspace{1cm}n\geq1,
\label{eqforsi}
\end{align}
where $\mathrm{Sym}$ symmetrises over the pairs $\{(j_k,q_k)\}_{k=1}^n$:
\begin{equation}
\underset{\{(j_k,q_k)\}_{k=1}^n}{\mathrm{Sym}}  A_{j_1...j_n}[q_1,...,q_n]= \frac{1}{n!}\sum_{\sigma \in S_n} A_{j_{\sigma(1)}...j_{\sigma(n)}}[q_{\sigma(1)},...,q_{\sigma(n)}].
\end{equation} 
The equations \refeq{eqforsi} only apply to on-shell momenta,  $\sum_{i=1}^n q_i=0$, since $\ssf_{j_1\ldots j_n}(q_1,\ldots,q_n)$ is only defined under this condition.\footnote{At first sight, the derivative $\frac{\partial}{\partial q_j^{\mu}}$ may seem to be affected by off-shell momenta. This is not the case, since an off-shell correction, $\ssf_{j_1...j_n} \to \ssf_{j_1...j_n}+f(\sum_{i=1}^n q_i)$ would give a vanishing contribution: $\sum_j \frac{\partial}{\partial q_j^{\mu}} f(\sum_{i=1}^n q_i)=0$ if $\sum_{i=1}^n q_i=0$.}
Guided by our discussion above, let us focus on the analytic solutions at the critical point of the potential ($\ssf_i=0$). The equation for the second-order coefficient $\ssf_{i_1i_2}(q,-q)=\ssf_{i_1i_2}(q)$ is,
\begin{equation}
2\ssf^i_{~(j_1}\ssf_{j_2)i}(q)-v_{j_1j_2}-\frac{q^2}{2}-\left(d-q_{\mu}\frac{\partial}{\partial q_{\mu}}\right)\ssf_{j_1j_2}(q)=0.
\end{equation} 
Working in a field basis with $v_{j_1j_2}=\frac{m_i^2}{2}\delta_{j_1j_2}$, this differential equation is solved by
\begin{equation}
\ssf_{ij}(q)=\ssf_{(i)}(q)\delta_{ij} =\left\{\frac{d}{4}+\frac{1}{2}\frac{q\left[K'_{\nu_{(i)}}(q)+c_i I'_{\nu_{(i)}}(q) \right]}{K_{\nu_{(i)}}(q)+c_i I_{\nu_{(i)}}(q)}\right\} \delta_{ij},
\end{equation} 
with $c_i$ an integration constant. Let us restrict ourselves to solutions that are also analytic in momenta (at zero). For generic $\nu_{(i)} \notin \mathbb{N}$, there are $2^n$ analytic solutions at $q^2=0$, corresponding to two values of $c_i$ for each $i$. If we choose $c_i=\infty$, we find the solution
\begin{align}
\ssf^{+}_{(i)}(q^2) &=\frac{d}{4}+q\frac{I'_{\nu_{(i)}}(q) }{ 2I_{\nu_{(i)}}(q)} \nonumber \\
&=\frac{\Delta^+_{(i)}}{2}+\frac{q^2}{4+4\nu_{(i)}}-\frac{q^4}{16[(1+\nu_{(i)})^2(2+\nu_{(i)})]}+\mathcal{O}(q^6),
  \label{Fsta}
\end{align} 
while $c_i=\frac{\pi \csc (\pi \nu_{(i)})}{2}$ leads to
\begin{align}
\ssf^{-}_{(i)}(q^2)&=\frac{d}{4}+q\frac{I'_{-\nu_{(i)}}(q) }{ 2I_{-\nu_{(i)}}(q)} \nonumber \\
&=\frac{\Delta^-_{(i)}}{2}+\frac{q^2}{4-4\nu_{(i)}}-\frac{q^4}{16[(1-\nu_{(i)})^2(2-\nu_{(i)})]}+\mathcal{O}(q^6).
 \label{Falt}
\end{align} 
These two solutions are related to the standard and alternate quantisation in AdS space, respectively~\cite{Klebanov:1999tb,Breitenlohner:1982bm,Breitenlohner:1982jf}. The fixed-point boundary actions provide a regulated Wilsonian version of the continuous fixed-point theory. The corresponding field theories are non-unitary when $\Delta_{(i)} < d/2-1$, so these solutions seem not admissible in the cutoff version of a unitary theory. On the other hand, there is no problem in using them in the action $W$ of Section~\ref{sec:schemes}, as $W$ is just a means of obtaining a parametrisation. Nevertheless, it should be noticed that in these cases and when the integer part of $\nu_i$ is odd, $\ssf^{-}_{(i)}(q^2)$ diverges at finite values of $q^2$.

 The higher-order equations are
\begin{align}
\left[\sum_{k=1}^n  \right.  & \left. \left( 2 \ssf^{\pm}_{(i_k)}(q_k) + q^{\mu}_k \frac{\partial}{\partial q^{\mu}_k}\right)-d\right]    \ssf_{i_1...i_n}(q_1,...,q_n) \nonumber \\
\mbox{} = & ~ v_{j_1...j_n} - \frac{1}{2}\sum_{k=2}^{n-2} (k+1)(n-k+1) \nonumber \\
& \times\underset{\{(j_k,q_k)\}_{k=1}^n}{\mathrm{Sym}} \ssf^i_{~ j_1...j_k}(-\scriptstyle\sum_{a=1}^k \displaystyle q_a, q_1,..., q_k )\ssf_{j_{k+1}... j_{n}i}(q_1,..., q_k,-\scriptstyle\sum_{a=1}^k q_a \displaystyle), \hspace{0.5cm} n \geq 3.
\end{align} 
This set of equations allows to find recursively all the orders in the expansion. Because it is a first order differential equation in $w_{i_1...i_n}$, there are infinitely many solutions, but only one of them is analytic at $q_i^{\mu}=0$. This can be shown expanding in powers of momenta and noticing than the whole expansion is determined once the second-order term is fixed. We need analytic solutions in momenta in order to get quasilocal Wilson actions with a well-defined derivative expansion.

Summarizing, there is a discrete set of $2^M$ fixed-point boundary actions that are analytic in both fields (at a critical point of the potential) and momenta (at zero). They are characterized by their quadratic terms:
\begin{equation}
\mathcal{S}^B=-\frac{v_0}{d} + \varphi^i \ssf^{\pm}_{(i)} [-\partial^2] \varphi_i + O(\varphi^3),
\end{equation}
with $\ssf^{\pm}_{(i)}$, as given in \refeq{Fsta} and \refeq{Falt}, and the higher-order terms determined by these choices. These fixed points are isolated points in theory space. As we show in Appendix~\ref{sec:AppendixB}, they admit a numerable set of independent eigenperturbations. Thus, they can be used to construct renormalisable theories, as described in the body of this paper. In addition to these special solutions, there are also continuous sets of analytic fixed-point solutions at non-critical points and non-analytic solutions, which we do not analyse further in this paper.

\subsection{Wilson Action}

The fixed-point Wilson actions can be calculated by a Legendre transform of the $S^B_*$ we have found above. Alternatively, it is possible to look directly for solutions of the corresponding fixed-point equation,
\beq
\hat{H}[-\frac{\delta S_*[\pi]}{\delta \pi},\pi] = 0.  \label{Wfixedpoint}
\eeq
This is what we do in the following.
For the scalar theory we are studying, 
\begin{equation}
\hat{H}[\frac{\delta S[\pi]}{\delta \pi},\pi] = -\int d^d x \left\{ \frac{1}{2}\pi_i^2- \frac{1}{2} \d_\mu \frac{\delta S}{\delta \pi_i} \d^\mu\frac{\delta S}{\delta \pi^i} -V\left(- \frac{\delta S}{\delta \pi} \right) +\frac{\delta S}{\delta \pi_i} x^{\mu} \partial_{\mu}\pi_i + d  \pi_i \frac{\delta S}{\delta \pi_i} \right\} .
\end{equation}
Writing $S_*=\int d^d x \mathcal{S}[\pi(x),\partial\pi(x),...]$, and ignoring again total derivatives,
\begin{equation}
0 = \mbox{} - \frac{1}{2} \pi_i^2 +\frac{1}{2} \left(\d^\mu \frac{\delta S_*}{\delta \pi_i}\right)\left( \d_\mu \frac{\delta S_*}{\delta \pi^i}\right)+V \left[ -\frac{\delta S_*}{\delta \pi} \right] 
  +d \mathcal{S}_* - d \pi_i \frac{\delta S_*}{\delta \pi_i}   - \mathcal{N}_\pi {S}_*  ,
\label{HJWexpanded}
\end{equation} 
with
\begin{equation}
\frac{\delta S_*}{\delta \pi_i}= \sum_{n=0}^{\infty} (-1)^n\partial^n \frac{\partial \mathcal{S}}{\partial [\partial^n \pi_i]}.
\end{equation} 

This equation has the same triangular property as (\ref{fixedHJexpanded}), so we could find the solution order by order in a derivative expansion if we wished. However,  as above, we shall see that it is possible to find compact formulas without resorting to such an expansion. To do this, let us work in momentum space and expand in conjugate momenta $\pi$:
\begin{align}
S[\pi]= &(2\pi)^d \delta(0) \tilde{\ssf}_0 +\tilde{\ssf}^i\check{\pi}_i(0)+ \int \frac{d^dq_1 d^dq_2}{(2\pi)^d} \delta(q_1+q_2) \tilde{\ssf}^{i_1i_2}(q_1,q_2) \check{\pi}_{i_1}(q_1) \check{\pi}_{i_2}(q_2) \nonumber \\
& + \mbox{} \sum_{n\geq3} \int \frac{d^dq_{i_1}}{(2\pi)^d}...\frac{d^dq_{i_n}}{(2\pi)^d} (2\pi)^d\delta\left( \sum_k q_{k} \right) \tilde{\ssf}^{i_1...i_n}(q_1,...,q_n) \check{\pi}_{i_1}(q_1)...\check{\pi}_{i_n}(q_n),
\end{align}
Note that this expansion about $\pi=0$ is conjugate to the expansion of $S^B$ at a critical point, with $\ssf_i=0$.
Eq.~(\ref{HJWexpanded}) can be written as the set of equations
\begin{align}
&\sum_{r=1}^n \sum_{\mathcal{P}^n_{r}} (k_1+1)...(n-k_{r-1}+1)  \underset{\{(j_s,q_s)\}_{s=1}^n}{\mathrm{Sym}}[ U_{i_1...i_r}(p_1,...,p_r) 
\tilde{\ssf}^{i_1 j_1...j_{k_1}}(p_1,q_1,...,q_{k_1}) \nn
& \hspace{4cm} \times \tilde{\ssf}^{i_2 j_{k_1+1}...j_{k_2}}(p_2,q_{k_1+1},...,q_{k_2})...
\tilde{\ssf}^{i_r...j_{n}} (p_{r},...,q_{n} ) ]\\
& ~~ \mbox{} =\frac{1}{2} \delta_{2n}+\left[ d(n-1) +  \sum_{s=1}^n q_s^{\mu} \frac{\partial}{q_s^{\mu}} \right]\tilde{\ssf}^{j_1...j_	n}(q_1,...,q_n), ~~~ n\geq0  \nonumber
\label{HJW2equation}
\end{align} 
where,
\begin{align}
& p_s =- \sum\limits_{k_{s-1}<l\leq k_s} q_l,  \\
& U_{i_1...i_r}(p_1,...,p_r)=v_{i_1...i_r}+\delta_{r2}\frac{p_1^2}{2},
\end{align} 
and $\mathcal{P}^n_{r}$ are all the strictly increasing sequences $\left\{ 0< k_1 < k_2 <...< k_{r-1}<k_r\equiv n \right\}$. This equation is to be evaluated only on shell, $\sum_{k=1}^{n}q_k=0$.
The equations at the first two orders, $n=0$ and $n=1$, read
\begin{align}
&\tilde{\ssf}_0=-\frac{v_0}{d},\\
&\sum_j\tilde{\ssf}^{ij}(0,0) v_j =0.\label{twoord}
\end{align}
Based on \refeq{twoord} we can distinguish two classes of solutions. First, the solutions with a singular $\tilde{\ssf}_{ij}$ have a non-analytic Legendre transform at the point $\varphi_0$ conjugate to $\pi=0$. These are conjugate to the non-analytic $S^B_*$ solutions at $\varphi_0$ that we have found above, with $dW_*^{(0)}[\varphi_0]+V[\varphi_0]=0$ and $\partial V(\varphi_0) \neq 0$. 
Second, there are solutions with a non singular Hessian matrix around $\pi=0$ when the potential has a critical point at $\tilde{\ssf}_{i}$, $v_i=\partial_i V(\tilde{\ssf}_{i})=0$. Their Legendre transforms are the special analytic $S^B_*$ with $w_i=0$ found above.  We continue discussing this class of solutions. The equation at order $n=2$ is 
\begin{equation}
\frac{1}{2}\left(m_{(i)}^ 2+q^2\right)\tilde{\ssf}_i^{~(j_1}\tilde{\ssf}^{j_2)i}(q)-\left[  d+q^{\mu}\frac{\partial}{\partial q^{\mu}} \right] \tilde{\ssf}^{j_1j_2}(q)+\frac{1}{2}=0,
\label{secwilhj}
\end{equation} 
where, as above, we are using a base in which the mass matrix is diagonalized $v_{ij}=2m^2_i \delta_{ij}$.
Eq.~ (\ref{secwilhj}) has two solutions $\tilde{\ssf}_{ij}^{\pm}(q)=\tilde{\ssf}^{\pm}_{(i)}(q)\delta_{ij}$ that are analytic at $q=0$, corresponding to the standard ($+$) and alternate ($-$) quantisation:
\begin{align}
\tilde{\ssf}^{\pm}_{(i)}(q) & =-\frac{I_{\pm\nu_{(i)}}(q)}{dI_{\pm\nu_{(i)}}(q)+2q I'_{\pm\nu_{(i)}}(q)} \nonumber \\
& =-\frac{1}{2 \Delta^{\pm}_{i}}+\frac{q^2}{4\Delta^{\pm2}_{i}(1\pm\nu_{(i)})}+\mathcal{O}(q^4).
\end{align}
Finally, the higher orders are given by the iterative equations
\begin{align}
& \left[d+  \sum_{k=1}^n\left[ m_{(i_k)}^2 \tilde{\ssf}^{\pm}_{(i_k)}(q_{k})-d+q^{\mu}_k\frac{\partial}{\partial q^{\mu}_k}\right] \right] \tilde{\ssf}^{i_1...i_n}(q_1,...,q_n)\nn
& ~~~ =\sum_{r=3}^n\sum_{\mathcal{P}^n_{r}} \bigg \{ (k_1+1)...(n-k_{r-1}+1)
 \underset{\{(i_s,q_s)\}_{s=1}^n}{\mathrm{Sym}} \left[  v_{j_1...j_r}
\tilde{\ssf}^{j_1 i_1...i_{k_1}}\left(- \ssum{l=1}{k_1} q_l,q_1,...,q_{k_1}\right)\right.\nn
& ~~~ \left. \left. \mbox{} \times\tilde{\ssf}^{j_2 i_{k_1+1}...i_{k_2}}\left(-\ssum{l=k_1+1}{k_2}q_l,q_{k_1+1},...,q_{k_2}\right)
...\tilde{\ssf}^{j_r...i_{n}} \left(-\ssum{l=k_{r-1}+1}{k_r}q_l,...,q_{n} \right) \right]\right\}, ~~~n\geq 3.
\end{align}
An expansion of $\tilde{\ssf}$ in power series around $q_0$ gives a unique solution, showing that, for each choice of $\{ \ssf_{(i)}^{\pm} \}$  there is exactly one analytic solution (in momenta, at $q_\mu=0$) to the differential equation, to any order $n$.

\section{Eigenperturbations of the fixed points}
\label{sec:AppendixB}

\subsection{Boundary Action}
In this appendix we study small deformations of the special fixed points we have found in Appendix~\ref{sec:AppendixA}. Recall that these fixed points have a boundary action $S_*^B$ that is analytic in fields and in momenta at a critical point $\varphi_0$ of both the potential and $S_*^B$ itself. We want to find the eigenperturbations, i.e. perturbations of the fixed point that, at the linearised order, diagonalize the Hamilton-Jacobi evolution. Consider the density of a perturbed boundary-action,  $\mathcal{S}^*[\varphi(x),\partial \varphi(x),...]+ \delta q(x) \, Q[\varphi(x),\partial \varphi(x),...]$. At order $\delta q$, the Hamilton-Jacobi equation reads
\begin{equation}
t \frac{\d}{\d t} Q_{\la t \ra} [\varphi(x),\partial \varphi(x),...]=\left(\hat{\Psi}-x^{\mu}\partial_{\mu}\right) Q_{\la t \ra} [\varphi(x),\partial \varphi(x),...],
\label{lineareq}
\end{equation} 
where we have defined the differential operator $\hat{\Psi}=\hat{\psi}-\mathcal{N}_\varphi$ with
\begin{equation}
 \hat{\psi}=- \sum_{m=0}^{\infty} \left( \partial^m  \left. \frac{\delta H[\varphi,\pi]}{\delta \pi_i} \right|_{\pi= \frac{S^B_*[\varphi]}{\delta \varphi} } \right) \frac{\partial}{\partial [\partial^m \varphi^i]},
\end{equation} 
As pointed out in Appendix~\ref{sec:AppendixA}, $\mathcal{N}_\varphi$ simply counts the number of derivatives of each term. The action of $\hat{\psi}$ on an arbitrary term of the Taylor expansion of $Q$ is
\begin{align}
&\hat{\psi}  \left[ (\d^{n_1} \varphi^{i_1})( \d^{n_2} \varphi^{i_2}) \cdots (\d^{n_r} \varphi^{i_r}) \right] \nn
& ~~~ \mbox{} =-\sum_{q=1}^r  (\d^{n_1} \varphi^{i_1})( \cdots \d^{n_{q-1}} \varphi^{i_{q-1}}) \left( \d^{n_q} \left. \frac{\partial H[\varphi,\pi]}{\partial \pi_i} \right|_{\pi=\frac{\delta S^B_*[\varphi]}{\delta \varphi}} \right)
(\d^{n_{q+1}} \varphi^{i_{q+1}} )\cdots (\d^{n_r} \varphi^{i_r}) , 
\end{align} 
It is obvious that $\hat{\psi}$ commutes with differentiation. Since $[\mathcal{N}_\varphi,\d^n]= n \d^n$, it follows that
\begin{equation}
\left[\hat{\Psi},\partial^n  \right]=- n \partial^n.
\label{comder}
\end{equation} 
Observe also that the operator $\hat{\Psi}$ satisfies the Leibniz's rule,
\begin{equation}
 \hat{\Psi} \left( Q_1 Q_2 \right)=Q_1 \hat{\Psi} Q_2+Q_2 \hat{\Psi} Q_1.
 \label{LeibnizRule}
\end{equation}
The eigenperturbations are by definition given by
\begin{equation}
\hat{\Psi} Q= - \Lambda Q.
\label{eigenform}
\end{equation}
Let us call $\Lambda$ the dimension of $Q$.
If $Q_1$ and $Q_2$ are eigendirections of $\hat{\Psi}$ with dimensions $\Lambda_1$ and $\Lambda_2$, then $Q_1 Q_2 $ will also be an eigendirection, with dimension $\Lambda_1+\Lambda_2$:
\begin{equation}
\begin{gathered}
 \hat{\Psi} (Q_1 Q_2) = Q_1 \hat{\Psi} Q_2 +  Q_2 \hat{\Psi} Q_1 =-(\Lambda_1+\Lambda_2) Q_1 Q_2. \label{factorisation}
\end{gathered}
\end{equation} 
This feature is dual to the factorisation of dimensions in the large N limit.
Moreover, from (\ref{comder}) we see that if $Q$ is an eigendirection with dimension $\Lambda$, $\partial^n Q$ will be an eigendirection with dimension $\Lambda+n$:
\begin{equation}
\hat{\Psi} \partial^n Q = \partial^n \hat{\Psi} Q - n \partial^n Q= -(\Lambda+n) \partial^n Q.  \label{derivatives}
\end{equation} 
Our strategy will be to find minimal solutions $\mathcal{T}_i$ to the eigenvalue problem, which can be used to construct general eigenperturbations by means of \refeq{factorisation} and \refeq{derivatives}. We make the ansatz
\begin{equation}
\mathcal{T}(x)= \alpha_i \varphi^i(x)+ O(\varphi^2)+O(\partial \varphi).
\end{equation} 
Inserting this expansion in (\ref{eigenform}) we get an iterative expression for the higher orders. 
To find their form, we Fourier transform $\mathcal{T}$ and work in momentum space:
\begin{align}
&\check{\mathcal{T}}(k)=\int \frac{dk}{(2\pi)^{d}} e^{-ikx} \mathcal{T}[\varphi(x),\partial \varphi(x),...],\nn
& \phantom{\check{\mathcal{T}}(k)} =\sum_{n\geq1} \check{\mathcal{T}}^{(n)}(k) \nn
& \phantom{\check{\mathcal{T}}(k)} = \sum_{n\geq1} \int \frac{d^dq_1}{(2\pi)^d}...\frac{d^dq_{n}}{(2\pi)^d}  t _{i_1...i_n}(q_1,...,q_n) \check{\varphi}^{i_1}(q_1)\cdots \check{\varphi}^{i_n}(q_n)(2\pi)^d\delta \left(\sum_{r=1}^n q_{r}-k \right).
\end{align}
Note that $\check{\mathcal{T}}(k)$ is a functional of the fields $\check{\varphi}$. The momentum-space version of the operators $\hat{\psi}$ and $\mathcal{N}$ is
\begin{align}
&\check{\psi} = - \int d^dq \left( \partial^m  \left. \frac{\delta H[\varphi,\pi]}{\delta \check{\pi}_i (q)} \right|_{\check{\pi}= \frac{S^B_*[\varphi]}{\delta \check{\varphi}} } \right)  
   \frac{\delta}{\delta \check{\varphi}^i (q)}   \\
& {\mathcal{N}}_{\check{\varphi}}=d+k^{\mu}\frac{\partial}{\partial k^{\mu}} +\int \frac{d^dq}{(2\pi)^d} \, \check{\varphi}^i(q)  q^\mu  \frac{\d}{\d q^ {\mu}} \frac{\delta}{\delta \check{\varphi}^i(-q)}  
\label{Qmomenta}
\end{align} 
For the scalar theory we are studying,
\begin{align}
&\check{\psi} \check{\mathcal{T}}^{(n)}(k)=-\sum_{m=1}^{n}(m+1)(n-m+1) \underset{\{(j_k,q_k)\}_{k=1}^n}{\mathrm{Sym}} \int  \frac{d^dq_1\dots d^dq_n}{(2\pi)^{nd}} \ssf^i_{~ j_1\ldots j_m}\left(- \scriptstyle \sum\limits_{r=1}^m \displaystyle q_r,q_1,\ldots,q_{m}\right) \nn
&\phantom{\check{\psi} \check{\mathcal{T}}^{(n)}} \mbox{} \times  t _{i j_{m+1}\ldots j_n}\left(\scriptstyle\sum\limits_{r=1}^m\displaystyle q_r,q_{m+1},\ldots,q_{n}\right)(2\pi)^d\delta \left(\scriptstyle \sum \limits_{r=1}^n \displaystyle q_{r}-k \right) \check{\varphi}^{j_1}(q_1)\cdots \check{\varphi}^{j_n}(q_n),\\
& \mathcal{N}_{\check{\varphi}} \check{\mathcal{T}}^{(n)}(k)=\int  \frac{d^dq_1\dots d^dq_n}{(2\pi)^{nd}} \left[ \sum_{r=1}^{n} q_r^{\mu}  \frac{\partial}{\partial q_r^{\mu}}  t _{j_1 \ldots j_n}(q_1, \ldots,q_n) \right]\nn
&\phantom{\mathcal{N}_{\check{\varphi}} \check{\mathcal{T}}^{(n)}(k)=} \times (2\pi)^d \delta \left(\ssum{r=1}{n} q_{r}-k \right)\check{\varphi}^{j_1}(q_1) \cdots \check{\varphi}^{j_n}(q_n).
\end{align}
At the leading order, (\ref{eigenform}) reads
\begin{equation}
\left[ q^{\mu} \frac{\partial}{\partial q^{\mu}}+ 2\ssf_{(i)}^{\pm}(q)-\Lambda\right]t_i (q)=0,\label{eqeigenpSB}
\end{equation} 
A general solution of this equation is
\begin{equation}
t_i(q)=C_i\exp\left[ \int \frac{d^dq^2}{q^2} \left(\frac{\Lambda}{2}- \ssf_{(i)}^{\pm}(q)  \right) \right].
\end{equation} 
Let us restrict ourselves to analytic solutions in momenta around $q=0$. Then we need $\frac{\Lambda}{2}-\ssf_{(i)}^{\pm}(0)\in \mathbb{N}$ $\forall i$, so the eigenvalues form a countable set. For generic field masses with non-exceptional $\Delta_{(i)}$, the only analytic solutions have
\begin{align}
&C_i=C_{(j)} \delta_{ij}\\
&\Lambda=\Delta_{(j)} +2 n_\d, \hspace{0.5cm} n_\d \in \mathbb{N},
\end{align} 
for some $j \in \{1,\ldots M\}.$ Explicitly, the solutions are
\begin{equation}
t^{(j,n_\d)}_i(q)=\delta_{i}^j\frac{q^{\pm \nu_{(j)}+2n_\d}}{\Gamma(1\pm \nu_{(j)})2^{\nu_{(j)}} I_{\pm \nu_{(j)}}(q)}=\delta_{i}^j\left[q^{2n_\d}-\frac{q^{2+2n_{\d}}}{4\pm 4\nu_{(j)}}+O(q^{4+2n_\d})\right]. \label{leading_t}
\end{equation} 
The eigenfunctions with $n_\d>0$ are descendants, which can be obtained from the ones with $n_\d=0$ using \refeq{derivatives}. We will call {\em basic functions\/} the minimal eigenperturbations with leading order given by  \refeq{leading_t} with $n_\d=0$. Their expansion in fields is of the form
\beq
\mathcal{T}^i(x)= \varphi^i(x)+ O(\varphi^2).
\eeq
The dimension of $\mathcal{T}^i$ is $\Delta_{(i)}$. These basic perturbations are in one-to-one correspondence with the fields $\phi^i$, and thereby with the single-trace primary operators of the dual theory.
The higher orders of \refeq{eigenform} are
\begin{align}
 &\left[\sum_{k=1}^n\left( q^{\mu}_k \frac{\partial}{\partial q^{\mu}_k}+ 2\ssf_{(i_k)}^{\pm}(q_{i_k})\right)-\Delta_{(a)}-2n_{\partial}\right]t^a_{i_1...i_n} (q_1,...,q_n) \nn
&~~ \mbox{} =-\sum_{m=2}^{n} \bigg[(m+1)(n-m+1) \times  \underset{\{(j_k,q_k)\}_{k=1}^n}{\mathrm{Sym}}  \ssf^{\pm\,i}_{~~\,\, j_1...j_m}\left(- \ssum{r=1}{m} q_r,q_1,...,q_{m}\right) \nn
&~~~ \phantom{=} \mbox{} \times
t^a _{i j_{m+1}...j_n}\left(\ssum{r=1}{m} q_r,q_{m+1},...,q_{n}\right) \bigg].\label{eigenpSB}
\end{align}
At the zero-momentum order and in the case of only one active field $\phi$, the solutions can be written in a closed form. Indeed, \refeq{eigenform} reduces to
\begin{equation}
\Lambda \mathcal{T}^{(0)}(\varphi) = \mathcal{W}_*^{(0)\,\prime}(\varphi) \mathcal{T}^{(0)\,\prime}(\varphi),
\end{equation}
which is readily solved:
\begin{equation}
 \mathcal{T}^{(0)}(\varphi)=C  \exp \left\{ {\Lambda \int \frac{d\varphi}{\mathcal{W}_*^{(0)\,\prime}(\varphi)}} \right\}.  \label{easybasic}
\end{equation}
Writing $\mathcal{W}_*^{(0)\,\prime}(\varphi)= \Delta \, \varphi \left[1+\varphi Z(\varphi) \right]$, with $Z$ analytic at $\varphi=\varphi_0=0$, we have
\begin{equation}
 \mathcal{T}^{(0)}(\varphi)=C \left[ \varphi e^{\int d\varphi \frac{Z(\varphi)}{1+\varphi Z(\varphi)}} \right]^{\frac{\Lambda}{\Delta}}=C \left[ \varphi +O(\varphi^2) \right]^{\frac{\Lambda}{\Delta}}.
\end{equation} 
This equation shows that the ratio $\Lambda/\Delta$ has to be an integer for $\mathcal{T}^{(0)}$ to be an analytic function of $\varphi$. So the allowed dimensions are $\Lambda=m \Delta$, $m\in \mathbb{N}$, in agreement with the general result above. Theres is only one basic perturbation, given by \refeq{easybasic} with $\Lambda=\Delta$.

\subsection{Wilson Action}
As explained in Section~\ref{sec:holography}, the eigenperturbations of the fixed-point Wilson actions can be obtained by a Legendre transform of the eigenperturbations calculated above for the boundary actions. Here we obtain them directly from the Hamilton-Jacobi equation for the Wilson action. 
At order $\delta g$, the Hamilton-Jacobi equation for a deformation $\mathcal{S}^*[\pi(x),\partial \pi(x),...]+ \delta g\, \mathcal{O}[\pi(x),\partial \pi(x),...]$ is
\begin{equation}
t\frac{\d}{\d t} \mathcal{O}_{\la t \ra}[\pi(x),\partial \pi(x),...]=\left(\tilde{\Psi}-x^{\mu}\frac{\partial}{\partial x^{\mu}}\right) \mathcal{O}_{\la t \ra}[\pi(x),\partial \pi(x),...] ,
\label{Wlineareq}
\end{equation} 
where $\tilde{\Psi}=\tilde{\psi}-\mathcal{N}_\pi- d \mathcal{D}_\pi$, with
\begin{align}
& \tilde{\psi}= \sum_{m=0}^{\infty} \left(\partial^m \left. \frac{\delta H[\varphi,\pi]}{\delta \varphi_i} \right|_{\varphi=-\frac{\delta S_*}{\delta \pi}} \right) \frac{\partial}{\partial [\partial^m \pi_i]},\\
& \mathcal{D}_\pi = \sum_{n=0}^\infty \partial^n \pi_i \frac{\partial }{\partial  \left[ \partial^n \pi_i \right]}.
\end{align}
The set of basic eigenoperators
\begin{equation}
\mathbf{O}_{i}= \pi_i+ O(\pi^2)+O(\partial \pi),
\end{equation} 
satisfying
\beq
\tilde{\Psi} \mathbf{O}_i = -\Lambda \mathbf{O}_i \label{Weigen}
\eeq
can be found working in momentum space:
\begin{align}
\mathcal{O}(k) &=\sum_{n\geq1} \mathcal{O}^{(n)}(k) \nn
& = \sum_{n\geq1} \int \frac{d^dq_1}{(2\pi)^n}...\frac{d^dq_{n}}{(2\pi)^n}  r^{i_1...i_n}(q_1,...,q_n) \pi_{i_1}(q_1)...\pi_{i_n}(q_n)(2\pi)^n\delta \left(\sum_{r=1}^n q_{r}-k \right).
\end{align}
We have
\begin{align}
&\tilde{\psi} \mathcal{O}^{(n)}(k)=\sum_{m=1}^{n}(n-m+1)  \underset{\{(j_k,q_k)\}_{k=1}^n}{\mathrm{Sym}}\int  \frac{d^dq_1}{(2\pi)^n}...\frac{d^dq_{n}}{(2\pi)^n} h_i^{~j_1...j_m}\left(q_1,...,q_{m}\right) \nn
&\phantom{\tilde{\psi} \mathcal{O}^{(n)}} \mbox{} \times  r^{i j_{m+1},...,j_n}\left(\scriptstyle\sum\limits_{r=1}^m\displaystyle q_r,q_{m+1},...,q_{n}\right)(2\pi)^n\delta \left(\scriptstyle \sum \limits_{r=1}^n \displaystyle q_{r}-k \right) \pi_{j_1}(q_1)...\pi_{j_n}(q_n),\\
& \left(\mathcal{N}_\pi + d \mathcal{D}_\pi\right) \mathcal{O}^{(n)}(k)=\int  \frac{d^dq_1}{(2\pi)^n}...\frac{d^dq_{n}}{(2\pi)^n}  \left[\sum_{s=1}^n \left(q_s^{\mu}  \frac{\partial}{\partial q_s^{\mu}}+d\right)    r^{i_1...i_n}(q_1,...,q_n) \right] \nn
& \phantom{\left(\mathcal{N}_\pi + d \mathcal{D}_\pi\right) \mathcal{O}^{(n)}(k)=} \mbox{} \times \pi_{i_1}(q_1)...\pi_{i_n}(q_n)(2\pi)^n\delta \left(\ssum{r=1}{n} q_{r}-k \right),
\end{align}
where $h_i^{~j_1j_2...}$ is defined by
\begin{equation}
\left. \frac{\delta\mathcal{H[\varphi,\pi]}}{\delta \varphi^i} \right|_{\varphi=-\frac{\delta S_*}{\delta \pi}} = \sum_{n\geq1}\int \frac{d^dq_1...d^dq_n}{(2\pi)^{d(n-1)}} h_i^{~j_1...j_n}(q_1,...,q_n)\pi_{j_1}(q_1)...\pi_{j_n}(q_n)\delta\left(p+ \sum_i q_i \right).
\end{equation}
The $h_{i}^{~j}$ term can be readily obtained: $h_{i}^{~j}(q)=h^{\pm}_{(i)}\delta_{i}^j$, with
\begin{equation}
h^{\pm}_{(i)}= 2(q^2+m_i^2)\tilde{\ssf}_{(i)}^{\pm}(q)=-\frac{2(q^2+m_i^2)I_{\pm\nu_{(i)}}(q)}{dI_{\pm\nu_{(i)}}(q)+2q I'_{\pm\nu_{(i)}}(q)}.
\end{equation} 
The first order of (\ref{Weigen}) becomes,
\begin{equation}
\left[ q^{\mu} \frac{\partial}{\partial q^{\mu}}+d-h^{\pm}_{(i)}(q)-\Lambda\right]r^i (q)=0,
\end{equation} 
which is solved by
\begin{equation}
r^i(q^2)=C^i\exp\left[ \int \frac{dq^2}{q^2}\frac{\Lambda-d+h^{\pm}_{(i)}(q)}{2} \right].
\end{equation} 
Again, analyticity at $q=0$ requires $\frac{\Lambda-d-h_{(i)}^{\pm}(0)}{2}=\frac{\Lambda-\Delta_{(i)}}{2}\in \mathbb{N}$ $\forall i$. For non-exceptional dimensions, the only possible analytic solutions have
\begin{align}
& C^i=C_{(j)} \delta_{j}^i\\
& \Lambda=\Delta_{(j)}+2 n_\d, \hspace{0.5cm} n_\d \in \mathbb{N}.
\end{align}
for some $j$.  The solutions read
\begin{align}
r_{(j,n_\d)}^i(q) & =\delta^i_{j}\frac{2 \Delta_{(j)}^\pm q^{\pm \nu_{(j)}+2n_\d}}{\Gamma(1\pm \nu_{(j)})2^{\nu_a} \left[ dI_{\pm \nu_{(j)}}(q) +2 q I_{\pm \nu_{(j)}}'(q)  \right]}\nn
& =\delta^i_j \left[q^{2n_\d}-\frac{\Delta_{(j)}^\pm+2}{\Delta_{(j)}^\pm \left(4\pm 4\nu_{(j)}\right)}q^{2+2n_\d}+O(q^{4+2n_\d})\right].
\end{align} 
The basic operators are the ones with $n_{\partial}=0$, while $n_{\partial}\neq0$ gives rise to their descendants.
The higher orders are
\begin{align}
& \left[\sum_{k=1}^n\left( q^{\mu}_k \frac{\partial}{\partial q^{\mu}_k}+d- 2h^{\pm}_{(i_k)}(q_{i_k})\right)-\Delta^{\pm}_{(a)}-2n \right]r_a^{i_1...i_n} (q_1,...,q_n) \nn
& ~~ \mbox{} = \sum_{m=2}^{n} \bigg[(n-m+1) 
\left.  \underset{\{(j_k,q_k)\}_{k=1}^n}{\mathrm{Sym}} h_j^{~i_1...i_m}\left(- \ssum{l=1}{m} q_l,q_1,...,q_{m}\right) \right. \nn
& \phantom{~~=} \mbox{} \times  r_a^{j i_{m+1}...i_n}\left(\ssum{l=1}{m} q_l,q_{m+1},...,q_{n}\right) \bigg].
 \end{align} 
The case of one single active field can be solved in a closed form in the zero momentum approximation. Eq.~(\ref{Weigen}) reduces to
\begin{equation}
\Lambda \mathcal{O}^{(0)}(\pi)  = \left[ V^\prime(-\mathcal{S}_*'(\pi))+d \pi   \right]  \mathcal{O}'^{(0)}(\pi).
\end{equation}
Its solution is
\begin{align}
\mathcal{O}^{(0)}(\pi) & = C \exp \, \left\{ \Lambda \int \frac{d \pi}{V^\prime[-\mathcal{S}_*'(\pi)]+d \pi   } \right\} \nn
 & = C \exp \, \left\{ - \Lambda \int \frac{d\pi }{\pi} \frac{\delta^2 S_*}{\delta \pi^2} \right\} \nn
 & = C \left[ \pi + O(\pi^2) \right]^{\Lambda/\Delta} \label{simpleO}
\end{align} 
Analyticity at $\pi=0$ requires $\Lambda/\Delta$ to be integer, so the allowed dimensions are $\Lambda=n \Delta$, corresponding to the basic perturbation ($n=1$) and its products.
%
It can be readily checked that \refeq{simpleO} is the perturbative Legendre transform of \refeq{easybasic}.

\section{Eigenperturbations with the method of characteristics}

There is a close relation between the set of basic eigenperturbations and the solutions in the gravity theory, as we have explained in Sections~\ref{sec:holography} and~\ref{sec:schemes} and illustrated in Section~\ref{sec:examples}. This relation is at the core of holographic renormalisation. Interestingly, another relation with solutions arises naturally when the eigenvalue problem is solved by the method of characteristics. 

Consider a solution $\hat{\phi}_t$ to the Hamilton-Jacobi equation, with a fixed point $S_*^B$ as principal Hamilton's function:
\begin{align}
t\frac{\d}{\d t}\hat{\phi}^i_t(x) & =\left. \frac{\delta \hat{H}[\hat{\phi}_t,\hat{\Pi}_t]}{\delta (\hat{\Pi}_t)_i(x)} \right|_{\hat{\Pi}_t=\frac{\delta S_*^B[\hat{\phi}_t]}{\delta \hat{\phi}_t}} \nn
& = \left. \frac{\delta {H}[\hat{\phi}_t,\hat{\Pi}_t]}{\delta (\hat{\Pi}_t)_i(x)} \right|_{\hat{\Pi}_t=\frac{\delta S_*^B[\hat{\phi}_t]}{\delta \hat{\phi}_t}} + x^{\mu}\frac{\partial}{\partial x^\mu} \hat{\phi}^i_t(x).
\label{eqchar}
\end{align} 
These solutions can play the role of characteristic curves of the Hamilton-Jacobi partial differential equation for small deformations of the fixed point $S_*^B$. To see this, consider
any local function of $\varphi$ and its derivatives at $x$, $Q[\varphi|x]= Q(\varphi(x),\partial \varphi(x),\ldots)$. The composition $Q\circ \hat{\phi}_t$ obeys the following equation:
\begin{align}
t\frac{\d}{\d t} Q[\hat{\phi}_t|x] &=
\sum_{n\geq0} \frac{\partial Q[\hat{\phi}_t|x]}{\partial (\partial^n\hat{\phi}^i_t)} t\frac{\d}{\d t}\partial^n\hat{\phi}_t^i(x)\nn
&= \sum_{n\geq0}\frac{\partial Q[\hat{\phi}_t|x]}{\partial (\partial^n\hat{\phi}^i_t)} \left\{ \partial^n \left. \frac{\delta {H}[\hat{\phi}_t,\hat{\Pi}_t]}{\delta (\hat{\Pi}_t)_i(x)} \right|_{\hat{\Pi}_t=\frac{\delta S_*^B[\hat{\phi}_t]}{\delta \hat{\phi}_t}}  +n \partial^n \hat{\phi}^i_t+ x^{\mu}\frac{\partial}{\partial x^{\mu}} \partial^n \hat{\phi}_t^i\right\} \nn
&=-\hat\Psi Q[\hat{\phi}_t|x] +x^{\mu}\frac{\partial}{\partial x^\mu}Q[\hat{\phi}_t|x] .
\label{Qeq}
\end{align} 
Taking (\ref{eigenform}) into account, we see that the eigenperturbations $Q_{\Lambda}$ of the fixed point $\mathcal{S}^B_*$ satisfy 
\begin{equation}
t \frac{\d}{\d t} Q_{\Lambda}[\hat{\phi}_t|x] =\Lambda Q_{\Lambda}[\hat{\phi}_t|x] + x^{\mu}\frac{\partial}{\partial x^\mu} Q_{\Lambda}[\hat{\phi}_t|x] .
\end{equation}
Analogously, we can work with Wilson actions and canonical momenta. Given a solution $\hat{\Pi}^t$ to the equation 
\begin{equation}
t\frac{\d}{\d t}(\hat{\Pi}_t)_i(x)=- \left. \frac{\delta \hat{H}[\hat{\phi}_t,\hat{\Pi}_t]}{\delta \hat{\phi}_t^i(x)} \right|_{\hat{\phi}_t=-\frac{\delta S_*[\hat{\Pi}_t]}{\delta \hat{\Pi}_t}}
+d (\hat{\Pi}_t)_i(x)+ x^{\mu} \frac{\partial}{\partial x^\mu}(\hat{\Pi}_t)_i(x),
\label{eqcharW}
\end{equation} 
the eigenoperators $\mathcal{O}_{\Lambda}$ around the fixed point $S_*$ satisfy
\begin{align}
t\frac{\d}{\d t} \mathcal{O}_{\Lambda}[\hat{\Pi}_t|x] &=- \tilde\Psi \mathcal{O}_{\Lambda}[\hat{\Pi}_t|x] +x^{\mu}\frac{\partial}{\partial x^\mu}   \mathcal{O}_{\Lambda}[\pi_t|x] \nn
& =\Lambda\mathcal{O}_{\Lambda}[\hat{\Pi}_t|x] +x^{\mu}\frac{\partial}{\partial x^\mu}   \mathcal{O}_{\Lambda}[\pi_t|x].
\label{Oeq}
\end{align}
The equations (\ref{eqchar}) and (\ref{eqcharW}) can be solved perturbativally. To do so, notice that (for the special analytic fixed points in Appendix~\ref{sec:AppendixA})
\begin{align}
&\left. \frac{\delta {H}[\varphi,\pi]}{\delta \pi_i(x)} \right|_{\pi=\frac{\delta S_*^B[\varphi]}{\delta \varphi}}=\Delta_{(i)} \varphi^i(x) + O(\varphi^2)+ O(\partial^2\varphi),\\
& \left. \frac{\delta {H}[\varphi,\pi]}{\delta \varphi^i(x)} \right|_{\varphi=-\frac{\delta S_*[\pi]}{\delta \pi}}+d\pi_i=\Delta_{(i)} \pi_i(x) + O(\pi^2)+ O(\partial^2\pi).
\end{align}
So, for $\hat{\phi}_t \approx 0$ and $\partial^n\hat{\phi}_t \approx 0$ (and similarly for $\hat{\Pi}_t$), the solutions of (\ref{eqchar}) and~(\ref{eqcharW}) are approximated by
\begin{align}
&\hat{\phi}^i_t \approx \sigma^{i}_t \equiv C^i(tx) t^{\Delta_{(i)}}, \label{simplesol1}\\
&(\hat{\Pi}_t)_i \approx (\tilde{\sigma}_t)_i \equiv \tilde{C}_i(tx) t^{\Delta_{(i)}}. \label{simplesol2}
\end{align}
The functions $\sigma^{i}_t$ and $\tilde{\sigma}^{i}_t$ are solutions of
\begin{equation}
t\frac{\d}{\d t}\sigma^{i}_t(x)=\Delta_{(i)} \sigma^{i}_t(x) +x^{\mu}\frac{\partial}{\partial x^\mu}  \sigma^{i}_t(x).
\label{sigmaeq}
\end{equation} 
Observe that this first-order equation is, crucially, identical to the equations \refeq{Qeq} and~\refeq{Oeq}, with $\Lambda=\Delta_{(i)}$. Thus, $\sigma^i$ and $\tilde{\sigma}_i$ must be compositions of eigenperturbations of dimension $\Delta_{(i)}$ and solutions.
The exact solutions can be found iteratively and written as 
\begin{align}
&\hat{\phi}^i_t(x)=\sigma^i_t(x)+\mathcal{F}(\sigma_t,\partial \sigma_t,...),\\
& \hat{\Pi}^i_t(x)=\tilde{\sigma}^i_t(x)+\tilde{\mathcal{F}}(\tilde{\sigma}_t,\partial \tilde{\sigma}_t,...),
\end{align}
with the functions
\begin{align}
& \mathcal{F}(\sigma_t,\partial \sigma_t,...)=O(\sigma^2)+O(\partial^2\sigma),
& \tilde{\mathcal{F}}(\tilde{\sigma}_t,\partial \tilde{\sigma}_t,...)=O(\tilde{\sigma}^2)+O(\partial^2\tilde{\sigma})
\end{align} 
capturing the corrections to \refeq{simplesol1} and~\refeq{simplesol2}.
Now, the point is that these functions can be inverted to find $\sigma_i$ and $\tilde{\sigma}_t$ (and thus $Q_{\Delta_{(i)}}$ and $\mathcal{O}_{\Delta_{(i)}}$ composed with the solutions)  as a function of $\hat{\phi}_i$ and $\hat{\Pi}_t$, respectively:
\begin{align}
&\sigma^i_t(x)=\hat{\phi}^i_t(x)+O(\hat{\phi}_t^2)+O(\partial^2\hat{\phi}_t) \label{niceinversion1}\\
&(\tilde{\sigma}_t)_i(x)=(\hat{\Pi}_t)_i(x)+O(\hat{\Pi}_t^2)+O(\partial^2\hat{\Pi}_t). \label{niceinversion2}
\end{align}
Comparing the first order, we see that $\sigma^i$ and $(\tilde{\sigma})$ are equal to basic perturbations composed with solutions:
\begin{align}
&\sigma^i_t(x)=\mathcal{T}^{i}[\hat{\phi}_t|x],\\
&(\tilde{\sigma}_t)_i(x)=\mathbf{O}_{i}[\hat{\Pi}_t|x].
\end{align}
Therefore, \refeq{niceinversion1} and \refeq{niceinversion2} show  that the functional form of the basic eigenperturbations $\mathcal{T}^{i}$ and $\mathbf{O}_{i}$ about a fixed point $S^{B}_*$ is determined by (and can be found from) the solutions generated by $S^{B}_*$.


\end{document}